\documentclass[aps,prx,10pt,amsmath,amssymb,floatfix,superscriptaddress,notitlepage,twocolumn]{revtex4-2}
\usepackage[T1]{fontenc}
\usepackage{hyperref}
\usepackage{graphicx}
\usepackage{dcolumn}
\usepackage{bm}
\usepackage{xcolor}
\usepackage{braket}
\usepackage{amsfonts}
\usepackage{amssymb}
\usepackage[normalem]{ulem}
\usepackage{dsfont}
\usepackage{soul}

\DeclareMathOperator{\csch}{csch}

\begin{document}

\title{Sweet-spot protection of hole spins in sparse arrays via spin-dependent magnetotunneling}

\author{Esteban A. Rodr\'iguez-Mena}
\affiliation{Univ. Grenoble Alpes, CEA, IRIG-MEM-L Sim, Grenoble, France.}
\author{Biel Mart\'inez}%
\affiliation{Univ. Grenoble Alpes, CEA, Leti, F-38000, Grenoble, France.}
\author{Ahmad Fouad Kalo}%
\affiliation{Univ. Grenoble Alpes, CEA, IRIG-MEM-L Sim, Grenoble, France.}
\author{Yann-Michel Niquet}
\affiliation{Univ. Grenoble Alpes, CEA, IRIG-MEM-L Sim, Grenoble, France.}
\author{Jos\'e C. Abadillo-Uriel}
\email{jc.abadillo.uriel@csic.es}
\affiliation{Instituto de Ciencia de Materiales de Madrid (ICMM), Consejo Superior de Investigaciones Científicas (CSIC), Sor Juana Inés de la Cruz 3, 28049 Madrid, Spain}%

\newcommand{\jc}[1]{{\color{green} #1}}
\newcommand{\BM}[1]{{\color{blue} #1}}
\newcommand{\BMC}[1]{{\color{blue} [BM: #1]}}
\newcommand{\WIP}[1]{{\color{cyan} #1}} 

\newcommand{\additions}[1]{{\color{red} #1}} 

\date{\today}

\begin{abstract}
Recent advances in the scaling of spin qubits have led to the development of sparse architectures where spin qubits are distributed across multiple quantum dots. This distributed approach allows for qubit manipulation through hopping and flopping modes and may enable spin shuttling protocols to entangle spins beyond nearest neighbors.
Here, we develop a microscopic theory of a minimal sparse array formed by a hole in a double quantum dot. We show the existence of spin-dependent magnetic corrections to the tunnel couplings that help preserve existing sweet spots, even for quantum dots with different $g$-factors, and introduce new ones that are not accounted for in the simplest models. Our analytical and numerical results explain observed sweet spots in state-of-the-art shuttling and cQED experiments, are relevant to hopping and flopping modes, and apply broadly as corrections to each interdot tunnel link in sparse array encodings of any size.
\end{abstract}

\maketitle

\section{\label{sec:intro}Introduction}
Hole spin qubits have emerged as a promising platform for quantum computation, owing to their intrinsic strong spin-orbit coupling (SOC), which enables efficient electrical manipulation of spins~\cite{maurand2016cmos,watzinger2018germanium,crippa2018electrical,hendrickx2020single,froning2021ultrafast,jirovec2021singlet,wang2022ultrafast, fang2023recent, liles2024singlet}. By electrostatically confining holes in Si and Ge quantum dots, the qubits can be tuned through gate voltages, and achieve excellent manipulability and long coherence times~\cite{piot2022single, hendrickx2024sweet, carballido2024compromise, bassi2024optimal}. Recent experiments have reported successful implementations of high-fidelity two-qubit gates \cite{hendrickx2020fast,geyer2024anisotropic}, multi-qubit processors~\cite{hendrickx2021four,john2024two}, and quantum simulators~\cite{wang2023probing, hsiao2024exciton}. 

In the path towards scaling hole spin qubits, there is an increasing interest in sparse arrays of spin qubits, where devices have more quantum dot sites than encoded qubits~\cite{boter2022spiderweb,wang2024operating,unseld2025baseband,rimbach2024spinless}. These sparse arrays enable shuttling protocols~\cite{sanchez2014long,mills2019shuttling,zwerver2023shuttling,van2024coherent,kunne2024spinbus,langrock2023blueprint,fernandez2024flying,ginzel2024scalable}, flopping-mode qubits with large dipolar couplings~\cite{benito2019electric, mutter2021natural, froning2021strong, hu2023flopping, sen2023classification, stastny2025singlet, teske2023flopping}, and manipulation by hopping spins~\cite{wang2024operating, rimbach2024spinless}. The minimal sparse array consists of a single hole spin delocalized across two quantum dots, which can be operated in two distinct modes: as a hopping spin manipulated via DC or ramped gate voltages, or as a flopping mode spin qubit controlled by AC gate voltages. On the one hand, hopping spins exploit the inherent variability between dots arising from local strains~\cite{liles2021electrical, abadillo2023hole}, interface roughness~\cite{rodriguez2023linear}, and electrostatic potential landscape differences~\cite{martinez2022hole, martinez2025variability}. On the other hand, flopping-mode qubits harness strong SOC to achieve large dipolar coupling allowing their integration with circuit quantum electrodynamics (cQED) systems~\cite{mi2018coherent, samkharadze2018strong, yu2023strong}. Recent experiments in silicon have achieved exceptionally high quality factors in flopping-mode qubits~\cite{noirot2025coherence} and demonstrated the first photon-mediated two-qubit gate with spin qubits~\cite{dijkema2025cavity}, underscoring their potential for scalable quantum information applications.

Despite these advantages, spreading the wavefunction across different quantum dots makes the qubits more susceptible to charge noise, and in particular detuning noise (fluctuations of the relative chemical potentials of the quantum dots). Therefore, it is essential to operate at sweet spot configurations that display minimal charge noise sensitivity~\cite{piot2022single, hendrickx2024sweet, carballido2024compromise, bassi2024optimal, reed2016reduced}. For symmetry reasons, a sweet spot naturally occurs at zero detuning in identical quantum dots; due to reciprocal sweetness~\cite{michal2023tunable}, this is also where the dipolar coupling is maximum, and is thus the optimal working point for flopping-mode qubits~\cite{yu2023strong}. However, this sweet spot is expected to disappear in sufficiently different quantum dots, thus limiting coherence and operational fidelity. The lack of charge-noise sweet spots in detuning space limits coherence in shuttling and hopping experiments as well~\cite{van2024coherent,wang2024operating}. Understanding and engineering sweet spots is thus essential to optimize the operation of sparse hole spin qubit arrays.

Interestingly, recent experiments with flopping-mode qubits in Si and spin shuttling in Ge have explored regimes where the differences between the gyromagnetic $g$-factors of quantum dots are sufficiently large so that the conventional theory predicts that no sweet spots should exist~\cite{yu2023strong,van2024coherent}. However, sweet spots were unexpectedly preserved in some cases, highlighting behaviors of $g$-factors beyond current theoretical understanding. Motivated by these observations, we develop a microscopic theory for single spins in minimal sparse arrays formed by double quantum dots. While microscopic details such as quantum dot shape~\cite{michal2021longitudinal, bosco2021squeezed,martinez2022hole}, inhomogeneous strains~\cite{liles2021electrical, abadillo2023hole, wang2024electrical}, interface roughness~\cite{martinez2025variability}, and local material variations~\cite{rodriguez2023linear} are known mechanisms that influence the detuning dependence of the different observables, we highlight that an unexplored spin-dependent magnetic correction to the tunnel coupling dominates the detuning correction to $g$-factors. We systematically explore how these microscopic effects influence key operational characteristics, focusing particularly on sweet spot robustness. We find that the spin-dependent magnetotunneling terms play a crucial role in preserving sweet spots even when quantum dots exhibit different $g$-factors, consistent with experimental observations~\cite{yu2023strong, van2024coherent}. Furthermore, we demonstrate that sweet spots can be made high-order in detuning, potentially enhancing coherence properties for all types of sparse array operations.

We first introduce in Sec.~\ref{sec:theory} the effective double-dot Hamiltonian that incorporates spin-dependent magnetotunneling corrections, outlining its derivation from our microscopic theoretical framework, and emphasizing the main contributions to the $g$-factors. In Sec.~\ref{sec:simulations}, we present realistic numerical simulations of the detuning dependence of the $g$-factors in a minimal sparse Ge array explicitly capturing these microscopic effects. Section~\ref{sec:sweet} quantifies the impact of the additional terms on relevant observables, such as the robustness of operational sweet spots against the variability of coupled quantum dots, thereby providing guidance for enhancing qubit stability. Finally, in Sec.~\ref{sec:discussion}, we discuss the broader implications of our results, highlighting strategies to improve both the robustness and scalability of hole spin qubits.

\begin{figure}
\includegraphics[width=1\columnwidth]{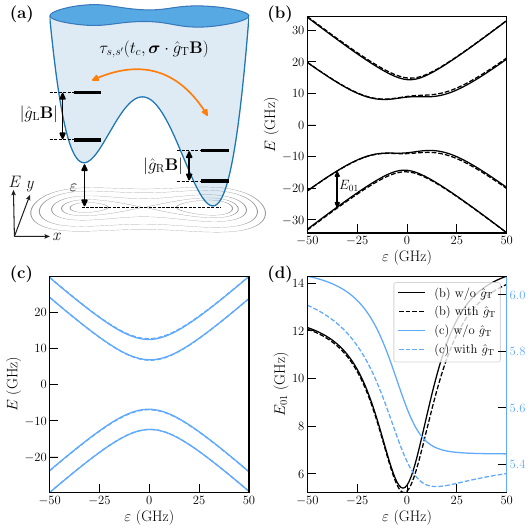}
\caption{Energy levels of a hole spin in a DQD. (a) Illustrative representation of the in-plane confinement potential and energy levels in a quasi-2D DQD strongly confined along $z$. The two dots, aligned along the $x$ axis, are detuned by the energy $\varepsilon$. Each dot is characterized by its respective $g$-matrix $\hat{g}_\mathrm{L,R}$. Tunneling between the two dots from spin $s$ to spin $s'$ ($\tau_{s,s'}$) acquires a magnetic dependence through the $g$-matrix $\hat{g}_\mathrm{T}$ (we take $\boldsymbol\mu_\mathrm{T}=\mathbf{0}$ for illustration). (b,c) Energy levels as a function of detuning for a case with large spin-charge mixing (b) and low spin-charge mixing (c). The parameters $t_c=9.6$ GHz, $t_0=1.12$ GHz, $\mathbf{t}_\mathrm{so}=(3.2,8.9,1.0)$ GHz are taken from the experiment in Ref.~\cite{yu2023strong}, $\hat{g}_\mathrm{L}$ and $\hat{g}_\mathrm{R}$ are explicitly given in ~\cite{pseudofootnote},
$\mathbf{B}=(0.78,0,0)$ T in (b), and $\mathbf{B}=(0.11,0.25,0)$ T in (c). (d) Energy splitting $E_{01}$ between the lowest two spin states as a function of detuning for (b) (solid black line) and (c) (solid blue line). In dashed lines, we show the corrected energies after the inclusion of the magnetotunneling mechanism with an isotropic $g$-matrix $\hat{g}_\mathrm{T}=0.2\mathds{1}$, recovering a sweet spot for the low spin-charge mixing case.}
\label{fig:intro}
\end{figure}

\section{\label{sec:theory}Theory of single-hole spins in a minimal array}
We start by outlining the basic low-energy physics of a double quantum dot (DQD) with a single spin, including the magnetotunneling correction, see Fig.~\ref{fig:intro}(a). This minimal array can be operated as a flopping mode spin qubit or as a hopping spin. We focus on hole spins and refer to Appendix~\ref{appendix:microscopic} for details on the derivation of the Hamiltonian from a microscopic theory, but we remark that our findings can be extended to electron spins with micro-magnets (Appendix~\ref{app:electron}).

A single hole spin within a DQD may be localized in the left ($\ket{L}$) and right ($\ket{R}$) quantum dot orbitals. The low-energy physics can be understood in terms of a four-level system with states $\{\ket{L\uparrow},\ket{L\downarrow},\ket{R\uparrow},\ket{R\downarrow}\}$, where the arrow indicates the pseudospin orientation \footnote{We assume here $\mathcal{T}\ket\uparrow=-\ket{\downarrow}$ and $\mathcal{T}\ket\downarrow=\ket{\uparrow}$, where $\mathcal{T}$ is the time-reversal symmetry operator.}. The difference in chemical potential between the two dots is characterized by the detuning energy $\varepsilon$, as shown in Fig.~\ref{fig:intro}(a), while the barrier between the two dots mainly controls the tunnel coupling $t_c$. By changing the ratio $\varepsilon/t_c$, it is possible to go from a single-dot regime $|\varepsilon|\gg |t_c|$, usual for hopping spins, to the flopping mode regime $|t_c|\gg |\varepsilon|$ where the $\ket{L}$ and $\ket{R}$ orbitals mix into bonding and antibonding states.

The magnetic response of the spin is characterized by the $g$-matrix $\hat{g}_{\mathrm{L},\mathrm{R}}$ of each dot. Furthermore, the SOC in the valence band of group IV semiconductors can mediate spin-flip tunneling between the two dots. There may also generally be magnetic field-dependent corrections to the tunneling Hamiltonian. We name these corrections, which have not been accounted for previously, \emph{spin-dependent magnetotunneling terms} in the following. With $\boldsymbol{\sigma}$ and $\boldsymbol{\tau}$ the Pauli matrices of the operators acting on the spin and orbital subspaces respectively, the most general effective Hamiltonian compatible with time-reversal symmetry constraints~\footnote{We assume that the Hamiltonian transforms under the time-reversal symmetry as $\mathcal{T}^{-1}H_\text{eff}(\mathbf{B})\mathcal{T}=H_\text{eff}(\mathbf{-B})$, which implies that the vector potential $\mathbf{A}$ must be odd with respect to the magnetic field $\mathbf{B}$ [$\mathbf{A}(\mathbf{r}, -\mathbf{B})=-\mathbf{A}(\mathbf{r}, \mathbf{B})$]. This is the case for all usual gauge choices (symmetric, Landau, ...). We emphasize that this constraint on the gauge simplifies the shape of Eq.~\eqref{eq:eff0} with no observable effect.} is, to first-order in the magnetic field:
\begin{equation}
\begin{aligned}
    H_{\text{eff}}&=\frac{\varepsilon}{2}\tau_z+t_0\tau_x+\tau_y(\mathbf{t}_{\mathrm{so}}\cdot\boldsymbol{\sigma})\\ &+\frac{\mu_B}{2}\tau_\mathrm{L}(\boldsymbol{\sigma}\cdot\hat{g}_\mathrm{L}\mathbf{B})+\frac{\mu_B}{2}\tau_\mathrm{R}(\boldsymbol{\sigma}\cdot\hat{g}_\mathrm{R}\mathbf{B})+H_\mathrm{MT},\\
    H_\mathrm{MT}&\equiv\frac{\mu_B}{2}\tau_x(\boldsymbol{\sigma}\cdot\hat{g}_\mathrm{T}\mathbf{B})+\frac{1}{2}\tau_y(\boldsymbol{\mu}_\mathrm{T}\cdot\mathbf{B}).
    \end{aligned}
    \label{eq:eff0}
\end{equation}
Here $\tau_{\mathrm{L},\mathrm{R}}=(\mathds{1}\pm\tau_z)/2$ and $\mathbf{B}=B\mathbf{b}$ is the external magnetic field with magnitude $B$ along the unit vector $\mathbf{b}$. The tunnel Hamiltonian features a spin-independent, $\propto t_0$ part and a spin-dependent, $\propto\mathbf{t}_{\mathrm{so}}\cdot\boldsymbol{\sigma}$ part due to spin-orbit coupling. The latter can be characterized by its strength $|\mathbf{t}_{\mathrm{so}}|$ and by a unit spin-orbit vector $\mathbf{n}_{\mathrm{so}}=\mathbf{t}_{\mathrm{so}}/|\mathbf{t}_{\mathrm{so}}|$. At zero magnetic field, the tunnel coupling splits the orbitals into bonding and antibonding branches separated by $\Delta=2t_c$ at $\varepsilon=0$, with $t_c\equiv\sqrt{t_0^2+|\mathbf{t}_{\mathrm{so}}|^2}$. The ratio $|\mathbf{t}_{\mathrm{so}}|/t_0$ is expected to be very small for strained Ge dots due to the large heavy-hole/light-hole splitting, but has been measured large in Si nanowires~\cite{froning2021strong,yu2023strong}. Finally, we have introduced the magnetotunneling term $H_\mathrm{MT}$ that describes magnetic field and spin-dependent tunneling events through the $g$-matrix $\hat{g}_\mathrm{T}$ and the vector $\boldsymbol{\mu}_\mathrm{T}$~\footnote{Microscopically, $\boldsymbol{\mu}_\mathrm{T}$ may arise as a Peierls phase $t_c\rightarrow t_c\exp(i\frac{e}{\hbar}\int^{\mathbf{x}_1}_{\mathbf{x}_0} \mathbf{A}\cdot d\mathbf{r})$ expanded up to first order in $\mathbf{B}$ in our linear-response theory, where $\mathbf{A}$ is the vector potential, see Appendix~\ref{appendix:microscopic}}.

The spin-dependent tunneling $\mathbf{t}_\mathrm{so}$ and the site-dependent $g$-matrices are well-known mechanisms that hybridize the spin and orbital degrees of freedom at finite magnetic field. As a consequence of the spin-charge hybridization, the spectrum gets renormalized, particularly near zero detuning. This is evidenced in Fig.~\ref{fig:intro}(b, c), where the different energy levels are mainly determined by the $\hat{g}$ matrices of each dot for large positive and negative detuning values. When there is a large spin-charge hybridization mechanism, as in Fig.~\ref{fig:intro}(b), the lower and upper branches of spin states interact with one another, leading to a strong renormalization of the spectrum, heavily contrasting with low spin-charge hybridization cases, as in Fig.~\ref{fig:intro}(c). The magnetotunneling terms $\hat{g}_\mathrm{T}$ and $\boldsymbol{\mu}_\mathrm{T}$ provide an additional mechanism for coupling spin and orbital degrees of freedom. They give rise to a renormalization of the effective $g$-factors near zero detuning that promotes the existence of sweet spots. This is illustrated in Fig.~\ref{fig:intro}(d), where we plot the splitting $E_{01}$ of the lowest Kramers pair for the cases in Fig.~\ref{fig:intro}(b) and (c). The solid lines are calculated without magnetotunneling corrections, while the dashed lines include the magnetotunneling term with a diagonal $g$-matrix $\hat{g}_\mathrm{T}=0.2\mathds{1}$. The magnetotunneling term introduces a sweet spot near zero detuning in the low spin-charge hybridization case of Fig.~\ref{fig:intro}(d).

To highlight the effect of $\hat{g}_\mathrm{T}$ and $\boldsymbol{\mu}_\mathrm{T}$ on the qubit spectrum near zero detuning, we go to the spin-orbit frame that lumps the spin-dependent tunneling terms into the $g$-matrices of the L and R dots~\cite{sen2023classification,geyer2024anisotropic}. For that purpose, we apply the unitary transform $\hat{U}=\exp\left[\mathrm{i}\tau_z\theta_{\mathrm{so}}\sigma_{\mathrm{so}}/2\right]$ to Eq.~\eqref{eq:eff0}, where $\sigma_{\mathrm{so}}={\mathbf{n}}_{\mathrm{so}}\cdot\boldsymbol{\sigma}$, and $\theta_{\mathrm{so}}=\arctan(|\mathbf{t}_{\mathrm{so}}|/t_0)$. This transformation gauges away the spin-dependent tunneling term and rotates the $\hat{g}$ matrices, which yields
\begin{equation}
\begin{aligned}
    H^{\text{eff}}_{\text{so}}&=\hat{U}H^{\text{eff}}\hat{U}^\dagger=\frac{\varepsilon}{2}\tau_z+t_c\tau_x+\frac{\mu_B}{2}\tau_\mathrm{L}(\boldsymbol{\sigma}\cdot\tilde{g}_\mathrm{L}\mathbf{B})\\ &+\frac{\mu_B}{2}\tau_\mathrm{R}(\boldsymbol{\sigma}\cdot\tilde{g}_\mathrm{R}\mathbf{B})+\frac{\mu_B}{2}\tau_x(\boldsymbol{\sigma}\cdot\tilde{g}_\mathrm{T}\mathbf{B})+\frac{1}{2}\tau_y(\tilde{\boldsymbol{\mu}}_\mathrm{T}\cdot\mathbf{B}),
    \end{aligned}
    \label{eq:effso}
\end{equation}
where $\tilde{g}_{\mathrm{L},\mathrm{R}}$ are related to the original matrices $\hat{g}_{\mathrm{L},\mathrm{R}}$ by rotations about the spin-orbit vector $\mathbf{\hat{n}}_{\mathrm{so}}$ with angles $\pm\theta_{\mathrm{so}}$. As an example of this rotation, for diagonal $\hat{g}$ matrices and $\mathbf{n}_\text{so}=\mathbf{e}_y$ (the unit vector along $y$), the localized $\tilde{g}$ matrices are:
\begin{equation}
    \tilde{g}_\mathrm{L/R}=\begin{pmatrix}
        \hat{g}_{\mathrm{L/R},xx}\cos\theta_\text{so} & 0 & \pm\hat{g}_{\mathrm{L/R},zz}\sin\theta_\text{so} \\
        0 & \hat{g}_{yy} & 0 \\
        \mp \hat{g}_{\mathrm{L/R},xx}\sin\theta_\text{so} & 0 & \hat{g}_{\mathrm{L/R},zz}\cos\theta_\text{so}
    \end{pmatrix}.
    \label{eq:rotatedg}
\end{equation}
Similarly, in the spin-orbit frame the magnetotunneling term $H_\text{MT}$ is modified and the spin-orbit phase is gauged into the $\tilde{g}_\mathrm{T}$ matrix $\tilde{g}_\mathrm{T}=\left[\mathds{1}-(1-\cos\theta_\mathrm{so})\mathbf{n}_\mathrm{so}\otimes \mathbf{n}_\mathrm{so}\right]\hat{g}_\mathrm{T}+\sin\theta_\mathrm{so}\mathbf{n}_\mathrm{so}\otimes\boldsymbol{\mu}_\mathrm{T}/\mu_B$ and with $\tilde{\boldsymbol{\mu}}_\mathrm{T}=\cos\theta_\mathrm{so}\boldsymbol{\mu}_\mathrm{T}-\mu_B\sin\theta_\mathrm{so}(^t\hat{g}_\mathrm{T} \mathbf{n}_\text{so})$, where $\otimes$ denotes the outer product. 

For a given external magnetic field $\mathbf{B}$, we define the Larmor vectors of each dot, $\boldsymbol{\omega}_{01}(\varepsilon\rightarrow-\infty)=\tilde{g}_\mathrm{L}\mathbf{B}$ and $\boldsymbol{\omega}_{01}(\varepsilon\rightarrow+\infty)=\tilde{g}_\mathrm{R}\mathbf{B}$. Without the magnetotunneling term, the Larmor vector at zero detuning is the average $\boldsymbol{\omega}_{\mathrm{01}}(\varepsilon=0)=(\tilde{g}_\mathrm{R}+\tilde{g}_\mathrm{L})\mathbf{B}/2$. From the triangle inequality, we deduce that
\begin{equation}
 |\boldsymbol{\omega}_{\mathrm{01}}(\varepsilon=0)|\leq\frac{1}{2}\left(|\boldsymbol{\omega}_{01}(\varepsilon\rightarrow-\infty)|+|\boldsymbol{\omega}_{01}(\varepsilon\rightarrow+\infty)|\right).   
 \label{eq:triangular}
\end{equation}
We conclude from this equation that the effective $g$-factor $g^{(\mathbf{b})}=|\tilde{g}\mathbf{b}|$ of the ground Kramers doublet satisfies $g^{(\mathbf{b})}(\varepsilon\rightarrow0)\leq (g_\mathrm{L}^{(\mathbf{b})}+g_\mathrm{R}^{(\mathbf{b})})/2$ for any magnetic field orientation $\mathbf{b}$. This inequality, however, breaks down once we introduce the magnetotunneling term, which results in:
\begin{equation}
 \boldsymbol{\omega}_{01}(\varepsilon=0)=(\tilde{g}_\mathrm{R}/2+\tilde{g}_\mathrm{L}/2-\tilde{g}_\mathrm{T})\mathbf{B}. 
\end{equation}
This underlines that $\tilde{g}_\mathrm{T}$ is particularly relevant near zero detuning where it unbounds the spectrum from the triangular inequality. As shown in Fig.~\ref{fig:intro}(d), this term influences the presence of a sweet spot or, even, the emergence of multiple sweet spots as we discuss in the next sections. 

We emphasize that $\tilde{\boldsymbol{\mu}}_\mathrm{T}$ has no effect on $\boldsymbol{\omega}_{01}$. This conclusion, which will be generalized to arbitrary detunings in section \ref{sec:sweet}, only holds in the spin-orbit frame (the original $\boldsymbol{\mu}_\mathrm{T}$ indeed contributes to $\boldsymbol{\omega}_{01}$, as shown by the expression of $\tilde{g}_T$). Furthermore, we discuss gauge invariance of the effective $g$-factors in Appendix~\ref{app:gauge}.

\begin{figure*}
\includegraphics[width=2\columnwidth]{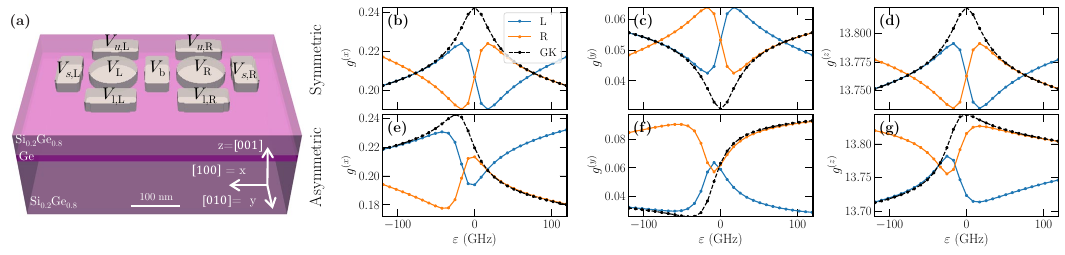}
\caption{Simulated $g$-factors of Ge DQD devices. (a) Top view of the simulated Germanium DQD device. $V_\mathrm{L}$ and $V_\mathrm{R}$ are the plunger gate voltages of left and right dots, while $V_{b}$ is the barrier gate voltage. The other side gates can be used to tune the asymmetry of the potential landscape. (b-d) Effective $g$-factors $g^{(\mathbf{b})}$ as a function of detuning for $\mathbf{b}=\mathbf{e}_x$ (b), $\mathbf{b}=\mathbf{e}_y$ (c), and $\mathbf{b}=\mathbf{e}_z$ (d), where $\mathbf{e}_j$ is the unit vector along the $j$ direction, in a symmetric configuration with $V_\mathrm{L}=V_\mathrm{R}=-50$ mV and $t_c=9.6$ GHz at $\varepsilon=0$. The black lines are the effective $g$-factors of the ground Kramers (GK) pair, while the blue and orange lines are those of the Wannierized left and right orbitals, respectively. Equivalently, (e-g) are the effective $g$-factors in an asymmetric configuration with $V_\mathrm{L}=-56.9$ mV, $V_\mathrm{R}=-43.1$ mV, $t_c=9.6$ GHz at zero detuning, and the top and bottom gates around the left plunger squeezing and displacing the dots ($V_{\mathrm{u,L}}=15$ mV and $V_{\mathrm{l,L}}=12$ mV).}
\label{fig:simexamples}
\end{figure*}

The elements of the spin-dependent magnetotunneling $g$-matrix $\hat{g}_\mathrm{T}$ can be estimated from the microscopic details of the quantum dots. We derive analytical expressions with the Luttinger-Kohn Hamiltonian, by assuming left and right-localized ansatz wavefunctions in a DQD confinement potential and using a Schrieffer-Wolff transformation to integrate out the light-hole states~\cite{winkler2001spin}. The details of this calculation are given in Appendix~\ref{appendix:microscopic}. The resulting non-zero magnetotunneling $g$-matrix elements of a heavy-hole DQD are given by
\begin{equation}
    \begin{aligned}
        \hat{g}_{\mathrm{T},xx}&=-\frac{4\gamma_2\left(\beta\gamma_3+2\kappa\right)m^\parallel_{h}}{m_0\Delta_{\mathrm{LH}}}t_c \\
       \hat{g}_{\mathrm{T},yy}&=\frac{8\left(2\gamma_3^2L_x^2-\gamma_2\left(\beta\gamma_3+2\kappa\right)L_y^2\right)m^\parallel_{h}}{m_0L_y^2\Delta_{\mathrm{LH}}}t_c \\
        \hat{g}_{\mathrm{T},zz}&=-\frac{64\gamma_3(\gamma_2+2\eta_R^2\gamma_3L_y^2)L_x^2m^\parallel_{h}}{3m_0L_y^2\Delta_{\mathrm{LH}}}t_c.
    \end{aligned}
    \label{eq:gt}
\end{equation}
They are thus proportional to the tunnel coupling $t_c$, and depend on the harmonic lengths $L_x$, $L_y$ of the quantum dots, the heavy-hole light-hole splitting $\Delta_{\mathrm{LH}}$, the heavy-hole effective mass $m^\parallel_{h}\approx m_0/(\gamma_1+2\gamma_2)$, the Luttinger parameters $\gamma_{1,2,3}$, the isotropic Zeeman parameter $\kappa$, and the Rashba parameters $\eta_R$, $\beta$ (see Appendix~\ref{appendix:microscopic} for specific definitions). The linear dependence on $t_c$ highlights the kinetic nature of the magnetotunneling term and the scaling with $1/\Delta_{\mathrm{LH}}$ shows that it originates from an interplay between heavy-hole and light-hole states. Corrections to the $g$-matrix $\hat{g}_\mathrm{T}$ as well as to the localized $g$-matrices $\hat{g}_{\mathrm{L},\mathrm{R}}$ due to strains are given in Appendix~\ref{app:strain}. The nature of $\boldsymbol{\mu}_\mathrm{T}$ is discussed in Appendix~\ref{appendix:microscopic}.

\begin{figure*}
\includegraphics[width=1.9\columnwidth]{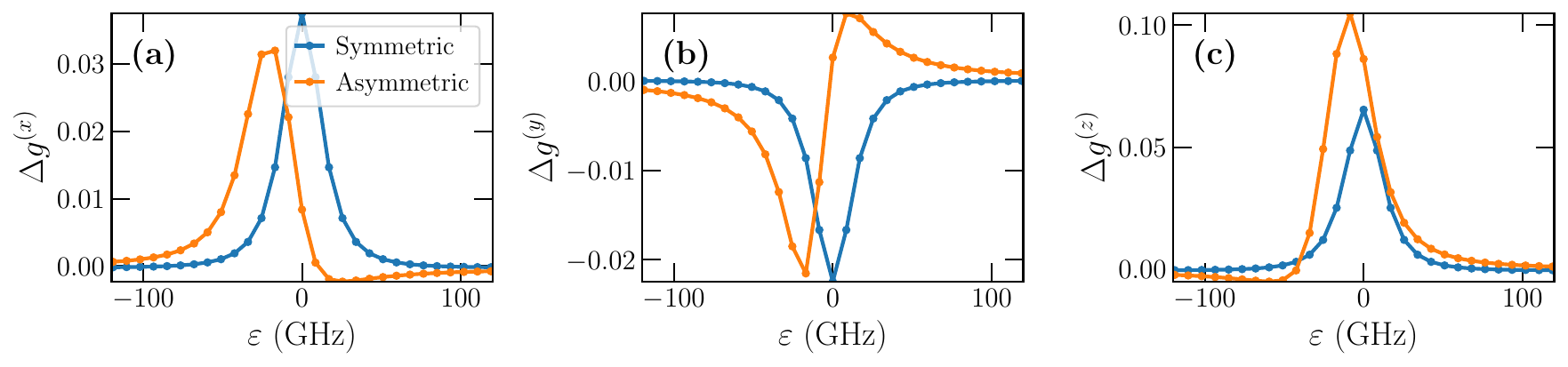}
\caption{Difference $\Delta g^{(\mathbf{b})}=g^{(\mathbf{b})}-g^{(\mathbf{b})}_\text{eff}[\tilde{g}_\mathrm{T}=0]$ between the numerical $g$-factor $g^{(\mathbf{b})}$ of Fig.~\ref{fig:simexamples} and the  effective model of Eq.~\eqref{eq:effso} without magnetotunneling terms, $g^{(\mathbf{b})}_\text{eff}[\tilde{g}_\mathrm{T}=0]$. $\Delta g^{(\mathbf{b})}$ is plotted as a function of detuning for (a) $\mathbf{b}=\mathbf{e}_x$, (b) $\mathbf{b}=\mathbf{e}_y$, and (c) $\mathbf{b}=\mathbf{e}_z$, for the symmetric and asymmetric configurations of Fig.~\ref{fig:simexamples}.}
\label{fig:deltags}
\end{figure*}

\section{\label{sec:simulations}Simulations in Ge devices}
We now compare the above effective theory with simulations of realistic devices. For illustration purposes, we focus on Ge heterostructures where the small value of $|\mathbf{t}_{\mathrm{so}}|$ makes the magnetotunneling term the dominant $g$-factor renormalization mechanism at zero detuning for identical dots. We discuss simulations of Si nanowires in Appendix~\ref{app:Si} for completeness. The simulated device is shown in Fig.~\ref{fig:simexamples}(a). Two plunger gates with radius $R=50$ nm, separated by a center-to-center distance $d=180$ nm, define the DQD by confining holes with their respective voltages $V_\mathrm{L}$ and $V_\mathrm{R}$. A barrier gate at voltage $V_b$ controls the potential between the dots, while the outer gates with voltages $V_\mathrm{u,L/R},\,V_\mathrm{l,L/R},\,V_\mathrm{s,L/R}$ may be used to squeeze the dot and manipulate the hole spin. The Ge well is 16 nm thick and the buffer and top 50-nm-thick barrier are $\mathrm{Si}_{0.2}\mathrm{Ge}_{0.8}$ alloys modeled in the virtual-crystal approximation by linearly interpolating Si and Ge parameters. The Al gates are 20 nm thick, and are separated by 5 nm of Al$_2$O$_3$ from the heterostructure. We account for the inhomogeneous strains imposed by the thermal contraction of the gate stack upon cool-down \cite{abadillo2023hole}. We compute the electrostatic potential in the device by solving Poisson's equation with a finite volumes method, the strains by minimizing the continuum elasticity energy, and then solve for the eigenstates and eigenfunctions of the system with a finite-differences discretization of the four-band Luttinger-Kohn Hamiltonian $H_{\mathrm{LK}}$. The effects of the magnetic field are described by the hole Zeeman Hamiltonian $H_{\mathrm{Z}}$ and by the action of the vector potential on the orbital motion (see Appendix~\ref{appendix:microscopic}). 

\subsection{$g$-factor renormalization}
The low-energy physics is described by the lowest two Kramers pairs, which are linear combinations of the left- and right-localized states with different spins. We numerically extract the $g$-matrix components of each Kramers pair \cite{venitucci2018electrical}. In general, we chose $V_b$ such that the tunnel coupling is $t_c=9.6$\,GHz at zero detuning and magnetic field. We discuss two examples: (i) a device in a symmetric configuration, where the properties of each dot are identical, and (ii) a device in an asymmetric configuration, where the dots are different. In the first case, we ground the outer gates, and vary the detuning by changing the voltage difference $V_\mathrm{L}-V_\mathrm{R}$. The detuning energy is indeed $\varepsilon=\alpha(V_\mathrm{R}-V_\mathrm{L})$, where $\alpha$ is the lever arm, which can be estimated from the gate voltages and eigenenergies of the charge states (usually in the range $\alpha=0.1-0.2$ eV/V in our device). For the asymmetric case (ii), we bias the outer gates around the left plunger to distort the dots. Their $g$-matrices are thus different, and the detuning energy becomes $\varepsilon=\varepsilon_0+\alpha_\mathrm{R}V_\mathrm{R}-\alpha_\mathrm{L}V_\mathrm{L}$, where the lever arms $\alpha_\mathrm{L,R}$ now depend on the gate. The zero detuning point is displaced from the symmetric bias $V_\mathrm{L}=V_\mathrm{R}$ by the shift $\varepsilon_0$, similarly to disordered dots.

The results of our simulations for these two cases are shown in Fig.~\ref{fig:simexamples}(b–g). In the top row, we display effective $g$-factors for magnetic fields aligned along the unit vectors $\mathbf{e}_x$, $\mathbf{e}_y$, and $\mathbf{e}_z$, $g^{(j)}=|\hat{g} \mathbf{e}_j|$, in the symmetric configuration. As long as the Zeeman splitting remains much smaller than the tunnel coupling, the low-energy physics is mainly governed by the ground Kramers pair, whose effective $g$-factors are shown in black. Notably, we observe features that can only be accounted for by the magnetotunneling term even in the symmetric case. The most prominent features are the maxima in the effective $g$-factors $g^{(x)}$ and $g^{(z)}$. As an illustrative example, a Rashba SOC would rotate the $g$-matrices about the $\sigma_y$-axis [Eq.~\eqref{eq:rotatedg}], resulting in $g^{(x)}$ and $g^{(z)}$ being smaller at zero detuning than the average $g$-factors of the dots in the absence of magnetotunneling~\footnote{Note that when we mention rotations about a spin axis, we are implicitly fixing a specific spin basis set. The observables, such as the effective $g$-factors of the eigenstates, remain the same independently of this choice.}, as argued before through the triangular inequality [Eq.~\eqref{eq:triangular}]. However, our simulations reveal maxima at $\varepsilon=0$, highlighting significant corrections beyond the physics of localized $g$-matrices and spin-orbit fields. For the same reason, and due to the symmetry of the dots, $g^{(y)}$ should be independent on detuning in the symmetric case, yet shows a minimum at $\varepsilon=0$. We observe similar features in the asymmetric configuration (bottom row, Fig.~\ref{fig:simexamples}(e–g)): maxima in $g^{(x)}$ and $g^{(z)}$ near zero detuning, and a minimum in $g^{(y)}$. Remarkably, the minimum of $g^{(y)}$ is a sweet spot that cannot arise without magnetotunneling, since a spin-orbit vector along $\mathbf{e}_{y}$ cannot renormalize this $g$-factor near zero detuning; thus, any $g$-factor difference between the dots would lead to a monotonic evolution of $g^{(y)}$ with detuning, from $g^{(y)}_\mathrm{L}$ to $g^{(y)}_\mathrm{R}$ with no sweet spots.

These numerical observations are consistent with the existence of a spin-dependent magnetotunneling term, which renormalizes the effective $g$-factors near zero detuning. However, the simulated device may display physical effects that go beyond the effective microscopic model as well, since inhomogeneous strain fields and intricate electrostatics may introduce extra detuning dependences. To disentangle the different effects, we use the Wannierization method outlined in Appendix~\ref{appendix:wannier}, which we summarize below. First, we obtain from the simulations the lowest four eigenstates $\{\ket{0},\ket{1},\ket{2},\ket{3}\}$ at a given value of detuning. The Wannierization rotates this eigenbasis to localized left and right orbitals $\{\ket{L\uparrow},\ket{L\downarrow},\ket{R\uparrow},\ket{R\downarrow}\}$. The spin basis is chosen such that $\ket\uparrow$ and $\ket\downarrow$ are the heavy-hole states with maximum spin projection $J_z=\pm 3/2$, respectively (to allow simpler analysis). We can then directly map the Luttinger-Kohn Hamiltonian in this basis set onto Eq.~\eqref{eq:eff0}, and extract the effective $g$-matrix $\hat{g}_{\mathrm{L},\mathrm{R}}$ of the localized quantum dots~\cite{venitucci2018electrical}, as well as the bare tunnel coupling $t_c$, the spin-dependent tunnel coupling $\mathbf{t}_{\mathrm{so}}$, the magnetotunneling $g$-matrix $\hat{g}_\mathrm{T}$, and the vector $\boldsymbol{\mu}_\mathrm{T}$. We note that $|\mathbf{t}_{\mathrm{so}}|/t_c\approx 10^{-5}$ in this Ge device; therefore $\mathbf{t}_{\mathrm{so}}$ has no sizable impact and will be neglected in the following analysis. The present spin basis is thus practically a spin-orbit frame, so that $\tilde{g}_{\mathrm{L,R,T}}\approx \hat{g}_{\mathrm{L,R,T}}$ and $\boldsymbol{\mu}_\mathrm{T}$ (also small) has no effect on the $g$-factors.

\begin{figure*}
\includegraphics[width=1.9\columnwidth]{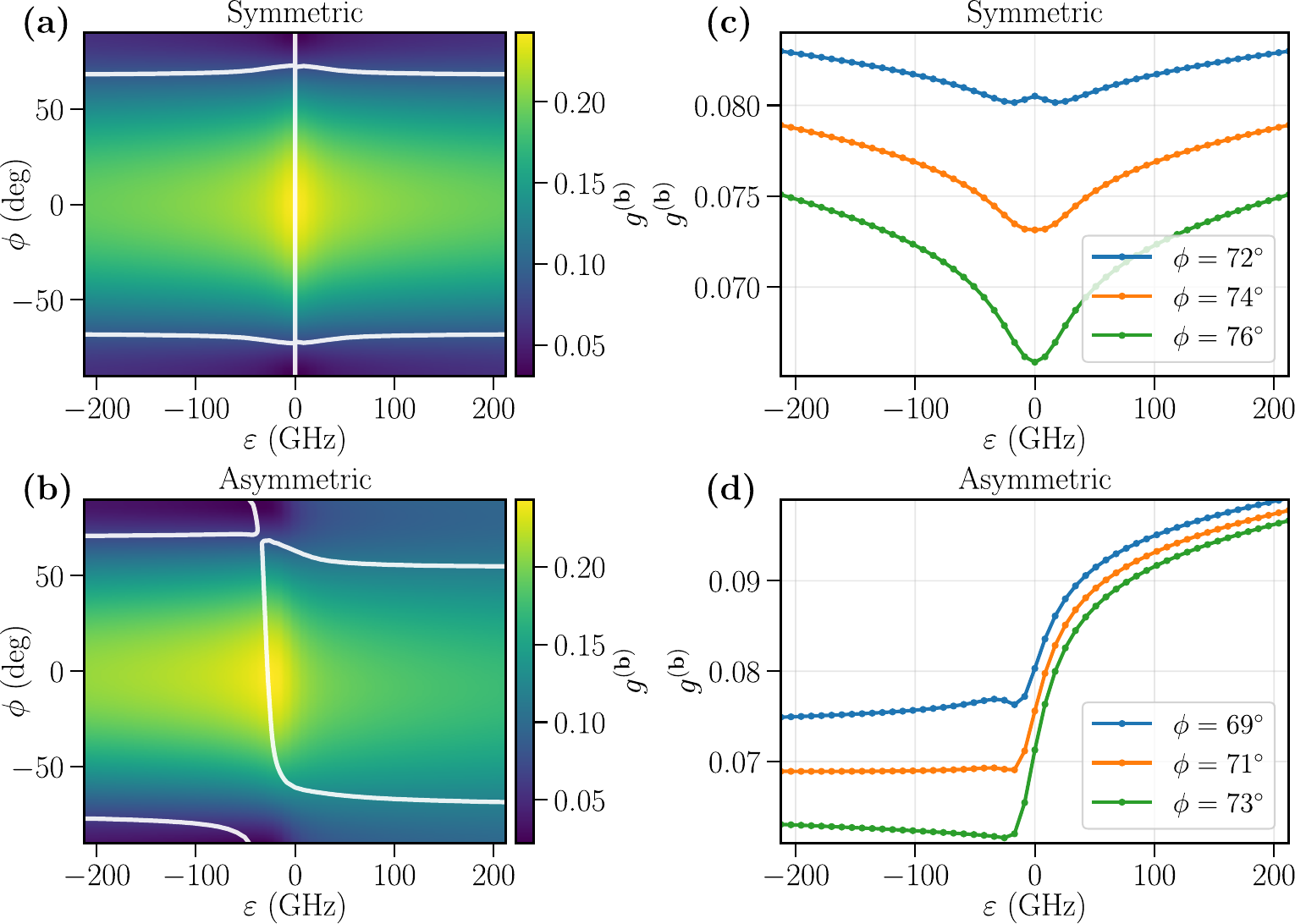}
\caption{Tunability of the $g$-factors. (a-c) Effective $g$-factors $g^{(\mathbf{b})}=|\hat{g} \mathbf{b}|$ as a function of detuning for $\mathbf{b}=\mathbf{e}_x$ (a), $\mathbf{b}=\mathbf{e}_y$ (b), and $\mathbf{b}=\mathbf{e}_z$ (c), and for different values of the tunnel coupling in the same asymmetric configuration as in Fig.~\ref{fig:simexamples}. We mark detuning sweet spots with vertical lines, where dotted lines indicate maxima and dashed lines indicate minima. (d-f) Effective in-plane $g$-factors, $g^{(x)}$ (d), $g^{(y)}$ (e), and $g^{(z)}$ (f) for symmetric configurations with different plunger gates voltages $\bar{V}=V_\mathrm{L}=V_\mathrm{R}$ with $t_c=9.6$ GHz.}
\label{fig:tunability}
\end{figure*}

The effective $g$-factors of the left and right quantum dots localized with the Wannierization method are shown together with those of the ground Kramers pair in Fig.~\ref{fig:simexamples}(b-g). As expected, when $|\varepsilon| \gg |t_c|$, the effective $g$-factors of the ground Kramers pair tend to the left and right $g$-factors for negative and positive detuning respectively. Conversely, when $\varepsilon\rightarrow 0$, the $g$-factors of the ground Kramers pair clearly deviate from any linear combination of the left and right $g$-factors, which, given the negligible spin-flip tunnel coupling in the device, can only be explained by the magnetotunneling mechanism. Besides this mechanism, we note that the localized $g$-matrices exhibit an intrinsic detuning dependence. This dependence is expected, since the gate voltages do not only change detuning but also the size, position, shape of the individual dots, and the strains experienced by the hole. We nonetheless note a sudden change of behavior of the localized $g$-matrices near zero detuning. This change can be tracked down to effects of the vector potential, anticorrelated for each dot ($\propto \tau_z$). This correction is indeed predicted by our microscopic model of Appendix~\ref{appendix:microscopic}, but it makes a negligible contribution to the $g$-factors of the lowest Kramers pair.

We can clearly quantify the magnetotunneling corrections by comparing the effective $g$-factors $g^{(\mathbf{b})}$ of the numerical simulations with those, $g^{(\mathbf{b})}_\text{eff}[\tilde{g}_\mathrm{T}=0]$, obtained from the model of Eq.~\eqref{eq:effso} in the absence of magnetotunneling terms. To compare both methods on an equal footing, we use the $\hat{g}_{\mathrm{L,R}}$ matrices obtained from the numerical simulations in Eq.~\eqref{eq:effso}. We plot the difference $\Delta g^{(\mathbf{b})}=g^{(\mathbf{b})}-g^{(\mathbf{b})}_\text{eff}[\tilde{g}_\mathrm{T}=0]$ as a function of detuning in Fig.~\ref{fig:deltags}. It displays a strong response near $\varepsilon=0$ where the wavefunction extends over the two dots, which maximizes the effects of the tunnel  Hamiltonian on the spectrum and emphasizes the signatures of magnetotunneling.

\subsection{Gate tunability}
\begin{figure}
\includegraphics[width=0.75\columnwidth]{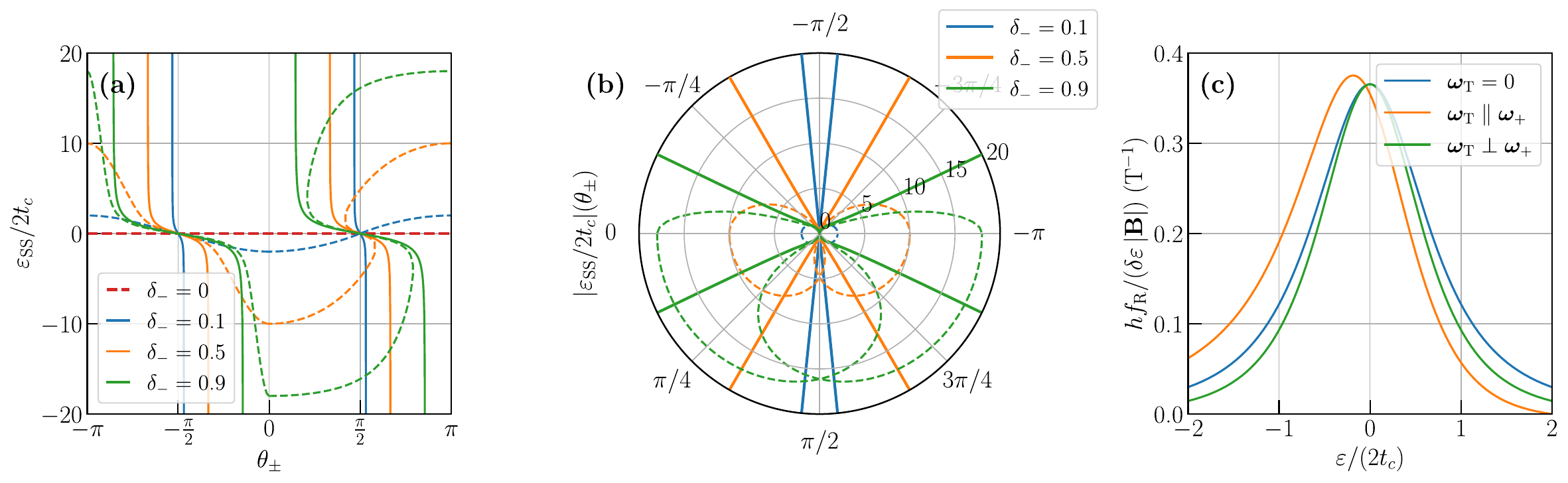}
\caption{Effective magnetotunneling $g$-factors $g^{(\mathbf{b})}_\mathrm{T}=|\hat{g}_\mathrm{T} \mathbf{b}|$ as a function of the tunnel coupling $t_c$ for the symmetric configuration described in the caption of Fig.~\ref{fig:simexamples} with $V_\text{L}=V_\text{R}=-50$ mV.}
\label{fig:tctunability}
\end{figure}
Overall, we find that the magnetotunneling term makes significant corrections to the $g$-factors near zero detuning. In Eq.~\eqref{eq:gt}, the $\hat{g}_\mathrm{T}$ matrix is proportional to the tunnel coupling and depends on the different characteristic lengths such as the interdot distance $d$ and the harmonic lengths $L_x$ and $L_y$. Since the plunger gates have a direct effect on the dot size and the barrier gate has a direct effect on the tunnel coupling, we can expect this mechanism to be gate-tunable as well. In Fig.~\ref{fig:tunability} we assess the tunability of the magnetotunneling term by varying the plunger gates and the tunnel coupling for the symmetric and asymmetric configurations. In Fig.~\ref{fig:tunability}(a-c), we plot the effective $g$-factors of the asymmetric configuration and we vary $V_b$ to change the tunnel coupling. The dependence of the effective $g$-factors on the ratio $\varepsilon/t_c$ is shown for $t_c=\{1,5,9,13\}$ GHz. The effect of changing $t_c$ can be observed near zero detuning where the renormalization of the effective $g$-factor is more prominent for higher tunnel coupling. In fact, we observe that for $t_c=1$ GHz, the sweet spots — marked with vertical lines — disappear for $g^{(z)}$ and are largely displaced for $g^{(x)}$. This is expected from the approximate linear dependence of $\hat{g}_\mathrm{T}$ on $t_c$ in Eq.~\eqref{eq:gt}, where the reduction of $t_c$ suppresses the renormalization near zero detuning. In this situation, if the $g$-factors at the two dots are too different, the magnetotunneling corrections may not be strong enough to preserve the sweet spot, and the $g$-factors just monotonously change with detuning from the localized $g$-factor in the left dot to the localized $g$-factor in the right dot. On the contrary, when $t_c$ is large enough the spin-dependent magnetotunneling correction provides a mechanism that breaks the monotonous $g$-factor behavior, potentially providing one or more sweet spots in some cases. This happens as an example for $g^{(y)}$ when $t_c\geq 5$ GHz, and we will explore this in more detail in the next section.  

We finally discuss the plunger gate dependence of the corrections for fixed tunnel coupling $t_c=9.6$ GHz in Fig.~\ref{fig:tunability}(d-f), where we show the changes in the in-plane $g$-factors. For simplicity, we focus on the symmetric case and find that the plunger gate has a stronger effect on the localized $g$-matrices than on the spin-dependent magnetotunneling terms. We observe that the plunger gate steadily shifts the effective $g$-factors independently of detuning, with little effect on $\hat{g}_\mathrm{T}$. This is expected from Eq.~\eqref{eq:gt}, where the dependence on size in Eq.~\eqref{eq:gt} is mostly ruled by the ratio of anisotropy $L_x^2/L_y^2$, which barely changes for the set of gate voltages used in this work, with quasi-circular quantum dots. We note the change in trend of the effective $g$-factor $g^{(y)}$ in Fig.~\ref{fig:tunability}(e) around $\bar{V}=-0.12$~V, consistent with a zero and reversal of $g\boldsymbol{e}_y$, as observed experimentally~\cite{wang2024operating}.

From these results, we conclude that the most efficient way to tune the $g_\mathrm{T}$ factors is to modulate the tunnel coupling. To quantify this more explicitly, the effective $g_\mathrm{T}$-factors are plotted as a function of $t_c$ in Fig.~\ref{fig:tctunability}. In our setup and chosen Wannier basis set, the $\hat{g}_\text{T}$-matrix is approximately diagonal but deviations may arise due to strain as discussed in Appendix~\ref{app:strain}. We find that the $\hat{g}_\mathrm{T}$ matrix tends to zero as $t_c\rightarrow 0$ and it increases with $t_c$, as expected from Eq.~\eqref{eq:gt}. The effective $g_\mathrm{T}$-factors are in the $0-0.1$ range, thus are in the same order of magnitude as the in-plane $g$-factors of the quantum dots ($0-0.5$), and therefore give rise to sizable corrections.

\subsection{Multiple and higher-order sweet spots}
\begin{figure}
\includegraphics[width=1\columnwidth]{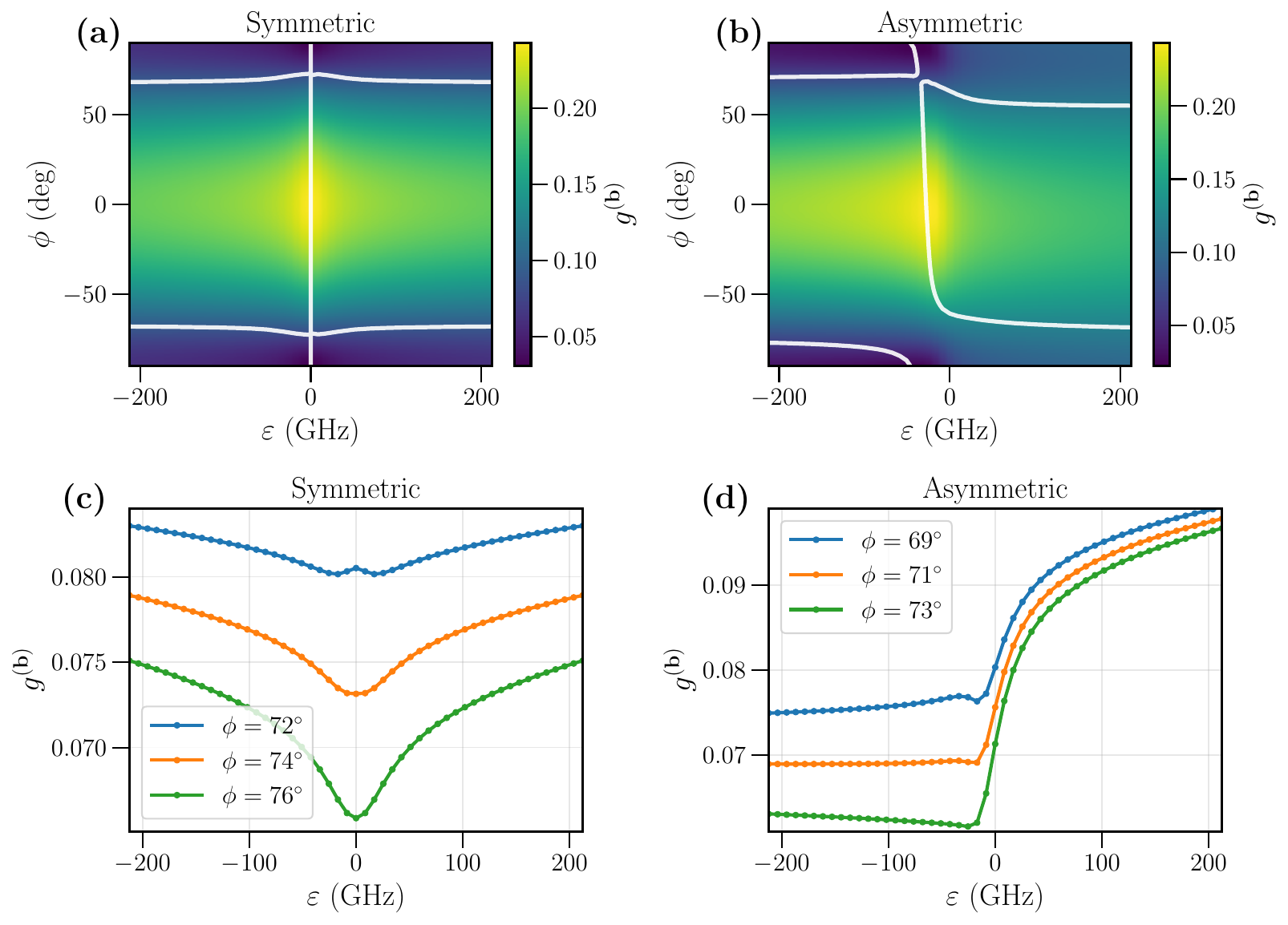}
\caption{Higher-order sweet spots and in-plane $g$-factor behavior. (a,b) Maps of the effective $g$-factor $g^{(\mathbf{b})}=|g \mathbf{b}|$ as a function of in-plane magnetic field orientation $\phi$ and detuning energy $\varepsilon$ for the symmetric and asymmetric configurations, respectively. Sweet spot positions are marked with white lines. (c,d) Effective $g$-factor as a function of detuning for magnetic field orientations where the $g$-factors exhibit a high-level of flatness.}
\label{fig:highsweetness}
\end{figure}

In Fig.~\ref{fig:simexamples} we observe opposite behavior of $g^{(x,z)}$ and $g^{(y)}$, displaying maxima or minima as a function of detuning, respectively. Furthermore, in Fig.~\ref{fig:tunability}(b), we find more than a single sweet spot for magnetic fields in the $y$ direction. Therefore, it is natural to investigate other magnetic field orientations that may exhibit intermediate behavior to understand whether multiple sweet spots can be engineered. To analyze this, we focus on the $x-y$ plane and characterize the detuning dependence of the effective $g$-factors as a function of the orientation of the magnetic field $\mathbf{B}=(\cos\phi,\sin\phi,0)$. We mark sweet spots in detuning with white lines in Fig.~\ref{fig:highsweetness}(a,b). For the symmetric case, we find a sweet sport at $\varepsilon=0$ for all $\phi$ as expected. Furthermore, we find other sweet lines for $\phi\approx \pm 74^\circ$ with little dependence on detuning. These sweet lines result from the competition between $g^{(x)}$ and $g^{(y)}$; as they exhibit opposite slopes with detuning, there must be an in-plane orientation where they cancel out. We take specific line cuts at fixed $\phi$ in Fig.~\ref{fig:highsweetness}(c), showing how three sweet spots for $\phi=72^\circ$ join into a third-order sweet spot at $\varepsilon=0$ and $\phi\approx74^\circ$. We observe similar behavior for the asymmetric quantum dot in Fig.~\ref{fig:highsweetness}(b,d). The asymmetry of electrostatics distorts the maps, reducing the number of sweet spots for a given $\phi$; however, there is still a double sweet spot around $\phi\approx 71^\circ$. The line cuts in Fig.~\ref{fig:highsweetness}(d) further show how the effective $g$-factors exhibit almost constant behavior at negative detunings for $\phi\approx 71^\circ$. Given that these higher-order sweet spots may occur near zero detuning, they exhibit a finite dipolar coupling and are, therefore, potentially useful working points. These results demonstrate that sweet lines and, more specifically, high-order and multiple sweet spots in hole quantum dot arrays can be engineered by appropriate choices of the tunnel coupling and the magnetic field orientation. 

\begin{figure*}
\includegraphics[width=2\columnwidth]{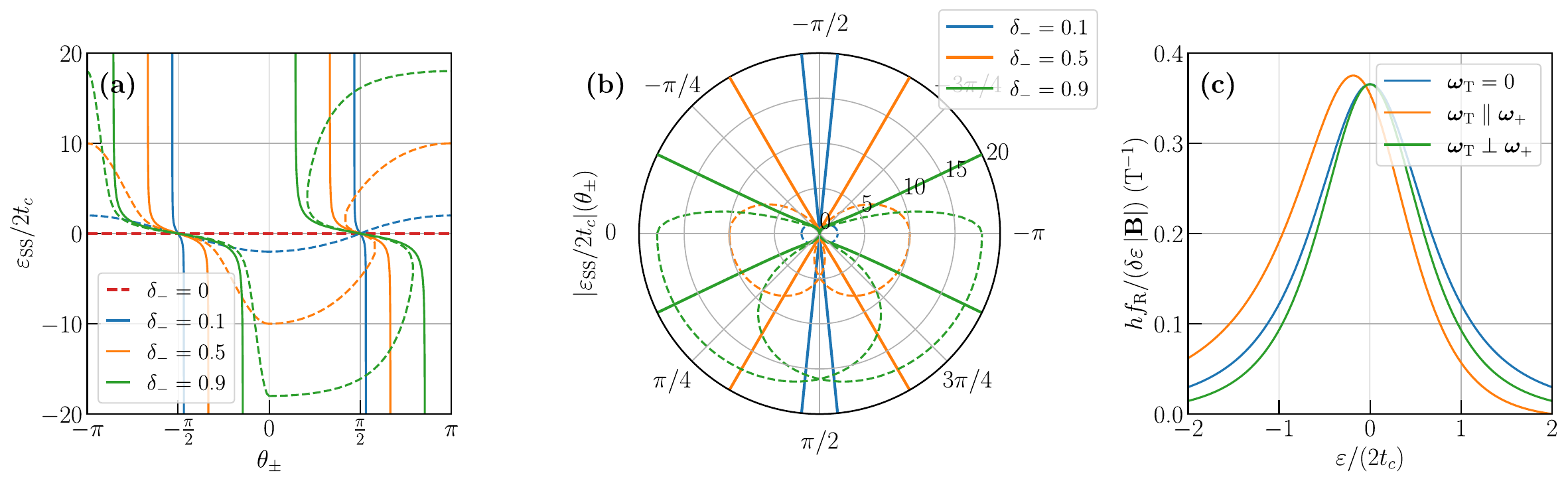}
\caption{Sweet spots and Rabi frequencies accounting for the magnetotunneling term. (a) Sweet spot position $\varepsilon_{\mathrm{SS}}/2t_c$ with $\boldsymbol{\omega}_+=|\boldsymbol{\omega}_+|\mathbf{e}_{z}$ and $\boldsymbol{\omega}_-=\delta_-|\boldsymbol{\omega}_+|(\cos\theta_\pm\mathbf{e}_{z}+\sin\theta_\pm \mathbf{e}_{x})$, as a function of the angle $\theta_\pm$. Solid lines are the case $\tilde{g}_\mathrm{T}=0$ and were solved using Eq.~(\ref{eq:toysweet}). Dashed lines are the case $\boldsymbol{\omega}_\mathrm{T}\approx0.05\boldsymbol{\omega}_+$ and were obtained with Eq.~(\ref{eq:sweetverygen}). (b) Same as (a) but as a polar coordinate plot. Note that the sweet spots for $\boldsymbol{\omega}_\mathrm{T}\approx-0.05\boldsymbol{\omega}_+$ can be obtained by the symmetry transformation $\theta_{\pm}\rightarrow \theta_{\pm}+\pi$. (c) Rabi frequency normalized to the driving amplitude $(\delta\varepsilon/h)$ and to the applied magnetic field $\mathbf{B}$, as a function of detuning. The chosen parameters are $t_c=9.6$ GHz, $\boldsymbol{\omega}_+=0.2\mathbf{e}_{x}$, and $\boldsymbol{\omega}_-=0.1\mathbf{e}_{y}$. For the case $\boldsymbol{\omega}_\mathrm{T}\parallel \boldsymbol{\omega}_+$, we choose $\boldsymbol{\omega}_\mathrm{T} \approx 0.05\mathbf{e}_x$, while for the case $\boldsymbol{\omega}_\mathrm{T}\perp \boldsymbol{\omega}_+$, we choose $\boldsymbol{\omega}_\mathrm{T}=0.05\mathbf{e}_{z}$.}
\label{fig:ssandrabi}
\end{figure*}

\section{\label{sec:sweet} Effects on spin qubit properties}
In previous sections, we have shown that the magnetotunneling mechanism has direct implications for spin qubit operation and performance. Flopping-mode qubits in double quantum dots exhibit effective $g$-factors that cannot be explained by the local $g$-matrices alone, and are potentially larger than expected from single-dot considerations. As an example of the implications for spin qubits, such a $g$-factor enhancement allows heavy-hole states to achieve resonance with superconducting cavities at lower magnetic fields. In this section we derive and assess the observable consequences of magnetotunneling corrections using the effective model of Eq.~\eqref{eq:effso}, focusing on sweet spot formation and Rabi frequencies. 

\subsection{Sweet spot robustness}
Having demonstrated through simulations that the magnetotunneling mechanism can stabilize sweet spots even in the presence of large local $g$-factor differences, we now provide a systematic analysis of the sweet spot robustness and the conditions under which they survive.

We start from the effective Hamiltonian in Eq.~\eqref{eq:effso} and focus on the spin dynamics in the double dot. First, we define $\Omega$ and $\Theta$ such that $\varepsilon=\Omega\cos\Theta$ and $2t_c=\Omega\sin\Theta$ and diagonalize the spin-independent part of the Hamiltonian. Physically, $\Omega$ is the charge qubit energy while $\Theta$ is the charge mixing angle quantifying the localization of the states. By applying a rotation with angle $\Theta$ around $\tau_y$ in the charge subspace, we diagonalize the Hamiltonian for $\mathbf{B}=\mathbf{0}$. To simplify the spin part of the Hamiltonian, we introduce the average and the difference of the $g$-matrices $\tilde{g}_\pm=(\tilde{g}_\mathrm{L}\pm\tilde{g}_\mathrm{R})/2$. In this way, the charge-diagonal effective Hamiltonian becomes
\begin{equation}
\begin{aligned}
    H^{\text{eff}}_c&=\frac{\Omega}{2}\tau_z+\frac{\mu_B}{2}(\boldsymbol{\sigma}\cdot\tilde{g}_+\mathbf{B})\\ &
    -\frac{\mu_B}{2}(\tau_z\cos\Theta+\tau_x\sin\Theta)(\boldsymbol{\sigma}\cdot\tilde{g}_-\mathbf{B})\\ &-\frac{\mu_B}{2}(\tau_x\cos\Theta-\tau_z\sin\Theta)(\boldsymbol{\sigma}\cdot\tilde{g}_\mathrm{T}\mathbf{B}) \\
    &+\frac{1}{2}\tau_y(\tilde{\boldsymbol{\mu}}_\mathrm{T}\cdot\mathbf{B}).
    \end{aligned}
    \label{eq:effcharge}
\end{equation}
To simplify the following analysis, we define the shorthand notation $\boldsymbol{\omega}_\pm=\mu_B \tilde{g}_\pm\mathbf{B}$ and $\boldsymbol{\omega}_\mathrm{T}=\mu_B \tilde{g}_\mathrm{T}\mathbf{B}$. In general, $\boldsymbol{\omega}_-$ has components parallel and perpendicular to $\boldsymbol{\omega}_+$, so that $\boldsymbol{\omega}_-\cdot\boldsymbol{\omega}_+=|\boldsymbol{\omega}_-||\boldsymbol{\omega}_+|\cos\theta_\pm$, where $\theta_\pm$ is the relative angle between the two vectors. We plot with solid lines in Fig.~\ref{fig:ssandrabi}(a,b) the position of the sweet spots in detuning, $\varepsilon_\text{SS}$, as a function of $\theta_\pm$, without the magnetotunneling terms. They are computed as the solutions of $\partial_\varepsilon E_{01}=0$ with $E_{01}$ the splitting between the lowest two eigenstates of Eq.~\eqref{eq:effcharge}. For a given ratio $\delta_-=|\boldsymbol{\omega}_-|/|\boldsymbol{\omega}_+|$, there is only a window of $\theta_\pm$ where there is a sweet spot solution. Outside this window, the effective $g$-factor difference between the dots exceeds the renormalization of the $g$-factor at zero detuning, preventing sweet spot formation.

To understand the formation of sweet spots, we extract the Larmor vector of the lowest Kramers pair, $\boldsymbol{\omega}_{01}=(\boldsymbol{\omega}_++\cos\Theta\boldsymbol{\omega}_--\sin\Theta\boldsymbol{\omega}_\mathrm{T})$, from the $\propto B\tau_z$ terms of the Hamiltonian. Note that $\tilde{\boldsymbol{\mu}}_\mathrm{T}$ only appears in an off-diagonal $\propto \tau_y$ term in the charge eigenbasis and, therefore, does not contribute to $\boldsymbol{\omega}_{01}$ whatever the detuning. The sweet spot condition $\partial_\varepsilon E_{01}=0$ can alternatively be written $\boldsymbol{\omega}_{01}\cdot(\partial_\varepsilon\boldsymbol{\omega}_{01})=0$, which means that the variation of the Larmor vector must be either zero or purely transverse. The derivative of the Larmor vector with respect to detuning is $\partial_\varepsilon\boldsymbol{\omega}_{01}=-(d\Theta/d\varepsilon)(\sin\Theta\boldsymbol{\omega}_-+\cos\Theta\boldsymbol{\omega}_\mathrm{T})$, with $d\Theta/d\varepsilon=-2t_c/\Omega^2$. For non-zero $\boldsymbol{\omega}_-$, the sweet spots without magnetotunneling term ($\tilde{g}_\mathrm{T}=0$) thus fulfill two possible conditions:
\begin{equation}
\begin{aligned}
    \sin\Theta&=0 \text{ or }\frac{d\Theta}{d\varepsilon}=0\Rightarrow  \varepsilon\rightarrow \pm \infty \\
    \cos\Theta&=-\frac{\boldsymbol{\omega}_+\cdot \boldsymbol{\omega}_-}{|\boldsymbol{\omega}_-|^2}\\ \Rightarrow \frac{\varepsilon_\text{SS}}{2t_c}&=-\frac{\boldsymbol{\omega}_+\cdot\boldsymbol{\omega}_-}{\sqrt{|\boldsymbol{\omega}_-|^4-\left(\boldsymbol{\omega}_+\cdot\boldsymbol{\omega}_-\right)^2}},
    \end{aligned}
    \label{eq:toysweet}
\end{equation}
where the first condition is the single-dot limit. There is, therefore, a sweet spot at $\varepsilon_\text{SS}=0$ only if $\boldsymbol{\omega}_-\perp\boldsymbol{\omega}_+$. Furthermore, there is no sweet spot when the denominator becomes imaginary, which happens if $|\boldsymbol{\omega}_-\cdot\boldsymbol{\omega}_+|>|\boldsymbol{\omega}_-|^2$. Such a situation occurs in Fig.~\ref{fig:intro}(c) and corresponds to the diverging cases in Fig.~\ref{fig:ssandrabi}(a). This condition actually defines a sweet spot window $|\cos\theta_{\pm}|< \delta_-$, particularly explicit in the polar plot of Fig.~\ref{fig:ssandrabi}(b) (solid lines). Outside this window, sweet spots can only be found in the single-dot regime. We note that even if there is no sweet spot but $\delta_-\ll1$, the spectrum becomes quite insensitive to detuning, potentially enabling long coherence times; in this situation, manipulation may become very slow, as discussed in Sec.~\ref{sec:rabi}. In particular, when $\boldsymbol{\omega}_-=\boldsymbol{\omega}_\mathrm{T}=0$, the derivative $\partial_\varepsilon\boldsymbol{\omega}_{01}$ is exactly zero for any value of $\varepsilon$. This is a special case where the spin splitting is unaffected by SOC, hence, the spin cannot be electrically manipulated.

The dashed lines of Figs.~\ref{fig:ssandrabi}(a,b) illustrate what happens when a finite magnetotunneling correction $\boldsymbol{\omega}_\mathrm{T}=0.05\boldsymbol{\omega}_+$ is added, showing contrasting behavior with the $\tilde{g}_\mathrm{T}=0$ case. The presence of this small correction indeed stabilizes the sweet spot position, preventing divergence in $\theta_\pm$ for given $\delta_-$. In this situation, there is at least one sweet spot in detuning for any value of $\theta_{\pm}$. In a polar plot as a function of $\theta_\pm$ (Fig.~\ref{fig:ssandrabi}(b)), the sweet spots draw two disconnected lines for $\boldsymbol{\omega}_\mathrm{T}=0$ but closed curves for finite $|\boldsymbol{\omega}_\mathrm{T}|$. Depending on the relative amplitude of $\delta_-$ and $|\boldsymbol{\omega}_\mathrm{T}|$, these closed curves can be single loops with one sweet spot per $\theta_\pm$ when $\delta_-<|\boldsymbol{\omega}_\mathrm{T}|$, or more complicated closed loops when $\delta_->|\boldsymbol{\omega}_\mathrm{T}|$, such as the butterfly patterns in Fig.~\ref{fig:ssandrabi}(b). Interestingly, there can be up to three sweet spots for a given value of $\theta_\pm$ when this pattern arises for $\delta_->|\boldsymbol{\omega}_\text{T}|$.

The multiple sweet spot condition corresponds to what we observe on the simulations in Fig.~\ref{fig:tunability}(b) as well as in Fig.~\ref{fig:highsweetness}. In the presence of a non-zero $\tilde{g}_\mathrm{T}$, the sweet spot equation admits more solutions since a non-zero $\tilde{g}_-$ implies that there are two mechanisms changing the effective $g$-factors as a function of detuning. In fact, the original solution $\sin\Theta=0$ is no longer valid and these sweet spots are displaced to finite detuning values. The more general sweet spot equation can again be obtained by imposing the condition $\boldsymbol{\omega}_{01}\cdot(\partial_\varepsilon\boldsymbol{\omega}_{01})=0$ to Eq.~\eqref{eq:effcharge}, which yields:
\begin{equation}
\begin{aligned}
    \boldsymbol{\omega}_-\cdot\boldsymbol{\omega}_\mathrm{T}\cos2\Theta+\tfrac{1}{2}\left(|\boldsymbol{\omega}_-|^2-|\boldsymbol{\omega}_\mathrm{T}|^2\right)\sin2\Theta\\ = -\boldsymbol{\omega}_+\cdot\left(\boldsymbol{\omega}_-\sin\Theta+\boldsymbol{\omega}_\mathrm{T}\cos\Theta\right).
    \end{aligned}
    \label{eq:sweetverygen}
\end{equation}
We note that the sign exchange $\boldsymbol{\omega}_\mathrm{T}\rightarrow-\boldsymbol{\omega}_\mathrm{T}$ is equivalent to $\boldsymbol{\omega}_-\rightarrow-\boldsymbol{\omega}_-$ and $\Theta\rightarrow \Theta+\pi$. Therefore, the sweet spot solutions for $\boldsymbol{\omega}_\mathrm{T}= 0.05\boldsymbol{\omega}_+$ in Fig.~\ref{fig:ssandrabi}(a,b) can be directly related to those for $\boldsymbol{\omega}_\mathrm{T}= -0.05\boldsymbol{\omega}_+$ by taking $\theta_\pm\rightarrow \theta_\pm+\pi$.

Eq.~\eqref{eq:sweetverygen} is difficult to solve analytically, but a non-trivial solution can be obtained under the assumption that there is a sweet spot near zero detuning such that $|\varepsilon_\text{SS}|\ll 2t_c$. The sweet spot detuning $\varepsilon_\text{SS}/2t_c=\cot\Theta$ then fulfills the condition
\begin{equation}
\begin{aligned}
\cos\Theta&\approx -\frac{\boldsymbol{\omega}_+\cdot\boldsymbol{\omega}_-}{|\boldsymbol{\omega}_-|^2+\left(\boldsymbol{\omega}_+-\boldsymbol{\omega}_\mathrm{T}\right)\cdot\boldsymbol{\omega}_\mathrm{T}}, 
    \end{aligned}
    \label{eq:sweetgen}
\end{equation}
which amounts to Eq.~(\ref{eq:toysweet}) when $\tilde{g}_\mathrm{T}=0$. The presence of $\tilde{g}_\mathrm{T}$ in the denominator provides robustness to the sweet spots. Interestingly, even when the previous condition requires $\cos\Theta >1$ and, hence, this specific sweet spot does not exist, other solutions to Eq.~\eqref{eq:sweetverygen} emerge, as can be seen in Fig.~\ref{fig:ssandrabi}(a,b).

Another illustrative situation that can be solved analytically is the case $\boldsymbol{\omega}_+\cdot\boldsymbol{\omega}_-=0$ and $\boldsymbol{\omega}_\mathrm{T}\cdot\boldsymbol{\omega}_-=0$. This occurs, for instance, when the dots are identical, have diagonal $g$-matrices, yet there is a net rotation when tunneling from one dot to the other, such as when there is a spin-orbit vector. Here, the spin-orbit vector can be lumped into $\tilde{g}_-$ in the spin-orbit frame, fulfilling the previous conditions, and we obtain a trivial solution $\cos\Theta=0$, corresponding to zero detuning and potentially two more sweet spot solutions 
\begin{equation}
\frac{\varepsilon_\text{SS}}{2t_c} = \pm \sqrt{\left(\frac{|\boldsymbol{\omega}_-|^2 - |\boldsymbol{\omega}_\mathrm{T}|^2}{\boldsymbol{\omega}_+ \cdot \boldsymbol{\omega}_\mathrm{T}}\right)^2-1},
\end{equation}
which, when the magnetic field is chosen such that those sweet spots coincide at $\varepsilon=0$, leads to a third order sweet spot, as in Fig.~\ref{fig:highsweetness}(b).

\subsection{Corrections to Rabi frequencies}
\label{sec:rabi}
We next focus on the effect of the magnetotunneling terms on the transverse dipolar coupling. This is particularly relevant for the manipulation of flopping mode spin qubits with an AC drive, and for spin-photon coupling experiments. We consider a detuning fluctuation $\delta\varepsilon$ (static pulse, drive field, or noise) described by the Hamiltonian $\delta H_\text{det}=\delta\varepsilon\tau_z$. In the charge diagonal basis set, this operator reads $H_\text{c,det}=-\delta\varepsilon(\cos\Theta\tau_z+\sin\Theta\tau_x)$. We next perform a second-order Schrieffer-Wolff transformation to work out the action of $H_\text{c,det}$ in the qubit subspace, which yields a correction to the Larmor vector
$\delta \boldsymbol{\omega}_{01}=-\frac{\delta\varepsilon}{2\Omega}\sin\Theta(\sin\Theta\boldsymbol{\omega}_-+\cos\Theta\boldsymbol{\omega}_\mathrm{T})$. In the $g$-matrix formalism, the Rabi frequency is given by
\begin{equation}
\begin{aligned}
        f_\mathrm{R}&=\frac{\mu_B}{2|\boldsymbol{\omega}_{01}|}(\boldsymbol{\omega}_{01}\times\delta\boldsymbol{\omega}_{01})\\&=\frac{\mu_B B\delta\varepsilon\sin\Theta}{4\Omega}\Bigg|\frac{(\boldsymbol{\omega}_+\times\boldsymbol{\omega}_-)\sin\Theta}{|\boldsymbol{\omega}_++\cos\Theta\boldsymbol{\omega}_--\sin\Theta\boldsymbol{\omega}_\mathrm{T}|} \\ &+\frac{(\boldsymbol{\omega}_\mathrm{T}\times\boldsymbol{\omega}_-)-(\boldsymbol{\omega}_\mathrm{T}\times\boldsymbol{\omega}_+)\cos\Theta}{|\boldsymbol{\omega}_++\cos\Theta\boldsymbol{\omega}_--\sin\Theta\boldsymbol{\omega}_\mathrm{T}|}\Bigg|,
\end{aligned}
\end{equation}
where the first line is the usual contribution to the dipolar coupling resulting from the interplay between $\boldsymbol\omega_+$ and $\boldsymbol\omega_-$, and the second line is the dipolar correction resulting from the magnetotunneling mechanism. Without the magnetotunneling term, the dipolar coupling shows an overall dependence on $\sin^2\Theta$, and thus tends to be larger for $\varepsilon=0$; yet this picture may be slightly modified by a non-zero $\boldsymbol\omega_\mathrm{T}$, bringing an extra dependence on other harmonics. This is illustrated in Fig.~\ref{fig:ssandrabi}(c), where a non-zero $\tilde{g}_\mathrm{T}$ slightly modifies the Rabi frequency as a function of detuning. In the scenario where $|\boldsymbol{\omega}_+|>|\boldsymbol{\omega}_-|$ we find that the projection of $\boldsymbol{\omega}_\mathrm{T}$ over $\boldsymbol{\omega}_+$ introduces the largest correction, which may create an imbalance between the contributions from $\boldsymbol{\omega}_-$ and $\boldsymbol{\omega}_+$, yielding an asymmetric correction to the Rabi frequency. Still, the first line will dominate in most scenarios due to the comparatively small elements of $\tilde{g}_\mathrm{T}$. For instance, the multiple and higher-order sweet spots in Fig.~\ref{fig:highsweetness} exhibit a finite dipolar coupling, particularly when they occur near $\varepsilon=0$. 

Interestingly, even when $\boldsymbol{\omega}_-\parallel \boldsymbol{\omega}_+$ (no dipolar coupling in the conventional theory), the magnetotunneling term gives rise to a finite dipolar coupling correction, which is not necessarily symmetric in detuning. This feature may be observed more clearly in flopping-mode qubits by choosing a  magnetic field orientation where $\boldsymbol{\omega}_-\parallel \boldsymbol{\omega}_+$, as the Rabi frequency could display an unexpected asymmetry as a function of detuning. 

\section{\label{sec:discussion}Discussion}
The magnetotunneling mechanism we have found has implications for the design and operation of hole spin qubits in quantum dot arrays. It provides an additional knob for engineering sweet spots, flattening the spectrum as a function of detuning, and enabling robust qubit operation even when device variability leads to non-identical quantum dots. Nevertheless, the $\tilde{g}_\mathrm{T}$ matrix introduces an extra detuning dependence in the Larmor vector, complicating the classification of spin properties in quantum dots based on localized $g$-matrices~\cite{sen2023classification}. Besides, it introduces further constraints for finding magic angle magnetic field orientations where the pseudospin is preserved along transport. These features may require extra analysis in the future but go beyond the scope of this work. 

Inhomogeneous strain fields~\cite{abadillo2023hole} can be another key contribution to the detuning dependence of the energy spectrum and $g$-factors. As the hole moves with detuning, the wavefunction samples different strains, which modifies the local $g$-matrices. Our simulations show that, near zero detuning, the effects of the strains induced by the differential thermal contraction of materials are negligible with respect to the spin-dependent magnetotunneling mechanism, consistently with variability studies in Ge~\cite{martinez2025variability}. However, this does not preclude strains from having a sizable effect on $g_\mathrm{T}$ more generally: as detailed in Appendix~\ref{app:strain}, shear strain rotates the $g_\mathrm{T}$ matrix, while uniaxial strains renormalize the diagonal components. Looking ahead, the interplay between magnetotunneling and controlled strain engineering~\cite{mauro2025strain} may offer further opportunities to tailor the detuning dependence of the qubit energies. 

An important question is how the magnetotunneling mechanism generalizes to larger quantum dot systems. Extending our effective model from a double quantum dot to a sparse array is actually straightforward. In a linear array of $N$ quantum dots, each nearest-neighbor tunneling link between sites $i$ and $i+1$ acquires its own magnetotunneling matrix $\hat{g}_{\mathrm{T},i}$ and vector $\boldsymbol{\mu}_{\mathrm{T},i}$. The effective Hamiltonian then contains terms of the form $\frac{1}{2}\mu_B \tau_{x,i}(\boldsymbol{\sigma}\cdot\hat{g}_{\mathrm{T},i}\mathbf{B})$ and $\frac{1}{2} \tau_{y,i}(\boldsymbol{\mu}_{\mathrm{T},i}\cdot\mathbf{B})$ for each link, where $\tau_{x,i}$ and $\tau_{y,i}$ represent the tunneling operators. The observable effect is the same as in the DQD: the effective $g$-factors get strongly renormalized near interdot transitions, potentially enabling sweet spots. This behavior can be extended to the case where multiple interdot transitions occur simultaneously in parameter space, as analyzed in Appendix~\ref{app:1Dchains}.

In two-dimensional arrays, similar corrections apply to all nearest-neighbor couplings, potentially enabling even richer sweet-spot engineering through the interplay of multiple magnetotunneling processes. However, it is important to note that in effectively 1D arrays we can always find a set of local pseudospin frames that lump spin-orbit and Peierls phases into the local $g$-matrices. This is not generally possible in 2D arrays where loops introduce gauge-invariant corrections coming from the spin-flip tunneling and Peierls phases~\cite{kolok2024protocols}, which strongly depend on the geometry of the array. Nevertheless, the central point of the present work still holds: regardless of array size or dimensionality, each interdot Hamiltonian acquires a magnetotunneling term characterized by a $\hat{g}_{\mathrm{T},i}$ matrix, which renormalizes the Zeeman splitting near charge transitions. This scalability suggests that the mechanisms we have identified will remain relevant as hole spin qubit architectures increase in size and complexity, providing continued opportunities for enhancing coherence and operational robustness in large-scale quantum processors.

Another important aspect to consider in sparse arrays is undesired fluctuations of the local electrostatic landscape induced by control cross-talk and long-range Coulomb interactions. In practice, remote gate cross-talk and occasional charge rearrangements while operating different spins in the device can shift the effective detuning $\varepsilon$ of a given link. These effects can, in principle, be calibrated and compensated. If not calibrated, operating at the sweet spots will strongly suppress the impact of these fluctuations on the qubit energy. Cross-talk may also induce small changes in $t_c$, which could displace the sweet spots. This can, likewise, be solved in principle by recalibrating the sweet spot position.

For shuttling experiments, it is important to keep the spectrum as flat as possible as a function of detuning to reduce extra dephasing~\cite{mills2019shuttling, van2024coherent, lin2025interplayzeemansplittingtunnel}. More explicitly, in protocols where the detuning $\varepsilon(t)$ is time dependent, and the evolution is approximately adiabatic, the qubit accumulates a dynamical phase $\phi=\int dt\,|\boldsymbol{\omega}_{01}[\varepsilon(t)]|$. Thus, low-frequency fluctuations of $\varepsilon$ translate into undesired phase noise $\delta\phi\propto\int dt(\partial_\varepsilon |\boldsymbol{\omega}_{01}|)\delta\varepsilon(t)$. This is particularly relevant near charge transitions where adiabatic transport requires to stay for longer times~\cite{guery2019shortcuts}, which is precisely where magnetotunneling enhances the robustness of sweet spots, possibly reducing this undesired dynamical phase.
However, this also leads to a potentially larger dynamical range of $g$-factors. Importantly, the existence of intermediate sweet spots in detuning space implies changes of sign of the derivative of the Larmor frequency with respect to detuning~\cite{bosco2024high, langrock2023blueprint}. Such changes of sign may be particularly beneficial for quasi-static or low-frequency detuning noise, where partial cancellation of phase errors can occur along opposite-slope segments (this is, however, less efficient for higher-frequency noise). A precise optimization of shuttling pulses should therefore explicitly account for the magnetotunneling contribution $\tilde{g}_T$ since it may have a direct impact on the resulting shuttling fidelity.

In flopping modes, it is important to operate near sweet spots close to zero detuning. The magnetotunneling mechanism enables the appearance of multiple sweet spots within experimentally accessible ranges, providing working points which may have different Larmor vectors, dipolar couplings, and coherence properties. This, in turn, opens up new possibilities for hopping-spin protocols: instead of transferring the spin between the two dots, intermediate sweet spots in detuning with different Larmor vectors can also serve as operational points.

\section{\label{sec:conclusions}Conclusion}
In this work, we have identified and characterized a fundamental correction to the effective physics of holes in double quantum dots: the spin-dependent magnetotunneling term. This mechanism emerges naturally from the interplay between the momentum-dependent heavy-hole-light-hole interaction and the spin $3/2$ Zeeman interaction of holes, revealing that conventional effective models have overlooked a $g$-matrix component.

Our theoretical analysis establishes that magnetotunneling introduces three observable effects: (i) an additional $g$-factor tunability with strong detuning dependence, (ii) modified sweet spot conditions that enhance robustness against device asymmetries, and (iii) corrections to Rabi frequencies and, therefore, spin-photon coupling. All of them provide experimental signatures that can be tested. These predictions are confirmed through comprehensive simulations of realistic Ge quantum dot devices using the four-band Luttinger-Kohn model.

The analytical framework we developed provides closed-form approximate expressions for all the key observable effects, applicable to any DQD system operating in the regime where the Zeeman energy is smaller than tunnel coupling. We provide the minimal model that contains this correction, which needs to be accounted for in any hole quantum dot array. Furthermore, electron quantum dot arrays may display similar corrections when considering general magnetic field gradients, see Appendix~\ref{app:electron}.

Our results are consistent with experimental observations, notably explaining the unexpected sweet spots reported in recent hole spin qubit experiments~\cite{yu2023strong, van2024coherent}. This work thus provides both fundamental understanding and practical guidance for the continued development of hole spin qubit technologies, establishing magnetotunneling as a relevant consideration for the understanding of hole spin qubits in quantum dot arrays.

\acknowledgments
Work supported by the Spanish Ministry of Science, innovation, and Universities through Grants RYC2022-037527-I and PID2023-148257NA-I00 funded by MICIU/AEI/10.13039/501100011033 and by FSE+ and FEDER, UE, “ERDF A way of making Europe” and European Union Next Generation EU/PRTR. JCAU acknowledges the support of the CSIC’s Quantum Technologies Platform (QTEP) and the Severo Ochoa Centres of Excellence program through Grant CEX2024-001445-S. The calculations were made possible through CESGA's FinisTerrae III supercomputer, which has been funded by the NextGeneration EU 2021 Recovery, Transformation and Resilience Plan, ICT2021-006904, and also from the Pluriregional Operational Programme of Spain 2014-2020 of the European Regional Development Fund (ERDF), ICTS-2019-02-CESGA-3, and from the State Programme for the Promotion of Scientific and Technical Research of Excellence of the State Plan for Scientific and Technical Research and Innovation 2013-2016 State subprogramme for scientific and technical infrastructures and equipment of ERDF, CESG15-DE-3114. YMN and EARM acknowledge support from the Horizon Europe Framework Program (grant agreement 101174557 QLSI2).

\appendix
\setcounter{figure}{0}
\setcounter{equation}{0}
\renewcommand{\thefigure}{\thesection.\arabic{figure}}
\renewcommand{\theequation}{\thesection.\arabic{equation}}

\section{Microscopic theory}
\label{appendix:microscopic}
In this Appendix, we develop the microscopic theory of a single hole spin in a DQD. We consider a DQD with axis $x=[100]$, formed in a hole gas strongly confined along $z=[001]$. We start from the Hamiltonian $H_{\mathrm{DQD}}=H_\mathrm{LK}+H_\mathrm{BP}+H_\mathrm{Z}+H_\mathrm{conf}$, where $H_\mathrm{LK}$ is the Luttinger-Kohn Hamiltonian, $H_\mathrm{BP}$ is the Bir-Pikus Hamiltonian, $H_\mathrm{Z}$ the Zeeman Hamiltonian of the holes, and $H_\mathrm{conf}$ the double-dot confinement potential. The Luttinger-Kohn Hamiltonian for the envelopes of the Bloch functions with angular momentum $J=3/2$, $J_z=\{+\tfrac{3}{2},\,+\tfrac{1}{2},\,-\tfrac{1}{2},\,-\tfrac{3}{2}\}$ reads:
\begin{equation}
H_\mathrm{LK}=\begin{pmatrix}
P+Q & -S & R & 0 \\
-S^\dagger & P-Q & 0 & R \\
R^\dagger & 0 & P-Q & S \\
0 & R^\dagger & S^\dagger & P+Q
\end{pmatrix}
\label{eq:LK},
\end{equation}
where the matrix elements are:
\begin{subequations}
\begin{align}
P&=\frac{1}{2m_0}\gamma_1(p_x^2+p_y^2+p_z^2) \\
Q&=\frac{1}{2m_0}\gamma_2(p_x^2+p_y^2-2p_z^2) \\
R&=\frac{1}{2m_0}\sqrt{3}\left[-\gamma_2(p_x^2-p_y^2)+2i\gamma_3\{p_x,\,p_y\}\right] \\
S&=\frac{1}{2m_0}2\sqrt{3}\gamma_3\{p_x-ip_y,\,p_z\}
\end{align}
\end{subequations}
with $\{A,\,B\}=\tfrac{1}{2}(AB+BA)$, $\mathbf{p}=-i\hbar\nabla_{\mathbf{r}}+e\mathbf{A}$ the momentum, $\mathbf{A}$ the vector potential, and $\gamma_i$ the Luttinger parameters. The Bir-Pikus Hamiltonian is
\begin{equation}
H_\mathrm{BP}=\begin{pmatrix}
P_\epsilon+Q_\epsilon & -S_\epsilon & R_\epsilon & 0 \\
-S_\epsilon^\dagger & P_\epsilon-Q_\epsilon & 0 & R_\epsilon \\
R_\epsilon^\dagger & 0 & P_\epsilon-Q_\epsilon & S_\epsilon \\
0 & R_\epsilon^\dagger & S_\epsilon^\dagger & P_\epsilon+Q_\epsilon
\end{pmatrix}
\label{eq:BP},
\end{equation}
where
\begin{subequations}
\begin{align}
P_\epsilon&=-a_v(\epsilon_{xx}+\epsilon_{yy}+\epsilon_{zz}) \\
Q_\epsilon&=-\frac{1}{2}b_v(\epsilon_{xx}+\epsilon_{yy}-2\epsilon_{zz}) \\
R_\epsilon&=\frac{\sqrt{3}}{2}b_v(\epsilon_{xx}-\epsilon_{yy})-id_v\epsilon_{xy} \\
S_\epsilon&=-d_v(\epsilon_{xz}-i\epsilon_{yz})
\end{align}
\end{subequations}
with $\epsilon_{ij}$ the strain tensor and $b_v$, $d_v$ the uniaxial and shear deformation potentials of the valence band. Besides, the Zeeman Hamiltonian of the holes is $H_\mathrm{Z}=2\mu_B(\kappa\mathbf{B}\cdot\mathbf{J}+q\mathbf{B}\cdot\mathbf{J}^3)$, with $\mathbf{J}$ the angular momentum operator and $\kappa$, $q$ the isotropic and cubic Zeeman parameters. We assume separable confinement in a quartic potential along $x$ (which describes the DQD), a harmonic potential along $y$, and a triangular well potential with electric field $F$ along $z$:
\begin{equation}
    H_\mathrm{conf}=\frac{1}{2d^2}m^\parallel_{h}\omega_x^2\left(\frac{x^2}{L_x^2}-\frac{d^2}{4L_x^2}\right)^2+\frac{1}{2}m^\parallel_{h}\omega_y^2y^2+eFz
\end{equation}
with $\omega_{x/y}$ the harmonic frequencies along $x$ and $y$, $L_x=\sqrt{\hbar/m^\parallel_{h}\omega_x}$ the harmonic length along $x$~\footnote{While the harmonic length depends on whether we have a heavy-hole or light-hole state through their in-plane effective masses $m^\parallel_{h,l}$, we find the difference to have no qualitative impact in our results except for complicating the expressions. Therefore, we approximate $L_x=L_x^{(h)}\approx L_x^{(l)}$ in the manuscript}, and $d$ the distance between the dots. 

\subsection{Ansatz wavefunctions}
\label{sec:ansatz}
We intend to calculate the gyromagnetic response of the DQD in a minimal subspace of left and right heavy- and light-hole states $\ket{\Psi^{(h,l)}_{L,R}}$ that captures the qualitative physics in a truncated orbital subspace. In this confinement potential, we can assume that the wave functions $\ket{\Psi^{(h,l)}_{L,R}(x,y,z)}=\ket{\psi_{L,R}^{h,l}(x)}\ket{Y^{h,l}_n(y)}\ket{Z^{h,l}(z)}$ are separable, and deal with vector potential corrections perturbatively. We find sufficient to consider single left and right states $\ket{\psi^{(h,l)}_{L,R}(x)}$ for the motion of heavy and light-holes along $x$. Two states $\ket{Y_0^{(h,l)}(y)}$, $\ket{Y_1^{(h,l)}(y)}$ with opposite parities must, however, be included, as the Hamiltonian contains important terms that flip the parity along $y$. Finally, a single heavy-hole and light-hole state $\ket{Z^{(h,l)}(z)}$ is enough to account for the Rashba interaction. We detail the choice of these wavefunctions in the following.

To ensure orthogonality and localization, the wavefunctions along $x$ are constructed in three steps. First, we build a harmonic ansatz in the local minima of the potential at $x=\pm d/2\pm\delta x^{(h,l)}$, $\ket{\psi^{0}_{L,R}(x)}=N\exp[-(x\mp d/2\mp \delta x^{(h,l)})^2/2L_x^2]$, with $N$ a normalization coefficient. The corrections $\delta x^{(h,l)}$ describe the displacement of the dots at finite detuning. We assume the same in-plane mass for heavy- and light-holes in a first approximation and thus use the same ansatz for both~\footnote{We go beyond this approximation to include detuning-dependent corrections.}. These localized wavefunctions are not orthogonal and have an overlap $s=\bra{\psi^{0}_{L}(x)}\ket{\psi^{0}_{R}(x)}=\exp(-d^2/4L_x^2)$. Yet we can construct orthonormal bonding and antibonding orbitals for the DQD as:
\begin{equation}
\begin{aligned}
    \ket{\psi_{\pm}(x)}=\frac{1}{\sqrt{2\pm 2s}}(\ket{\psi^{0}_{L}(x)}\pm\ket{\psi^{0}_{R}(x)}).
    \end{aligned}
\end{equation}
These orbitals are indeed orthogonal owing to their opposite parities along $x$. Finally, we diagonalize the $\hat{x}$ operator in the basis of the above bonding and antibonding states. The two opposite eigenvalues describe orthonormal left and right orbitals
\begin{equation}
    \begin{pmatrix}
        \psi_L \\
        \psi_R
    \end{pmatrix}=V\begin{pmatrix}
        \psi_+ \\
        \psi_-
    \end{pmatrix}
\end{equation}
with $V=\exp(i\pi\sigma_y/4)$.

We use the harmonic eigenstates $\ket{Y_n(y)}$ with quantum number $n=0,1$ as in-plane wavefunctions along $y$ (the ground even and odd states as explained above):
\begin{equation}
    \begin{aligned}
        \ket{Y_0(y)}&=N_0e^{\frac{-y^2}{2L_y^2}}, \\
        \ket{Y_1(y)}&=N_1ye^{\frac{-y^2}{2L_y^2}},
    \end{aligned}
\end{equation}
where $N_n$ are normalization coefficients and $L_y =\sqrt{\hbar/m^\parallel_{h}\omega_y}$.

Finally, the vertical $\ket{Z^{(h,l)}(z)}$ can be approximated (depending on the choice of boundary conditions) by Bastard wavefunctions in a quantum well with finite electric field or by Airy wavefunctions in a semi-infinite triangular well. For the present discussion, we choose to be agnostic for the vertical confinement potential and want to describe its qualitative effects only. Unlike for the in-plane wavefunctions, here it is key to take into account the difference between the heavy- and light-hole confinement masses. The lack of inversion symmetry introduced by the vertical electric field indeed leads to a non-zero Rashba coefficient $i\hbar\eta_R=\bra{Z^{h,l}(z)}p_z\ket{Z^{l,h}(z)}$, whose value depends on the details of the potential. We will also need to introduce the parameters $z_0=\bra{Z^{h,l}(z)}z\ket{Z^{l,h}(z)}$ as well as $i\hbar\beta=\bra{Z^{h,l}(z)}zp_z+p_zz\ket{Z^{l,h}(z)}$.

\subsection{Projection to the ansatz basis set}
\label{sec:projection}
We project the total Hamiltonian $H_{\mathrm{DQD}}$ in this subspace, then integrate out the left and right light-hole states using quasi-degenerate perturbation theory to second order. We get that way an effective Hamiltonian $H^{\text{eff}}$ in the heavy-hole subspace,
\begin{equation}
\begin{aligned}
    H^{\text{eff}}_{ij}&\approx \bra{\Psi^{(h)}_{i}}H_{\mathrm{DQD}}\ket{\Psi^{(h)}_{j}} +\frac{1}{2}\sum_{k} \left(\frac{1}{E_i-E_k}+\frac{1}{E_j-E_k}\right)\\ &\times\bra{\Psi^{(h)}_{i}}H_{\mathrm{DQD}}\ket{\Psi^{(l)}_{k}}\bra{\Psi^{(l)}_{k}}H_{\mathrm{DQD}}\ket{\Psi^{(h)}_{j}},
    \end{aligned}
\end{equation}
where $E_i$ is the energy of state $i$, and $k$ runs through the left and right light-hole orbitals with different spin projections and quantum numbers $n$ along $y$. In the $\{\ket{L\uparrow},\ket{L\downarrow},\ket{R\uparrow},\ket{R\downarrow}\}$ heavy-hole manifold (with $\ket{\uparrow}\equiv\ket{+\tfrac{3}{2}}$ and $\ket{\downarrow}\equiv\ket{-\tfrac{3}{2}}$), this Hamiltonian can be factorized as
\begin{equation}
\begin{aligned}
    H_{\text{eff}}&=\frac{\varepsilon}{2}\tau_z+t_0\tau_x+\tau_y(\mathbf{t}_{\mathrm{so}}\cdot\boldsymbol{\sigma})+\frac{\mu_B}{2}\tau_L(\boldsymbol{\sigma}\cdot\hat{g}_\mathrm{L}\mathbf{B})\\ &+\frac{\mu_B}{2}\tau_R(\boldsymbol{\sigma}\cdot\hat{g}_\mathrm{R}\mathbf{B})+H_\mathrm{MT},\\
    H_\mathrm{MT}&=\frac{\mu_B}{2}\tau_x(\boldsymbol{\sigma}\cdot\hat{g}_\mathrm{T}\mathbf{B})+\frac{1}{2}\tau_y(\boldsymbol{\mu}_\mathrm{T}\cdot\mathbf{B})
    \end{aligned}
    \label{eq:eff0appendix},
\end{equation}
where, compared to the usual toy models, there is a new $g$-matrix term $\frac{1}{2}\mu_B\tau_x(\boldsymbol{\sigma}\cdot\hat{g}_\mathrm{T}\mathbf{B})$ coupling the dots (along with the $\propto \tau_y\,\boldsymbol{\mu}_\mathrm{T}\cdot\mathbf{B}$ term). 

We can expand the Hamiltonian in series of the overlap parameter $s=\exp(-d^2/4L_x^2)$ and split each $g$-matrix into an overlap-independent part and an overlap-dependent part $\hat{g}_i=\hat{g}_i^{0}+\delta\hat{g}_i(s)$. Thus $\hat{g}_i^{0}$ is the $g$-matrix in the single dot limit $s=0$, and $\delta\hat{g}_i(s)$ are microscopic DQD corrections. The $\hat{g}_i^{0}$'s we obtain in the left and right dots are diagonal:
\begin{equation}
\begin{aligned}
    \hat{g}^{0}_{\mathrm{L}/\mathrm{R},xx}&=3q+\frac{3\hbar^2\gamma_2(\beta\gamma_3+2\kappa+4\eta_Rz_0\gamma_3)}{4m_0L_x^2\Delta_{\mathrm{LH}}} \\ &-\frac{3\hbar^2\gamma_2(\beta\gamma_3+2\kappa+4\eta_Rz_0\gamma_3)}{4m_0L_y^2\Delta_{\mathrm{LH}}}\\
    \hat{g}^{0}_{\mathrm{L}/\mathrm{R},yy}&=-3q-\frac{3\hbar^2(\beta\gamma_2\gamma_3+2\gamma_2\kappa+4\eta_Rz_0\gamma_3^2)}{4m_0L_y^2\Delta_{\mathrm{LH}}} \\ &+\frac{3\hbar^2\gamma_2(\beta\gamma_3+2\kappa+4\eta_Rz_0\gamma_3)}{4m_0L_x^2\Delta_{\mathrm{LH}}} \\
    \hat{g}^{0}_{\mathrm{L}/\mathrm{R},zz}&=6\kappa+\frac{27}{2}q-6\frac{\hbar^2\eta_R^2\gamma_3^2}{m_0\Delta_{\mathrm{LH}}},
    \end{aligned}
    \label{eq:glocal0}
\end{equation}
where $\Delta_\mathrm{LH}\approx\frac{2\pi^2\hbar^2\gamma_2}{m_0L_\mathrm{W}^2}+2b_v(\epsilon_\parallel-\epsilon_\perp)$ is the heavy-hole/light-hole splitting, with $L_\mathrm{W}$ the vertical confinement length, $\epsilon_\parallel=\epsilon_{xx}=\epsilon_{yy}$ and $\epsilon_\perp=\epsilon_{zz}$ in biaxially strained devices. For the sake of simplicity, we have assumed identical dots $L_x=L_{x,L}=L_{x,R}$, $L_y=L_{y,L}=L_{y,R}$, $\eta_R=\eta_{R,L}=\eta_{R,R}$, $z_0=z_{0,L}=z_{0,R}$, and $\beta=\beta_{L}=\beta_{R}$, but the expressions can be easily generalized to non-equivalent dots. They are qualitatively consistent with those of Ref.~\cite{martinez2022hole}, but differ quantitatively due to the truncation of the light-hole subspace.

We also find specific DQD corrections
\begin{equation}
    \begin{aligned}
        \delta\hat{g}_{\mathrm{L}/\mathrm{R},xx}&=\frac{3d^2\hbar^2\left(\beta\gamma_2\gamma_3+2\kappa\gamma_2+4\gamma_3^2\eta_Rz_0\right)}{4m_0L_x^4\Delta_{\mathrm{LH}}}s^2 \\
        \delta\hat{g}_{\mathrm{L}/\mathrm{R},yy}&=\frac{3d^2\hbar^2\gamma_2\left(\beta\gamma_3+2\kappa+4\gamma_3\eta_Rz_0\right)}{4m_0L_x^4\Delta_{\mathrm{LH}}}s^2 \\
        \delta\hat{g}_{\mathrm{L}/\mathrm{R},zz}&=-\frac{3d^2\hbar^2\gamma_2\gamma_3}{2m_0L_x^4m_0\Delta_{\mathrm{LH}}}s^2
    \end{aligned}
\end{equation}
to second order in $s$, as well as extra terms that depend on the motion $\delta x_\pm=\tfrac{1}{2}(\delta x^{(h)}\pm\delta x^{(l)})$ of the dots upon detuning:
\begin{equation}
    \begin{aligned}
        \delta\hat{g}^\varepsilon_{xx}&=-\frac{3d^3\hbar^2\gamma_3\left(\beta\gamma_2+4\gamma_3(1-2\eta_Rz_0)\right)}{4m_0L_x^6\Delta_{\mathrm{LH}}}\delta x_-s^2\tau_z \\
        \delta\hat{g}^\varepsilon_{yy}&=-\frac{3d\hbar^2\gamma_2\gamma_3}{4m_0L_x^6m_0\Delta_{\mathrm{LH}}}s^2\tau_z\\ &\times\left(d^2\delta x_+-\frac{(L_y^2L_x^2+(\beta-1)d^2L_y^2-L_x^4)\delta x_-}{L_y^2}\right) \\
        \delta\hat{g}^\varepsilon_{zz}&=-\frac{3d\hbar^2\gamma_2\gamma_3}{4m_0L_x^6m_0\Delta_{\mathrm{LH}}}s^2\tau_z\\ &\times\left(d^2\delta x_+-\frac{(L_y^2L_x^2+d^2L_y^2-L_x^4)\delta x_-}{L_y^2}\right),
    \end{aligned}
    \label{eq:gdetuning}
\end{equation}
where the superscript $\varepsilon$ indicates the relationship with detuning. At zero detuning, $\delta x_\pm=0$.

The $\hat{g}_\mathrm{T}$ matrix is also diagonal and its elements are given by
\begin{equation}
    \begin{aligned}
        \delta\hat{g}_{\mathrm{T},xx}&=-\frac{4\gamma_2\left(\beta\gamma_3+2\kappa\right)m^\parallel_{h}}{m_0\Delta_{\mathrm{LH}}}t_c \\
        \delta\hat{g}_{\mathrm{T},yy}&=\frac{8\left(2\gamma_3^2L_x^2-\gamma_2\left(\beta\gamma_3+2\kappa\right)L_y^2\right)m^\parallel_{h}}{m_0L_y^2\Delta_{\mathrm{LH}}}t_c \\
        \delta\hat{g}_{\mathrm{T},zz}&=-\frac{64\gamma_3(\gamma_2+2\eta_R^2\gamma_3L_y^2)L_x^2m^\parallel_{h}}{3m_0L_y^2\Delta_{\mathrm{LH}}}t_c,
    \end{aligned}
    \label{eq:gTappendix}
\end{equation}
up to first order in $\eta_R$ and $1/\Delta_{\mathrm{LH}}$. Therefore, we find that the $\hat{g}_\mathrm{T}$-matrix elements can be tracked down to the interplay between the kinetic-dependence of the heavy-hole light-hole interactions and the Zeeman term, with small vector potential corrections. 

Finally, the tunnel couplings in our theory are $t_0\approx 3\hbar^2d^2\csch(d^2/4L_x^2)/(4m_h^\parallel L_x^4)$ and
\begin{equation}
    \mathbf{t}_{\mathrm{so}}\approx -\frac{3\hbar^2\eta_R\gamma_3\left((\gamma_2+2\gamma_3)L_x^2/L_y^2-\gamma_2\right)m^\parallel_{h}}{dm_0^2\Delta_{\mathrm{LH}}}t_0\mathbf{e}_y.
\end{equation}
The spin-orbit vector thus points in the $y$ direction as expected when the DQD axis is along $x$ and there is a Rashba spin-orbit interaction.

\subsection{Peierls phase and the \texorpdfstring{$\bm{\mu}_{\mathrm{T}}$}{mu\_T} vector}
After integrating out the light-hole states, our effective Hamiltonian to $O(1/\Delta_\mathrm{LH})$ has the form
\begin{equation}
\begin{aligned}
    H_{\text{eff}}^{(0)}&=\frac{\varepsilon}{2}\tau_z+t_0\tau_x+\tau_y(\mathbf{t}_{\mathrm{so}}\cdot\boldsymbol{\sigma})+\frac{\mu_B}{2}\tau_L(\boldsymbol{\sigma}\cdot\hat{g}_\mathrm{L}\mathbf{B})\\ &+\frac{\mu_B}{2}\tau_R(\boldsymbol{\sigma}\cdot\hat{g}_\mathrm{R}\mathbf{B})+H_\mathrm{MT}^{(0)},\\
    H_\mathrm{MT}^{(0)}&=\frac{\mu_B}{2}\tau_x(\boldsymbol{\sigma}\cdot\hat{g}^{(0)}_\mathrm{T}\mathbf{B})
    \end{aligned},
    \label{eq:eff1appendix}
\end{equation}
which, compared to Eq.~\eqref{eq:eff0appendix} misses the $\propto(\boldsymbol{\mu}_\mathrm{T}\cdot\mathbf{B})\tau_y$ term. To find such a term in our microscopic theory, we must account for the action of the vector potential on the orbital motion of heavy-holes. Eqs.~\eqref{eq:glocal0}-\eqref{eq:gTappendix} indeed only include the effect of the vector potential on the heavy-hole/light-hole mixing. The proper way to include such an orbital contribution to all-orders in $\mathbf{B}$ within the tight-binding approximation is through the Peierls phase~\cite{graf1995electromagnetic, ismail2001coupling}. This modifies the tunneling matrix elements such that $\tau_x\rightarrow e^{i\varphi}\tau_++e^{-i\varphi}\tau_-$ and $\tau_y\rightarrow -ie^{i\varphi}\tau_++ie^{-i\varphi}\tau_-$, with $\varphi=\frac{e}{\hbar}\int_{\mathbf{x}_L}^{\mathbf{x}_R}\mathbf{A}\cdot d\mathbf{r}$ and $\mathbf{x}_{L/R}$ the position of left and right dots. We next expand $\varphi=\varphi_1(\mathbf{B})+O(B^2)$ in powers of the magnetic field and map to Eq.~\eqref{eq:eff0appendix} with:
\begin{equation}
\begin{aligned}
    \boldsymbol{\mu}_\mathrm{T}\cdot\mathbf{B}&=-2t_0\varphi_1(\mathbf{B}), \\
    \hat{g}_\mathrm{T}&=\hat{g}_\mathrm{T}^{(0)}-\frac{1}{\mu_B}\tan\theta_\mathrm{so}\mathbf{n}_\mathrm{so}\otimes\boldsymbol{\mu}_\mathrm{T},
\end{aligned}
\end{equation}
where we have used $\theta_\mathrm{so}=\arctan(|\mathrm{t}_\mathrm{so}|/t_0)$ and $\mathbf{n}_\mathrm{so}=\mathbf{t}_\mathrm{so}/|\mathbf{t}_\mathrm{so}|$. We have assumed for simplicity that $\mathbf{A}$ is odd with respect to $\mathbf{B}$ so that there is no zeroth-order term \cite{Note2}. 
Therefore, the Peierls phase leads to a non-zero $\boldsymbol{\mu}_\mathrm{T}$ and introduces a gauge-dependent correction to $\hat{g}_\mathrm{T}$ by substracting $\tan\theta_\mathrm{so}\mathbf{n}_\text{so}\otimes\boldsymbol{\mu}_\mathrm{T}/\mu_B$. Interestingly, by going to the spin-orbit frame with the unitary $\hat{U}=\exp\left[\mathrm{i}\tau_z\theta_{\mathrm{so}}\sigma_{\mathrm{so}}/2\right]$, this gauge-dependent correction exactly cancels out, leading to $\tilde{g}_\mathrm{T}=\left[\mathds{1}-(1-\cos\theta_\mathrm{so})\mathbf{n}_\mathrm{so}\otimes \mathbf{n}_\mathrm{so}\right]\hat{g}_\mathrm{T}^{(0)}$. In this frame, $\tilde{\boldsymbol{\mu}}_\mathrm{T}=\sec\theta_\mathrm{so}\boldsymbol{\mu}_\mathrm{T}-\sin\theta_\mathrm{so}(^t\hat{g}_\mathrm{T}^{(0)} \mathbf{n}_\text{so})$, where $\sec\theta_\mathrm{so}\boldsymbol{\mu}_\mathrm{T}\cdot\mathbf{B}=-2t_c\varphi_1$. Therefore, we find that the magnetotunneling $\tilde{g}_\mathrm{T}$ matrix in the spin-orbit frame does not depend on the choice of gauge for $\boldsymbol{\mu}_\mathrm{T}$, unlike $\hat{g}_\mathrm{T}$. This is expected since $\tilde{g}_\mathrm{T}$ is directly connected to the observable $g$-factors as shown in Sec.~\ref{sec:sweet}. We further discuss gauge-invariance in Appendix~\ref{app:gauge}.

Finally, we note that for a given working point, one may always choose a gauge such that $\boldsymbol{\mu}_\mathrm{T}=\mathbf{0}$ even if this vector is not only given by the Peierls phase. To see this, let us start from an arbitrary $\boldsymbol{\mu}_\mathrm{T}$. We can now change the gauge to $\mathbf{A}'=\mathbf{A}+\nabla_\mathbf{r}\chi(\mathbf{r},\mathbf{B})$ with $\chi(\mathbf{r},\mathbf{B})\equiv\chi_x(\mathbf{r})B_x+\chi_y(\mathbf{r})B_y+\chi_z(\mathbf{r})B_z$ linear in $\mathbf{B}$. In this new gauge, we have
\begin{equation}
    \boldsymbol{\mu}'_\mathrm{T}\cdot\mathbf{B}=\boldsymbol{\mu}_\mathrm{T}\cdot\mathbf{B}-2t_0\frac{e}{\hbar}\left[\chi(\mathbf{x}_R,\mathbf{B})-\chi(\mathbf{x}_L,\mathbf{B})\right].
\end{equation}
Since the $\chi_\alpha$'s may be any smooth functions, they can be chosen to cancel out $\boldsymbol{\mu}_\mathrm{T}$, leading to $\boldsymbol{\mu}'_\mathrm{T}=\mathbf{0}$. However, the ability to gauge out the $\boldsymbol{\mu}_\mathrm{T}$ vector does not guarantee that $\tilde{\boldsymbol{\mu}}_\mathrm{T}$ cancels out in the spin–orbit frame. Conversely, one could choose a gauge that removes $\tilde{\boldsymbol{\mu}}_\mathrm{T}$  in the spin–orbit frame. Yet, as long as both magnetotunneling and spin–orbit terms are present, $\boldsymbol{\mu}_\mathrm{T}$ and $\boldsymbol{\tilde\mu}_\mathrm{T}$ cannot be simultaneously chosen to be zero by a gauge choice.

\section{Electron double-quantum dots}
\label{app:electron}
While the magnetotunneling effect naturally occurs in hole DQDs due to the interplay between the kinetic, the Zeeman terms of the Luttinger-Kohn Hamiltonian, and the vector potential, a similar effect may occur in electron DQDs as well. Let a general magnetic field gradient be 
\begin{equation}
    H_\text{grad}=\frac{1}{2}\mu_B\boldsymbol{\sigma}\cdot\hat{g}\mathbf{B}(\mathbf{r}),
\end{equation}
where $\hat{g}\approx 2$ is in a first approximation a scalar for electrons, and $\mathbf{B}(\mathbf{r})$ is a position-dependent magnetic field. The projected Zeeman field at each dot is then given by
\begin{equation}
    \begin{aligned}
    H&_\text{eff}^{(s,s')}=\sum_{i=L,R}\bra{i,s}H_\text{grad}\ket{i,s'}\ket{i,s}\bra{i,s'}\\
    &=\frac{1}{4}\mu_B\boldsymbol{\sigma}\cdot\hat{g}\big[(\langle\mathbf{B}(\mathbf{r})\rangle_L+\langle\mathbf{B}(\mathbf{r})\rangle_R)\\ &+(\langle\mathbf{B}(\mathbf{r})\rangle_L-\langle\mathbf{B}(\mathbf{r})\rangle_R)\tau_z\big]\\
    &+\frac{1}{2}\mu_B\boldsymbol{\sigma}\cdot\hat{g}\left[\bra{L}\mathbf{B}(\mathbf{r})\ket{R}\tau_-+\bra{R}\mathbf{B}(\mathbf{r})\ket{L}\tau_+\right],
    \end{aligned}
\end{equation}
where $s$ and $s'$ indicate spin. This gradient first introduces a difference in Zeeman field in each dot (second line). As an example, a linear gradient $H_\text{grad}=\alpha_zx\boldsymbol{\sigma}\mathbf{B}$ leads to opposite expectation values for left and right dots, since $\mathbf{x}_L \approx -\mathbf{x}_R$, see, for instance, Ref.~\cite{benito2019electric}. Apart from this well-known contribution, a magnetotunneling effect connecting left and right eigenstates may arise if $\bra{L,s}H_\text{grad}\ket{R,s'}\neq 0$. If we assume identical dots with $\mathbf{x}_L \approx -\mathbf{x}_R$, a constant gradient cannot give rise to such a correction. A quadratic term, however, would lead to a magnetotunneling term. As an example, we provide an estimate using the envelopes for the left and right states given in Appendix~\ref{appendix:microscopic}. Taking $H_\text{grad}=\alpha_{z2}x^2\boldsymbol{\sigma}\mathbf{B}$, we get
\begin{equation}
    \delta H_\text{grad}=\frac{\alpha_{z2}d^2}{4}\csch(d^2/4L_x^2)\boldsymbol{\sigma}\cdot\mathbf{B}\tau_x.
\end{equation}
Overall, any position-dependent even contribution along $x$ would introduce a magnetotunneling correction. 

\section{Gauge invariance in linear response} 
\label{app:gauge}

Calculations in finite basis sets may break gauge invariance as the basis set is not able to accomodate arbitrary phase variations. In the numerical simulations of the manuscript, we deal with the magnetic field to first order in $B$ with the $g$-matrix formalism \cite{venitucci2018electrical}. This implies that the magnetic corrections are calculated perturbatively starting from the wavefunctions at zero magnetic field. It has been proven that this formalism is gauge invariant as long as the analysis is restricted to time-reversal symmetric eigenstates (Kramers pairs)~\cite{venitucci2018electrical}. In the present DQDs, we compute the lowest four eigenstates $\{\ket{0},\ket{1},\ket{2},\ket{3}\}$ at $\mathbf{B}=\mathbf{0}$ and each value of detuning and Wannierize this eigenbasis to localized left and right orbitals $\{\ket{L\uparrow},\ket{L\downarrow},\ket{R\uparrow},\ket{R\downarrow}\}$. In the $\{\ket{0},\ket{1},\ket{2},\ket{3}\}$ basis set, the effects of the magnetic field are described to first-order by the matrix
 \begin{equation} 
    H_{nm}=\left\langle n\left|\frac{\partial H}{\partial \mathbf{B}}\cdot \mathbf{B}\right|m\right\rangle\equiv\begin{pmatrix}
        H_\mathrm{AA} & H_\mathrm{AB} \\
        H_\mathrm{AB}^\dagger & H_\mathrm{BB}
    \end{pmatrix}.
    \label{eq:B}
\end{equation}
The set A contains the ground, degenerate $\{\ket{0},\ket{1}\}$ eigenstates and the set B the excited, degenerate $\{\ket{2},\ket{3}\}$ eigenstates. While the diagonal $2\times2$ blocks $H_\mathrm{AA}$ and $H_\mathrm{BB}$ are gauge-invariant (because $\{\ket{0},\ket{1}\}$, as well as $\{\ket{2},\ket{3}\}$ are time-reversal symmetric eigenstates), the off-diagonal block $H_\mathrm{AB}$ may include gauge-dependent contributions. However, this off-diagonal block only gives rise to second-order corrections to the Zeeman splittings of the A and B states. Therefore, we can discard $H_\mathrm{AB}$ and thus remove any gauge-dependent contribution without any effect on the $g$-factors of the ground and first excited pairs.

We thus do so before the rotation to the Wannierized basis set $\{\ket{L\uparrow},\ket{L\downarrow},\ket{R\uparrow},\ket{R\downarrow}\}$. Consequently, we remove gauge-dependent terms from $\hat{g}_\mathrm{T}$ and $\boldsymbol{\mu}_\mathrm{T}$. We have explicitly checked that eliminating the off-diagonal block $H_\mathrm{AB}$ does not affect the observable $g$-factors and, indeed, removes spurious contributions to the $g$-matrices in the Wannierized basis set.

\section{Strain corrections}
\label{app:strain}
Any local strain in a given dot gives rise to corrections to the corresponding $g$-factors \cite{abadillo2023hole}, given by:
\begin{equation}
\label{eq:gHH}
\begin{aligned}
\delta \hat{g}_{\mathrm{L}/\mathrm{R},xx}&=\delta \hat{g}_{\mathrm{L}/\mathrm{R},yy}=\frac{6b_v\kappa}{\Delta_\mathrm{LH}}\left(\langle\epsilon_{yy}\rangle_{\mathrm{L},\mathrm{R}}-\langle\epsilon_{xx}\rangle_{\mathrm{L},\mathrm{R}}\right)\\
\delta \hat{g}_{\mathrm{L}/\mathrm{R},zy}&=-\frac{4\sqrt{3}\kappa d_v}{\Delta_\mathrm{LH}}\langle\epsilon_{yz}\rangle_{\mathrm{L},\mathrm{R}} \\
\delta \hat{g}_{\mathrm{L}/\mathrm{R},zx}&=-\frac{4\sqrt{3}\kappa d_v}{\Delta_\mathrm{LH}}\langle\epsilon_{xz}\rangle_{\mathrm{L},\mathrm{R}} \\
\delta \hat{g}_{\mathrm{L}/\mathrm{R},xy}&=-\delta \hat{g}_{\mathrm{L}/\mathrm{R},yx}=\frac{4\sqrt{3}d_v\kappa}{\Delta_\mathrm{LH}}\langle\epsilon_{xy}\rangle_{\mathrm{L},\mathrm{R}}.
\end{aligned}
\end{equation}
Furthermore, there can be strain-induced corrections to the tunneling $g$-matrix as well:
\begin{equation}
\label{eq:gHHt}
\begin{aligned}
\delta \hat{g}_{\mathrm{T},xx}&=\delta \hat{g}_{\mathrm{L}/\mathrm{R},yy}=\frac{6b_v\kappa}{\Delta_\mathrm{LH}}\bra{L}\epsilon_{yy}-\epsilon_{xx}\ket{R}\\
\delta \hat{g}_{\mathrm{T},zy}&=-\frac{4\sqrt{3}\kappa d_v}{\Delta_\mathrm{LH}}\bra{L}\epsilon_{yz}\ket{R} \\
\delta \hat{g}_{\mathrm{T},zx}&=-\frac{4\sqrt{3}\kappa d_v}{\Delta_\mathrm{LH}}\bra{L}\epsilon_{xz}\ket{R} \\
\delta \hat{g}_{\mathrm{T},xy}&=-\delta \hat{g}_{\mathrm{T},yx}=\frac{4\sqrt{3}d_v\kappa}{\Delta_\mathrm{LH}}\bra{L}\epsilon_{xy}\ket{R}.
\end{aligned}
\end{equation}
Finally, we emphasize that inhomogeneous strains may contribute to the spin-dependent tunneling mechanism through strain-induced Rashba and Dresselhaus-like spin-orbit interactions~\cite{abadillo2023hole}. This correction to the tunneling Hamiltonian is $\delta H_{\mathrm{so}}=\tau_y(\delta t_{\mathrm{so},x},\delta t_{\mathrm{so},y},0)\cdot\boldsymbol{\sigma}$, with
\begin{equation}
\label{eq:tso-strain}
\begin{aligned}
\delta t_{\mathrm{so},x}&=-\frac{d_vt_0m^\parallel_{h}}{2\sqrt{3}m_0\Delta_{\mathrm{LH}}} \\
&\times(8\gamma_3\eta_RL_x^2\bar{\epsilon}_{xy}/d+d\gamma_2\Delta\epsilon_{yz}) \\
\delta t_{\mathrm{so},y}&=\frac{t_0m^\parallel_{h}}{6m_0\Delta_{\mathrm{LH}}} \\
&\times\left(12b_v\gamma_3\eta_RL_x^2(\bar{\epsilon}_{xx}-\bar{\epsilon}_{yy})/d+\gamma_2\Delta\epsilon_{yz}\right),
\end{aligned}
\end{equation}
where we have introduced the dot average and difference $\bar{\epsilon}_{ij}=\langle \tfrac{1}{2}(\epsilon_{ij}\rangle_\mathrm{L}+\langle\epsilon_{ij}\rangle_\mathrm{R})$ and $\Delta\epsilon_{ij}=\langle \epsilon_{ij}\rangle_\mathrm{L}-\langle\epsilon_{ij}\rangle_\mathrm{R}$, which characterize the inhomogeneous strains in the truncated subspace of the $\ket{L}$ and $\ket{R}$ orbitals.

\section{Simulation of Si nanowire devices}
\label{app:Si}
\begin{figure}
\includegraphics[width=1\columnwidth]{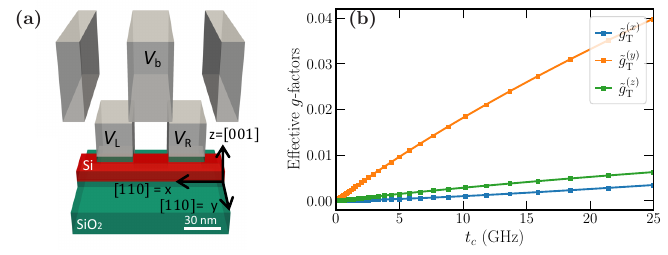}
\caption{Simulation of a Si nanowire device. (a) Simulated device with barrier gates and left and right plunger gates defining the DQD. (b) Effective tunneling $g$-factors $\tilde{g}_\mathrm{T}^{(\mathbf{b})}=|\tilde{g}_\mathrm{T}\mathbf{b}|$ as a function of the tunnel coupling $t_c$.}
\label{fig:Si}
\end{figure}
To validate the broader applicability of our magnetotunneling theory, we perform simulations in the SiMOS nanowire device shown in Fig.~\ref{fig:Si}(a). These devices represent an important complementary platform to the Ge quantum dots studied above, as they are widely used for hole spin qubits and exhibit distinct material and confinement properties that allow to test our theoretical predictions in a different regime. The device includes a set of front plunger gates with bias $V_L$ and $V_R$ together with a barrier gate with bias $V_b$ in a second metallization level. The gates are 30 nm long and are separated by 60 nm center to center. The nanowire is oriented along the $x=[110]$ crystallographic direction.

Si nanowire devices differ significantly from the planar Ge heterostructures studied previously. They do not, in particular, undergo biaxial strain due to lattice mismatch, so that the heavy-hole light-hole splitting $\Delta_{\mathrm{LH}}$ is around one order of magnitude smaller than in Ge~\cite{venitucci2018electrical, michal2021longitudinal,wang2024electrical}. In fact, we find $\Delta_\mathrm{LH}\approx 3$ meV in our Si simulations. Moreover, the hole wavefunction is squeezed against the side and top facets by the electric field from the front gates~\cite{Voisin_2014}. Both the smaller $\Delta_{\mathrm{LH}}$ and the strong lateral confinement along $y$ and $z$ enhance the heavy-hole light-hole hybridization compared to planar Ge devices, which results in larger in-plane $g$-factors~\cite{piot2022single, yu2023strong, bassi2024optimal}. Additionally, Si has a much smaller split-off energy $\Delta_\mathrm{so}$ than Ge (see Table~\ref{tab}), and thus requires simulations with the six bands $\mathbf{k}\cdot\mathbf{p}$ model rather than the four bands Luttinger-Kohn Hamiltonian.

Figure~\ref{fig:Si}(b) shows the effective magnetotunneling $g$-factors $\tilde{g}_\mathrm{T}^{(\mathbf{b})}$ as a function of the tunnel coupling in the Si device. Across all simulations they remain $\lesssim 0.04$. These values are smaller, yet of the same order of magnitude as the magnetotunneling $g$-factors we obtain in Ge. However, while for Ge these corrections were comparable to the in-plane localized $g$-factors ($g^{\mathbf{b}}\approx0-0.3$), for Si the localized $g$-factors are larger, making the magnetotunneling terms smaller in comparison ($g^{\mathbf{b}}\approx0.5-2$)~\cite{bassi2024optimal}.

To understand the underlying physics, we examine the numerator and denominator of Eqs.~\eqref{eq:gt} for $\hat{g}_\mathrm{T}$. Although these formulas were obtained by treating light-holes perturbatively, which may not be fully correct for Si nanowires, they shall approximately reproduce the main trends. While the ratio $m^\parallel_{h}/(m_0\Delta_{\mathrm{LH}})$ is roughly 100 times larger in Si, the numerators are quadratic in $\gamma_2$, $\gamma_3$, and $\kappa$. Moreover, $\gamma_2\leftrightarrow \gamma_3$ must be exchanged in the analytical expressions for devices oriented along $[110]$. Therefore, all magnetotunneling $g$-factors of the Si nanowire are proportional to either $\gamma_2$ or $\kappa$, which are nearly an order of magnitude smaller than in Ge (see Table~\ref{tab}). As a consequence, the numerators of Eqs.~\eqref{eq:gt} are a few tens of times smaller in Si than in Ge, so that the $\hat{g}_\mathrm{T}$ matrices are approximately of the same order of magnitude. Interestingly, $\kappa$ is negative in Si, which potentially leads to partial cancellations in the numerator in $\hat{g}_{{\mathrm{T}},xx}$.

Given these scalings, and for the parameter range considered, the magnetotunneling contribution is expected to be of lower importance in Si than in Ge. Nevertheless, it remains compatible with Ref.~\cite{yu2023strong}, where a multi-hole Si nanowire spin couples to a cavity photon: the measured spin-orbit field and $g$-matrices alone were not sufficient to explain the observed $g$-factor renormalization near zero detuning in the low spin-orbit-mixing cases. We therefore suspect that the magnetotunneling term is larger in that device than in our simulations, possibly due to strain and/or the distinct voltage and hole occupation regime. Therefore, in cases where the magnetic field is chosen such that the spin-orbit hybridization is low, the magnetotunneling term may play a role even in Si.

\begin{table}[h]
\centering
\begin{tabular}{|l|d|d|}
\hline
& \multicolumn{1}{c|}{Si} & \multicolumn{1}{c|}{Ge} \\ 
\hline
$\Delta_\mathrm{so}$ (eV) & 0.044 & 0.29 \\
$\gamma_1$ & 4.285 & 13.38 \\
$\gamma_2$ & 0.339 & 4.24 \\
$\gamma_3$ & 1.446 & 5.69 \\
$\kappa$ & -0.42 & 3.41 \\
$q$ & 0.06 & 0.01 \\
\hline
\end{tabular}
\caption{Spin-orbit splitting energy $\Delta$ in the valence band, Luttinger parameters, $\kappa$, and $q$ parameters~\cite{winkler2001spin} for Si and Ge materials.}
\label{tab}
\end{table}

\section{Wannierization and pseudospin basis}
\label{appendix:wannier}

In this Appendix, we explain how we rotate the basis of the four lowest eigenstates of the simulated DQD device to left and right localized orbitals with consistent pseudospin definitions. These orbitals are used to map the DQD Hamiltonian onto Eq.~\eqref{eq:eff0}. First, the output of the simulation includes (at zero magnetic field) the degenerate eigenstates $\ket{0}$ and $\ket{1}$ of the lowest Kramers pair, and the degenerate eigenstates $\ket{2}$ and $\ket{3}$ of the first excited Kramers pair. They are linear combinations of the desired L and R orbitals; to identify these orbitals, we define a virtual detuning gate operator in the basis set $\{\ket{0},\ket{1},\ket{2},\ket{3}\}$:
\begin{equation}
    H_\text{det}^{nm}=e\bra{n}\alpha v_L(x,y,z)-\beta v_R(x,y,z)\ket{m},
    \label{eq:det}
\end{equation}
where $v_{L,R}(x,y,z)$ are the potentials created by the L and R gates (with all other gates grounded), and $\alpha$, $\beta$ are two scalars such that $\alpha^2+\beta^2=1$. We choose $\alpha$ and $\beta$ such that $H_\text{det}$ has two opposite, degenerate eigenvalues \footnote{Specifically, $\alpha=\mathrm{Tr}(D_R)/\sqrt{\mathrm{Tr}(D_L)^2+\mathrm{Tr}(D_R)^2}$ and $\beta=\mathrm{Tr}(D_L)/\sqrt{\mathrm{Tr}(D_L)^2+\mathrm{Tr}(D_R)^2}$, where $\mathrm{Tr}$ is the trace and $D_L$, $D_R$ are the matrices of $v_L$, $v_R$ in the $\{\ket{0},\ket{1},\ket{2},\ket{3}\}$ basis set.}, which thus describe orbitals that are detuned symetrically by the virtual gate $\alpha v_L-\beta v_R$. They are the localized left and right orbitals $\{\ket{L\Uparrow},\ket{L\Downarrow},\ket{R\Uparrow'},\ket{R\Downarrow'}\}$, and the parameters $\alpha$ and $\beta$ define the ``pure detuning'' axis of the DQD in the $(v_L,\,v_R)$ plane.

At this point, it is already possible to extract localized $g$-matrices for given detuning values; however, the pseudospin axis remains random (since the Kramers pairs are degenerate) and not even consistent across different dots (which complicates the analysis of these $g$-matrices). To solve this issue we take another step and rotate the pseudospin axis in each quantum dot so that $\ket{\Uparrow}$, $\ket{\Downarrow}$, $\ket{\Uparrow'}$, $\ket{\Downarrow'}$ are the purest possible $m_J=\pm 3/2$ states (in order to ease comparisons with perturbation theories such as Appendix \ref{appendix:microscopic} that start from real $m_J=\pm 3/2$ envelopes). Namely, we apply a rotation matrix $W_s$ in each quantum dot so that the real part of the $m_J=+3/2$ envelope of the new spin state $\ket{\uparrow}$ has the largest possible norm (and likewise for the $m_J=-3/2$ envelope of $\ket{\downarrow}$). This transformation provides a $\{\ket{L\uparrow},\ket{L\downarrow},\ket{R\uparrow},\ket{R\downarrow}\}$ basis, allowing for a consistent analysis of the localized as well as tunneling spin physics.

\section{Extension to larger 1D arrays}
\label{app:1Dchains}
Along the manuscript we have focused on the minimal element of a sparse array, namely a double quantum dot. The central result is the emergence of a magnetotunneling $g_\mathrm{T}$-matrix, which renormalizes the Zeeman splitting in the vicinity of interdot charge transitions. In larger arrays operated in the single-charge regime, one may tune gate voltages such that more than one neighboring transition becomes relevant simultaneously. For example, when the chemical potentials of several connected dots are aligned, the orbital ground state can delocalize over multiple sites. In this Appendix, we analyze how magnetotunneling terms enter in this multi-site situation and how they modify the spin splitting near such multi-dot charge degeneracies.

We start by extending Eq.~\eqref{eq:effso} to an array of $N$ dots. In 1D arrays, it is always possible to find a local choice of pseudospin axes that eliminates the spin-flip tunneling, such that the model becomes
\begin{equation}
\begin{aligned}
    H^{\text{eff}}_{\text{so},N}&=\sum_{i=1}^N\frac{\varepsilon_i}{2}\ket{i}\bra{i}+\frac{\mu_B}{2}(\boldsymbol{\sigma}\cdot\tilde{g}_i\mathbf{B})\ket{i}\bra{i}+\\ &+\sum_{j=1}^{N-1}\left[t_{c,j}+\frac{\mu_B}{2}(\boldsymbol{\sigma}\cdot\tilde{g}_{\mathrm{T},j}\mathbf{B})-\frac{i}{2}(\tilde{\boldsymbol{\mu}}_{\mathrm{T},j}\cdot\mathbf{B})\right]\\
    &\times\ket{j}\bra{j+1}+\mathrm{H.c.},
    \end{aligned}
    \label{eq:multieffso}
\end{equation}
where $\varepsilon_i$ is the chemical potential of site $i$, $t_{c,j}$ is the tunnel coupling between sites $j$ and $j+1$, $\tilde{g}_i$ is the local $g$-matrix of site $i$, $\tilde{g}_{\mathrm{T},j}$ is the $g_\mathrm{T}$-matrix linking sites $j$ and $j+1$, and $\tilde{\boldsymbol{\mu}}_{\mathrm{T},j}$ is the corresponding vector for each link $j$. To achieve a charge transition involving $N$ dots, we can choose $\varepsilon_i=0\ \forall\ i$. For simplicity, we consider all tunnel couplings to be the same $t_{c,j}=t_c$. Analogously to the DQD case, we can now find a unitary transformation that exactly diagonalizes the orbital part for this zero-detuning scenario with identical tunnel couplings~\cite{hu1996analytical}. In the charge-diagonal frame, we obtain the Larmor vector $\boldsymbol{\omega}_{01,N}=\boldsymbol{\omega}_{+,N}-\boldsymbol{\omega}_{\mathrm{T},N}$, where $\boldsymbol{\omega}_{+,N}$ and $\boldsymbol{\omega}_{\mathrm{T},N}$ are the site-weighted vectors:
\begin{equation}
\begin{aligned}
    \boldsymbol{\omega}_{+,N}&=\sum_{k=1}^N\frac{2\mu_B}{N+1}\sin^2\left(\frac{k\pi}{N+1}\right)\tilde{g}_k\mathbf{B}\\ \boldsymbol{\omega}_{\mathrm{T},N}&=\sum_{k=1}^{N-1}\frac{4\mu_B}{N+1}\sin\left(\frac{k\pi}{N+1}\right)\sin\left(\frac{(k+1)\pi}{N+1}\right)\tilde{g}_{\mathrm{T},k}\mathbf{B}.
    \end{aligned}
    \label{eq:multilarmor}
\end{equation}
Therefore, we conclude that in the multi-charge transition regime, the magnetotunneling also renormalizes the Larmor vector. This renormalization is, however, weighted over multiple links yet its effect on the spectrum is qualitatively similar to that of the DQD regime. In the more general scenario with link-dependent tunnel couplings, these weights do not have the simple closed forms of Eq.~\eqref{eq:multilarmor} and may need to be computed numerically.

\bibliography{bib}

\begin{thebibliography}{72}%
\makeatletter
\providecommand \@ifxundefined [1]{%
 \@ifx{#1\undefined}
}%
\providecommand \@ifnum [1]{%
 \ifnum #1\expandafter \@firstoftwo
 \else \expandafter \@secondoftwo
 \fi
}%
\providecommand \@ifx [1]{%
 \ifx #1\expandafter \@firstoftwo
 \else \expandafter \@secondoftwo
 \fi
}%
\providecommand \natexlab [1]{#1}%
\providecommand \enquote  [1]{``#1''}%
\providecommand \bibnamefont  [1]{#1}%
\providecommand \bibfnamefont [1]{#1}%
\providecommand \citenamefont [1]{#1}%
\providecommand \href@noop [0]{\@secondoftwo}%
\providecommand \href [0]{\begingroup \@sanitize@url \@href}%
\providecommand \@href[1]{\@@startlink{#1}\@@href}%
\providecommand \@@href[1]{\endgroup#1\@@endlink}%
\providecommand \@sanitize@url [0]{\catcode `\\12\catcode `\$12\catcode `\&12\catcode `\#12\catcode `\^12\catcode `\_12\catcode `\%12\relax}%
\providecommand \@@startlink[1]{}%
\providecommand \@@endlink[0]{}%
\providecommand \url  [0]{\begingroup\@sanitize@url \@url }%
\providecommand \@url [1]{\endgroup\@href {#1}{\urlprefix }}%
\providecommand \urlprefix  [0]{URL }%
\providecommand \Eprint [0]{\href }%
\providecommand \doibase [0]{https://doi.org/}%
\providecommand \selectlanguage [0]{\@gobble}%
\providecommand \bibinfo  [0]{\@secondoftwo}%
\providecommand \bibfield  [0]{\@secondoftwo}%
\providecommand \translation [1]{[#1]}%
\providecommand \BibitemOpen [0]{}%
\providecommand \bibitemStop [0]{}%
\providecommand \bibitemNoStop [0]{.\EOS\space}%
\providecommand \EOS [0]{\spacefactor3000\relax}%
\providecommand \BibitemShut  [1]{\csname bibitem#1\endcsname}%
\let\auto@bib@innerbib\@empty
\bibitem [{\citenamefont {Maurand}\ \emph {et~al.}(2016)\citenamefont {Maurand}, \citenamefont {Jehl}, \citenamefont {Kotekar-Patil}, \citenamefont {Corna}, \citenamefont {Bohuslavskyi}, \citenamefont {Lavi{\'e}ville}, \citenamefont {Hutin}, \citenamefont {Barraud}, \citenamefont {Vinet}, \citenamefont {Sanquer} \emph {et~al.}}]{maurand2016cmos}%
  \BibitemOpen
  \bibfield  {author} {\bibinfo {author} {\bibfnamefont {R.}~\bibnamefont {Maurand}}, \bibinfo {author} {\bibfnamefont {X.}~\bibnamefont {Jehl}}, \bibinfo {author} {\bibfnamefont {D.}~\bibnamefont {Kotekar-Patil}}, \bibinfo {author} {\bibfnamefont {A.}~\bibnamefont {Corna}}, \bibinfo {author} {\bibfnamefont {H.}~\bibnamefont {Bohuslavskyi}}, \bibinfo {author} {\bibfnamefont {R.}~\bibnamefont {Lavi{\'e}ville}}, \bibinfo {author} {\bibfnamefont {L.}~\bibnamefont {Hutin}}, \bibinfo {author} {\bibfnamefont {S.}~\bibnamefont {Barraud}}, \bibinfo {author} {\bibfnamefont {M.}~\bibnamefont {Vinet}}, \bibinfo {author} {\bibfnamefont {M.}~\bibnamefont {Sanquer}}, \emph {et~al.},\ }\bibfield  {title} {\bibinfo {title} {A cmos silicon spin qubit},\ }\href {https://doi.org/10.1038/ncomms13575} {\bibfield  {journal} {\bibinfo  {journal} {Nature communications}\ }\textbf {\bibinfo {volume} {7}},\ \bibinfo {pages} {13575} (\bibinfo {year} {2016})}\BibitemShut {NoStop}%
\bibitem [{\citenamefont {Watzinger}\ \emph {et~al.}(2018)\citenamefont {Watzinger}, \citenamefont {Kuku{\v{c}}ka}, \citenamefont {Vuku{\v{s}}i{\'c}}, \citenamefont {Gao}, \citenamefont {Wang}, \citenamefont {Sch{\"a}ffler}, \citenamefont {Zhang},\ and\ \citenamefont {Katsaros}}]{watzinger2018germanium}%
  \BibitemOpen
  \bibfield  {author} {\bibinfo {author} {\bibfnamefont {H.}~\bibnamefont {Watzinger}}, \bibinfo {author} {\bibfnamefont {J.}~\bibnamefont {Kuku{\v{c}}ka}}, \bibinfo {author} {\bibfnamefont {L.}~\bibnamefont {Vuku{\v{s}}i{\'c}}}, \bibinfo {author} {\bibfnamefont {F.}~\bibnamefont {Gao}}, \bibinfo {author} {\bibfnamefont {T.}~\bibnamefont {Wang}}, \bibinfo {author} {\bibfnamefont {F.}~\bibnamefont {Sch{\"a}ffler}}, \bibinfo {author} {\bibfnamefont {J.-J.}\ \bibnamefont {Zhang}},\ and\ \bibinfo {author} {\bibfnamefont {G.}~\bibnamefont {Katsaros}},\ }\bibfield  {title} {\bibinfo {title} {A germanium hole spin qubit},\ }\href {https://doi.org/10.1038/s41467-018-06418-4} {\bibfield  {journal} {\bibinfo  {journal} {Nature communications}\ }\textbf {\bibinfo {volume} {9}},\ \bibinfo {pages} {3902} (\bibinfo {year} {2018})}\BibitemShut {NoStop}%
\bibitem [{\citenamefont {Crippa}\ \emph {et~al.}(2018)\citenamefont {Crippa}, \citenamefont {Maurand}, \citenamefont {Bourdet}, \citenamefont {Kotekar-Patil}, \citenamefont {Amisse}, \citenamefont {Jehl}, \citenamefont {Sanquer}, \citenamefont {Lavi{\'e}ville}, \citenamefont {Bohuslavskyi}, \citenamefont {Hutin} \emph {et~al.}}]{crippa2018electrical}%
  \BibitemOpen
  \bibfield  {author} {\bibinfo {author} {\bibfnamefont {A.}~\bibnamefont {Crippa}}, \bibinfo {author} {\bibfnamefont {R.}~\bibnamefont {Maurand}}, \bibinfo {author} {\bibfnamefont {L.}~\bibnamefont {Bourdet}}, \bibinfo {author} {\bibfnamefont {D.}~\bibnamefont {Kotekar-Patil}}, \bibinfo {author} {\bibfnamefont {A.}~\bibnamefont {Amisse}}, \bibinfo {author} {\bibfnamefont {X.}~\bibnamefont {Jehl}}, \bibinfo {author} {\bibfnamefont {M.}~\bibnamefont {Sanquer}}, \bibinfo {author} {\bibfnamefont {R.}~\bibnamefont {Lavi{\'e}ville}}, \bibinfo {author} {\bibfnamefont {H.}~\bibnamefont {Bohuslavskyi}}, \bibinfo {author} {\bibfnamefont {L.}~\bibnamefont {Hutin}}, \emph {et~al.},\ }\bibfield  {title} {\bibinfo {title} {Electrical spin driving by g-matrix modulation in spin-orbit qubits},\ }\href {https://doi.org/10.1103/PhysRevLett.120.137702} {\bibfield  {journal} {\bibinfo  {journal} {Physical review letters}\ }\textbf {\bibinfo {volume} {120}},\ \bibinfo {pages} {137702} (\bibinfo {year} {2018})}\BibitemShut
  {NoStop}%
\bibitem [{\citenamefont {Hendrickx}\ \emph {et~al.}(2020{\natexlab{a}})\citenamefont {Hendrickx}, \citenamefont {Lawrie}, \citenamefont {Petit}, \citenamefont {Sammak}, \citenamefont {Scappucci},\ and\ \citenamefont {Veldhorst}}]{hendrickx2020single}%
  \BibitemOpen
  \bibfield  {author} {\bibinfo {author} {\bibfnamefont {N.}~\bibnamefont {Hendrickx}}, \bibinfo {author} {\bibfnamefont {W.}~\bibnamefont {Lawrie}}, \bibinfo {author} {\bibfnamefont {L.}~\bibnamefont {Petit}}, \bibinfo {author} {\bibfnamefont {A.}~\bibnamefont {Sammak}}, \bibinfo {author} {\bibfnamefont {G.}~\bibnamefont {Scappucci}},\ and\ \bibinfo {author} {\bibfnamefont {M.}~\bibnamefont {Veldhorst}},\ }\bibfield  {title} {\bibinfo {title} {A single-hole spin qubit},\ }\href {https://doi.org/10.1038/s41467-020-17211-7} {\bibfield  {journal} {\bibinfo  {journal} {Nature communications}\ }\textbf {\bibinfo {volume} {11}},\ \bibinfo {pages} {3478} (\bibinfo {year} {2020}{\natexlab{a}})}\BibitemShut {NoStop}%
\bibitem [{\citenamefont {Froning}\ \emph {et~al.}(2021{\natexlab{a}})\citenamefont {Froning}, \citenamefont {Camenzind}, \citenamefont {van~der Molen}, \citenamefont {Li}, \citenamefont {Bakkers}, \citenamefont {Zumb{\"u}hl},\ and\ \citenamefont {Braakman}}]{froning2021ultrafast}%
  \BibitemOpen
  \bibfield  {author} {\bibinfo {author} {\bibfnamefont {F.~N.}\ \bibnamefont {Froning}}, \bibinfo {author} {\bibfnamefont {L.~C.}\ \bibnamefont {Camenzind}}, \bibinfo {author} {\bibfnamefont {O.~A.}\ \bibnamefont {van~der Molen}}, \bibinfo {author} {\bibfnamefont {A.}~\bibnamefont {Li}}, \bibinfo {author} {\bibfnamefont {E.~P.}\ \bibnamefont {Bakkers}}, \bibinfo {author} {\bibfnamefont {D.~M.}\ \bibnamefont {Zumb{\"u}hl}},\ and\ \bibinfo {author} {\bibfnamefont {F.~R.}\ \bibnamefont {Braakman}},\ }\bibfield  {title} {\bibinfo {title} {Ultrafast hole spin qubit with gate-tunable spin--orbit switch functionality},\ }\href {https://doi.org/10.1038/s41565-020-00828-6} {\bibfield  {journal} {\bibinfo  {journal} {Nature Nanotechnology}\ }\textbf {\bibinfo {volume} {16}},\ \bibinfo {pages} {308} (\bibinfo {year} {2021}{\natexlab{a}})}\BibitemShut {NoStop}%
\bibitem [{\citenamefont {Jirovec}\ \emph {et~al.}(2021)\citenamefont {Jirovec}, \citenamefont {Hofmann}, \citenamefont {Ballabio}, \citenamefont {Mutter}, \citenamefont {Tavani}, \citenamefont {Botifoll}, \citenamefont {Crippa}, \citenamefont {Kukucka}, \citenamefont {Sagi}, \citenamefont {Martins} \emph {et~al.}}]{jirovec2021singlet}%
  \BibitemOpen
  \bibfield  {author} {\bibinfo {author} {\bibfnamefont {D.}~\bibnamefont {Jirovec}}, \bibinfo {author} {\bibfnamefont {A.}~\bibnamefont {Hofmann}}, \bibinfo {author} {\bibfnamefont {A.}~\bibnamefont {Ballabio}}, \bibinfo {author} {\bibfnamefont {P.~M.}\ \bibnamefont {Mutter}}, \bibinfo {author} {\bibfnamefont {G.}~\bibnamefont {Tavani}}, \bibinfo {author} {\bibfnamefont {M.}~\bibnamefont {Botifoll}}, \bibinfo {author} {\bibfnamefont {A.}~\bibnamefont {Crippa}}, \bibinfo {author} {\bibfnamefont {J.}~\bibnamefont {Kukucka}}, \bibinfo {author} {\bibfnamefont {O.}~\bibnamefont {Sagi}}, \bibinfo {author} {\bibfnamefont {F.}~\bibnamefont {Martins}}, \emph {et~al.},\ }\bibfield  {title} {\bibinfo {title} {A singlet-triplet hole spin qubit in planar ge},\ }\href {https://doi.org/10.1038/s41563-021-01022-2} {\bibfield  {journal} {\bibinfo  {journal} {Nature Materials}\ }\textbf {\bibinfo {volume} {20}},\ \bibinfo {pages} {1106} (\bibinfo {year} {2021})}\BibitemShut {NoStop}%
\bibitem [{\citenamefont {Wang}\ \emph {et~al.}(2022)\citenamefont {Wang}, \citenamefont {Xu}, \citenamefont {Gao}, \citenamefont {Liu}, \citenamefont {Ma}, \citenamefont {Zhang}, \citenamefont {Wang}, \citenamefont {Cao}, \citenamefont {Wang}, \citenamefont {Zhang} \emph {et~al.}}]{wang2022ultrafast}%
  \BibitemOpen
  \bibfield  {author} {\bibinfo {author} {\bibfnamefont {K.}~\bibnamefont {Wang}}, \bibinfo {author} {\bibfnamefont {G.}~\bibnamefont {Xu}}, \bibinfo {author} {\bibfnamefont {F.}~\bibnamefont {Gao}}, \bibinfo {author} {\bibfnamefont {H.}~\bibnamefont {Liu}}, \bibinfo {author} {\bibfnamefont {R.-L.}\ \bibnamefont {Ma}}, \bibinfo {author} {\bibfnamefont {X.}~\bibnamefont {Zhang}}, \bibinfo {author} {\bibfnamefont {Z.}~\bibnamefont {Wang}}, \bibinfo {author} {\bibfnamefont {G.}~\bibnamefont {Cao}}, \bibinfo {author} {\bibfnamefont {T.}~\bibnamefont {Wang}}, \bibinfo {author} {\bibfnamefont {J.-J.}\ \bibnamefont {Zhang}}, \emph {et~al.},\ }\bibfield  {title} {\bibinfo {title} {Ultrafast coherent control of a hole spin qubit in a germanium quantum dot},\ }\href {https://doi.org/10.1038/s41467-021-27880-7} {\bibfield  {journal} {\bibinfo  {journal} {Nature Communications}\ }\textbf {\bibinfo {volume} {13}},\ \bibinfo {pages} {206} (\bibinfo {year} {2022})}\BibitemShut {NoStop}%
\bibitem [{\citenamefont {Fang}\ \emph {et~al.}(2023)\citenamefont {Fang}, \citenamefont {Philippopoulos}, \citenamefont {Culcer}, \citenamefont {Coish},\ and\ \citenamefont {Chesi}}]{fang2023recent}%
  \BibitemOpen
  \bibfield  {author} {\bibinfo {author} {\bibfnamefont {Y.}~\bibnamefont {Fang}}, \bibinfo {author} {\bibfnamefont {P.}~\bibnamefont {Philippopoulos}}, \bibinfo {author} {\bibfnamefont {D.}~\bibnamefont {Culcer}}, \bibinfo {author} {\bibfnamefont {W.}~\bibnamefont {Coish}},\ and\ \bibinfo {author} {\bibfnamefont {S.}~\bibnamefont {Chesi}},\ }\bibfield  {title} {\bibinfo {title} {Recent advances in hole-spin qubits},\ }\href {https://doi.org/10.1088/2633-4356/acb87e} {\bibfield  {journal} {\bibinfo  {journal} {Materials for Quantum Technology}\ }\textbf {\bibinfo {volume} {3}},\ \bibinfo {pages} {012003} (\bibinfo {year} {2023})}\BibitemShut {NoStop}%
\bibitem [{\citenamefont {Liles}\ \emph {et~al.}(2024)\citenamefont {Liles}, \citenamefont {Halverson}, \citenamefont {Wang}, \citenamefont {Shamim}, \citenamefont {Eggli}, \citenamefont {Jin}, \citenamefont {Hillier}, \citenamefont {Kumar}, \citenamefont {Vorreiter}, \citenamefont {Rendell} \emph {et~al.}}]{liles2024singlet}%
  \BibitemOpen
  \bibfield  {author} {\bibinfo {author} {\bibfnamefont {S.}~\bibnamefont {Liles}}, \bibinfo {author} {\bibfnamefont {D.}~\bibnamefont {Halverson}}, \bibinfo {author} {\bibfnamefont {Z.}~\bibnamefont {Wang}}, \bibinfo {author} {\bibfnamefont {A.}~\bibnamefont {Shamim}}, \bibinfo {author} {\bibfnamefont {R.}~\bibnamefont {Eggli}}, \bibinfo {author} {\bibfnamefont {I.~K.}\ \bibnamefont {Jin}}, \bibinfo {author} {\bibfnamefont {J.}~\bibnamefont {Hillier}}, \bibinfo {author} {\bibfnamefont {K.}~\bibnamefont {Kumar}}, \bibinfo {author} {\bibfnamefont {I.}~\bibnamefont {Vorreiter}}, \bibinfo {author} {\bibfnamefont {M.}~\bibnamefont {Rendell}}, \emph {et~al.},\ }\bibfield  {title} {\bibinfo {title} {A singlet-triplet hole-spin qubit in mos silicon},\ }\href {https://doi.org/10.1038/s41467-024-51902-9} {\bibfield  {journal} {\bibinfo  {journal} {Nature Communications}\ }\textbf {\bibinfo {volume} {15}},\ \bibinfo {pages} {7690} (\bibinfo {year} {2024})}\BibitemShut {NoStop}%
\bibitem [{\citenamefont {Piot}\ \emph {et~al.}(2022)\citenamefont {Piot}, \citenamefont {Brun}, \citenamefont {Schmitt}, \citenamefont {Zihlmann}, \citenamefont {Michal}, \citenamefont {Apra}, \citenamefont {Abadillo-Uriel}, \citenamefont {Jehl}, \citenamefont {Bertrand}, \citenamefont {Niebojewski} \emph {et~al.}}]{piot2022single}%
  \BibitemOpen
  \bibfield  {author} {\bibinfo {author} {\bibfnamefont {N.}~\bibnamefont {Piot}}, \bibinfo {author} {\bibfnamefont {B.}~\bibnamefont {Brun}}, \bibinfo {author} {\bibfnamefont {V.}~\bibnamefont {Schmitt}}, \bibinfo {author} {\bibfnamefont {S.}~\bibnamefont {Zihlmann}}, \bibinfo {author} {\bibfnamefont {V.}~\bibnamefont {Michal}}, \bibinfo {author} {\bibfnamefont {A.}~\bibnamefont {Apra}}, \bibinfo {author} {\bibfnamefont {J.}~\bibnamefont {Abadillo-Uriel}}, \bibinfo {author} {\bibfnamefont {X.}~\bibnamefont {Jehl}}, \bibinfo {author} {\bibfnamefont {B.}~\bibnamefont {Bertrand}}, \bibinfo {author} {\bibfnamefont {H.}~\bibnamefont {Niebojewski}}, \emph {et~al.},\ }\bibfield  {title} {\bibinfo {title} {A single hole spin with enhanced coherence in natural silicon},\ }\href {https://doi.org/10.1038/s41565-022-01196-z} {\bibfield  {journal} {\bibinfo  {journal} {Nature Nanotechnology}\ }\textbf {\bibinfo {volume} {17}},\ \bibinfo {pages} {1072} (\bibinfo {year} {2022})}\BibitemShut {NoStop}%
\bibitem [{\citenamefont {Hendrickx}\ \emph {et~al.}(2024)\citenamefont {Hendrickx}, \citenamefont {Massai}, \citenamefont {Mergenthaler}, \citenamefont {Schupp}, \citenamefont {Paredes}, \citenamefont {Bedell}, \citenamefont {Salis},\ and\ \citenamefont {Fuhrer}}]{hendrickx2024sweet}%
  \BibitemOpen
  \bibfield  {author} {\bibinfo {author} {\bibfnamefont {N.}~\bibnamefont {Hendrickx}}, \bibinfo {author} {\bibfnamefont {L.}~\bibnamefont {Massai}}, \bibinfo {author} {\bibfnamefont {M.}~\bibnamefont {Mergenthaler}}, \bibinfo {author} {\bibfnamefont {F.}~\bibnamefont {Schupp}}, \bibinfo {author} {\bibfnamefont {S.}~\bibnamefont {Paredes}}, \bibinfo {author} {\bibfnamefont {S.}~\bibnamefont {Bedell}}, \bibinfo {author} {\bibfnamefont {G.}~\bibnamefont {Salis}},\ and\ \bibinfo {author} {\bibfnamefont {A.}~\bibnamefont {Fuhrer}},\ }\bibfield  {title} {\bibinfo {title} {Sweet-spot operation of a germanium hole spin qubit with highly anisotropic noise sensitivity},\ }\href {https://doi.org/10.1038/s41563-024-01857-5} {\bibfield  {journal} {\bibinfo  {journal} {Nature Materials}\ }\textbf {\bibinfo {volume} {23}},\ \bibinfo {pages} {920} (\bibinfo {year} {2024})}\BibitemShut {NoStop}%
\bibitem [{\citenamefont {Carballido}\ \emph {et~al.}(2025)\citenamefont {Carballido}, \citenamefont {Svab}, \citenamefont {Eggli}, \citenamefont {Patlatiuk}, \citenamefont {Chevalier~Kwon}, \citenamefont {Schuff}, \citenamefont {Kaiser}, \citenamefont {Camenzind}, \citenamefont {Li}, \citenamefont {Ares} \emph {et~al.}}]{carballido2024compromise}%
  \BibitemOpen
  \bibfield  {author} {\bibinfo {author} {\bibfnamefont {M.~J.}\ \bibnamefont {Carballido}}, \bibinfo {author} {\bibfnamefont {S.}~\bibnamefont {Svab}}, \bibinfo {author} {\bibfnamefont {R.~S.}\ \bibnamefont {Eggli}}, \bibinfo {author} {\bibfnamefont {T.}~\bibnamefont {Patlatiuk}}, \bibinfo {author} {\bibfnamefont {P.}~\bibnamefont {Chevalier~Kwon}}, \bibinfo {author} {\bibfnamefont {J.}~\bibnamefont {Schuff}}, \bibinfo {author} {\bibfnamefont {R.~M.}\ \bibnamefont {Kaiser}}, \bibinfo {author} {\bibfnamefont {L.~C.}\ \bibnamefont {Camenzind}}, \bibinfo {author} {\bibfnamefont {A.}~\bibnamefont {Li}}, \bibinfo {author} {\bibfnamefont {N.}~\bibnamefont {Ares}}, \emph {et~al.},\ }\bibfield  {title} {\bibinfo {title} {Compromise-free scaling of qubit speed and coherence},\ }\href {https://doi.org/10.1038/s41467-025-62614-z} {\bibfield  {journal} {\bibinfo  {journal} {Nature Communications}\ }\textbf {\bibinfo {volume} {16}},\ \bibinfo {pages} {7616} (\bibinfo {year} {2025})}\BibitemShut {NoStop}%
\bibitem [{\citenamefont {Bassi}\ \emph {et~al.}(2026)\citenamefont {Bassi}, \citenamefont {Rodríguez-Mena}, \citenamefont {Brun}, \citenamefont {Zihlmann}, \citenamefont {Nguyen}, \citenamefont {Champain}, \citenamefont {Abadillo-Uriel}, \citenamefont {Bertrand}, \citenamefont {Niebojewski}, \citenamefont {Maurand}, \citenamefont {Niquet}, \citenamefont {Jehl}, \citenamefont {De~Franceschi},\ and\ \citenamefont {Schmitt}}]{bassi2024optimal}%
  \BibitemOpen
  \bibfield  {author} {\bibinfo {author} {\bibfnamefont {M.}~\bibnamefont {Bassi}}, \bibinfo {author} {\bibfnamefont {E.~A.}\ \bibnamefont {Rodríguez-Mena}}, \bibinfo {author} {\bibfnamefont {B.}~\bibnamefont {Brun}}, \bibinfo {author} {\bibfnamefont {S.}~\bibnamefont {Zihlmann}}, \bibinfo {author} {\bibfnamefont {T.}~\bibnamefont {Nguyen}}, \bibinfo {author} {\bibfnamefont {V.}~\bibnamefont {Champain}}, \bibinfo {author} {\bibfnamefont {J.~C.}\ \bibnamefont {Abadillo-Uriel}}, \bibinfo {author} {\bibfnamefont {B.}~\bibnamefont {Bertrand}}, \bibinfo {author} {\bibfnamefont {H.}~\bibnamefont {Niebojewski}}, \bibinfo {author} {\bibfnamefont {R.}~\bibnamefont {Maurand}}, \bibinfo {author} {\bibfnamefont {Y.-M.}\ \bibnamefont {Niquet}}, \bibinfo {author} {\bibfnamefont {X.}~\bibnamefont {Jehl}}, \bibinfo {author} {\bibfnamefont {S.}~\bibnamefont {De~Franceschi}},\ and\ \bibinfo {author} {\bibfnamefont {V.}~\bibnamefont {Schmitt}},\ }\bibfield  {title} {\bibinfo {title} {Optimal operation of hole spin qubits},\
  }\href {https://doi.org/10.1038/s41567-025-03106-1} {\bibfield  {journal} {\bibinfo  {journal} {Nature Physics}\ }\textbf {\bibinfo {volume} {22}},\ \bibinfo {pages} {75} (\bibinfo {year} {2026})}\BibitemShut {NoStop}%
\bibitem [{\citenamefont {Hendrickx}\ \emph {et~al.}(2020{\natexlab{b}})\citenamefont {Hendrickx}, \citenamefont {Franke}, \citenamefont {Sammak}, \citenamefont {Scappucci},\ and\ \citenamefont {Veldhorst}}]{hendrickx2020fast}%
  \BibitemOpen
  \bibfield  {author} {\bibinfo {author} {\bibfnamefont {N.}~\bibnamefont {Hendrickx}}, \bibinfo {author} {\bibfnamefont {D.}~\bibnamefont {Franke}}, \bibinfo {author} {\bibfnamefont {A.}~\bibnamefont {Sammak}}, \bibinfo {author} {\bibfnamefont {G.}~\bibnamefont {Scappucci}},\ and\ \bibinfo {author} {\bibfnamefont {M.}~\bibnamefont {Veldhorst}},\ }\bibfield  {title} {\bibinfo {title} {Fast two-qubit logic with holes in germanium},\ }\href {https://doi.org/10.1038/s41586-019-1919-3} {\bibfield  {journal} {\bibinfo  {journal} {Nature}\ }\textbf {\bibinfo {volume} {577}},\ \bibinfo {pages} {487} (\bibinfo {year} {2020}{\natexlab{b}})}\BibitemShut {NoStop}%
\bibitem [{\citenamefont {Geyer}\ \emph {et~al.}(2024)\citenamefont {Geyer}, \citenamefont {Het{\'e}nyi}, \citenamefont {Bosco}, \citenamefont {Camenzind}, \citenamefont {Eggli}, \citenamefont {Fuhrer}, \citenamefont {Loss}, \citenamefont {Warburton}, \citenamefont {Zumb{\"u}hl},\ and\ \citenamefont {Kuhlmann}}]{geyer2024anisotropic}%
  \BibitemOpen
  \bibfield  {author} {\bibinfo {author} {\bibfnamefont {S.}~\bibnamefont {Geyer}}, \bibinfo {author} {\bibfnamefont {B.}~\bibnamefont {Het{\'e}nyi}}, \bibinfo {author} {\bibfnamefont {S.}~\bibnamefont {Bosco}}, \bibinfo {author} {\bibfnamefont {L.~C.}\ \bibnamefont {Camenzind}}, \bibinfo {author} {\bibfnamefont {R.~S.}\ \bibnamefont {Eggli}}, \bibinfo {author} {\bibfnamefont {A.}~\bibnamefont {Fuhrer}}, \bibinfo {author} {\bibfnamefont {D.}~\bibnamefont {Loss}}, \bibinfo {author} {\bibfnamefont {R.~J.}\ \bibnamefont {Warburton}}, \bibinfo {author} {\bibfnamefont {D.~M.}\ \bibnamefont {Zumb{\"u}hl}},\ and\ \bibinfo {author} {\bibfnamefont {A.~V.}\ \bibnamefont {Kuhlmann}},\ }\bibfield  {title} {\bibinfo {title} {Anisotropic exchange interaction of two hole-spin qubits},\ }\href {https://doi.org/10.1038/s41567-024-02481-5} {\bibfield  {journal} {\bibinfo  {journal} {Nature Physics}\ }\textbf {\bibinfo {volume} {20}},\ \bibinfo {pages} {1152} (\bibinfo {year} {2024})}\BibitemShut {NoStop}%
\bibitem [{\citenamefont {Hendrickx}\ \emph {et~al.}(2021)\citenamefont {Hendrickx}, \citenamefont {Lawrie}, \citenamefont {Russ}, \citenamefont {Van~Riggelen}, \citenamefont {De~Snoo}, \citenamefont {Schouten}, \citenamefont {Sammak}, \citenamefont {Scappucci},\ and\ \citenamefont {Veldhorst}}]{hendrickx2021four}%
  \BibitemOpen
  \bibfield  {author} {\bibinfo {author} {\bibfnamefont {N.~W.}\ \bibnamefont {Hendrickx}}, \bibinfo {author} {\bibfnamefont {W.~I.}\ \bibnamefont {Lawrie}}, \bibinfo {author} {\bibfnamefont {M.}~\bibnamefont {Russ}}, \bibinfo {author} {\bibfnamefont {F.}~\bibnamefont {Van~Riggelen}}, \bibinfo {author} {\bibfnamefont {S.~L.}\ \bibnamefont {De~Snoo}}, \bibinfo {author} {\bibfnamefont {R.~N.}\ \bibnamefont {Schouten}}, \bibinfo {author} {\bibfnamefont {A.}~\bibnamefont {Sammak}}, \bibinfo {author} {\bibfnamefont {G.}~\bibnamefont {Scappucci}},\ and\ \bibinfo {author} {\bibfnamefont {M.}~\bibnamefont {Veldhorst}},\ }\bibfield  {title} {\bibinfo {title} {A four-qubit germanium quantum processor},\ }\href {https://doi.org/10.1038/s41586-021-03332-6} {\bibfield  {journal} {\bibinfo  {journal} {Nature}\ }\textbf {\bibinfo {volume} {591}},\ \bibinfo {pages} {580} (\bibinfo {year} {2021})}\BibitemShut {NoStop}%
\bibitem [{\citenamefont {John}\ \emph {et~al.}(2025)\citenamefont {John}, \citenamefont {Yu}, \citenamefont {{van Straaten}}, \citenamefont {{Rodr{\'i}guez-Mena}}, \citenamefont {Rodr{\'i}guez}, \citenamefont {Oosterhout}, \citenamefont {Stehouwer}, \citenamefont {Scappucci}, \citenamefont {{Rimbach-Russ}}, \citenamefont {Bosco}, \citenamefont {Borsoi}, \citenamefont {Niquet},\ and\ \citenamefont {Veldhorst}}]{john2024two}%
  \BibitemOpen
  \bibfield  {author} {\bibinfo {author} {\bibfnamefont {V.}~\bibnamefont {John}}, \bibinfo {author} {\bibfnamefont {C.~X.}\ \bibnamefont {Yu}}, \bibinfo {author} {\bibfnamefont {B.}~\bibnamefont {{van Straaten}}}, \bibinfo {author} {\bibfnamefont {E.~A.}\ \bibnamefont {{Rodr{\'i}guez-Mena}}}, \bibinfo {author} {\bibfnamefont {M.}~\bibnamefont {Rodr{\'i}guez}}, \bibinfo {author} {\bibfnamefont {S.~D.}\ \bibnamefont {Oosterhout}}, \bibinfo {author} {\bibfnamefont {L.~E.~A.}\ \bibnamefont {Stehouwer}}, \bibinfo {author} {\bibfnamefont {G.}~\bibnamefont {Scappucci}}, \bibinfo {author} {\bibfnamefont {M.}~\bibnamefont {{Rimbach-Russ}}}, \bibinfo {author} {\bibfnamefont {S.}~\bibnamefont {Bosco}}, \bibinfo {author} {\bibfnamefont {F.}~\bibnamefont {Borsoi}}, \bibinfo {author} {\bibfnamefont {Y.-M.}\ \bibnamefont {Niquet}},\ and\ \bibinfo {author} {\bibfnamefont {M.}~\bibnamefont {Veldhorst}},\ }\bibfield  {title} {\bibinfo {title} {Robust and localised control of a 10-spin qubit array in germanium},\ }\href
  {https://doi.org/10.1038/s41467-025-65577-3} {\bibfield  {journal} {\bibinfo  {journal} {Nature Communications}\ }\textbf {\bibinfo {volume} {16}},\ \bibinfo {pages} {10560} (\bibinfo {year} {2025})}\BibitemShut {NoStop}%
\bibitem [{\citenamefont {Wang}\ \emph {et~al.}(2023)\citenamefont {Wang}, \citenamefont {D{\'e}prez}, \citenamefont {Tidjani}, \citenamefont {Lawrie}, \citenamefont {Hendrickx}, \citenamefont {Sammak}, \citenamefont {Scappucci},\ and\ \citenamefont {Veldhorst}}]{wang2023probing}%
  \BibitemOpen
  \bibfield  {author} {\bibinfo {author} {\bibfnamefont {C.-A.}\ \bibnamefont {Wang}}, \bibinfo {author} {\bibfnamefont {C.}~\bibnamefont {D{\'e}prez}}, \bibinfo {author} {\bibfnamefont {H.}~\bibnamefont {Tidjani}}, \bibinfo {author} {\bibfnamefont {W.~I.}\ \bibnamefont {Lawrie}}, \bibinfo {author} {\bibfnamefont {N.~W.}\ \bibnamefont {Hendrickx}}, \bibinfo {author} {\bibfnamefont {A.}~\bibnamefont {Sammak}}, \bibinfo {author} {\bibfnamefont {G.}~\bibnamefont {Scappucci}},\ and\ \bibinfo {author} {\bibfnamefont {M.}~\bibnamefont {Veldhorst}},\ }\bibfield  {title} {\bibinfo {title} {Probing resonating valence bonds on a programmable germanium quantum simulator},\ }\href {https://doi.org/10.1038/s41534-023-00727-3} {\bibfield  {journal} {\bibinfo  {journal} {npj Quantum Information}\ }\textbf {\bibinfo {volume} {9}},\ \bibinfo {pages} {58} (\bibinfo {year} {2023})}\BibitemShut {NoStop}%
\bibitem [{\citenamefont {Hsiao}\ \emph {et~al.}(2024)\citenamefont {Hsiao}, \citenamefont {Cova~Fari{\~n}a}, \citenamefont {Oosterhout}, \citenamefont {Jirovec}, \citenamefont {Zhang}, \citenamefont {van Diepen}, \citenamefont {Lawrie}, \citenamefont {Wang}, \citenamefont {Sammak}, \citenamefont {Scappucci} \emph {et~al.}}]{hsiao2024exciton}%
  \BibitemOpen
  \bibfield  {author} {\bibinfo {author} {\bibfnamefont {T.-K.}\ \bibnamefont {Hsiao}}, \bibinfo {author} {\bibfnamefont {P.}~\bibnamefont {Cova~Fari{\~n}a}}, \bibinfo {author} {\bibfnamefont {S.~D.}\ \bibnamefont {Oosterhout}}, \bibinfo {author} {\bibfnamefont {D.}~\bibnamefont {Jirovec}}, \bibinfo {author} {\bibfnamefont {X.}~\bibnamefont {Zhang}}, \bibinfo {author} {\bibfnamefont {C.~J.}\ \bibnamefont {van Diepen}}, \bibinfo {author} {\bibfnamefont {W.~I.}\ \bibnamefont {Lawrie}}, \bibinfo {author} {\bibfnamefont {C.-A.}\ \bibnamefont {Wang}}, \bibinfo {author} {\bibfnamefont {A.}~\bibnamefont {Sammak}}, \bibinfo {author} {\bibfnamefont {G.}~\bibnamefont {Scappucci}}, \emph {et~al.},\ }\bibfield  {title} {\bibinfo {title} {Exciton transport in a germanium quantum dot ladder},\ }\href {https://doi.org/10.1103/PhysRevX.14.011048} {\bibfield  {journal} {\bibinfo  {journal} {Physical Review X}\ }\textbf {\bibinfo {volume} {14}},\ \bibinfo {pages} {011048} (\bibinfo {year} {2024})}\BibitemShut {NoStop}%
\bibitem [{\citenamefont {Boter}\ \emph {et~al.}(2022)\citenamefont {Boter}, \citenamefont {Dehollain}, \citenamefont {Van~Dijk}, \citenamefont {Xu}, \citenamefont {Hensgens}, \citenamefont {Versluis}, \citenamefont {Naus}, \citenamefont {Clarke}, \citenamefont {Veldhorst}, \citenamefont {Sebastiano} \emph {et~al.}}]{boter2022spiderweb}%
  \BibitemOpen
  \bibfield  {author} {\bibinfo {author} {\bibfnamefont {J.~M.}\ \bibnamefont {Boter}}, \bibinfo {author} {\bibfnamefont {J.~P.}\ \bibnamefont {Dehollain}}, \bibinfo {author} {\bibfnamefont {J.~P.}\ \bibnamefont {Van~Dijk}}, \bibinfo {author} {\bibfnamefont {Y.}~\bibnamefont {Xu}}, \bibinfo {author} {\bibfnamefont {T.}~\bibnamefont {Hensgens}}, \bibinfo {author} {\bibfnamefont {R.}~\bibnamefont {Versluis}}, \bibinfo {author} {\bibfnamefont {H.~W.}\ \bibnamefont {Naus}}, \bibinfo {author} {\bibfnamefont {J.~S.}\ \bibnamefont {Clarke}}, \bibinfo {author} {\bibfnamefont {M.}~\bibnamefont {Veldhorst}}, \bibinfo {author} {\bibfnamefont {F.}~\bibnamefont {Sebastiano}}, \emph {et~al.},\ }\bibfield  {title} {\bibinfo {title} {Spiderweb array: A sparse spin-qubit array},\ }\href {https://doi.org/10.1103/PhysRevApplied.18.024053} {\bibfield  {journal} {\bibinfo  {journal} {Physical review applied}\ }\textbf {\bibinfo {volume} {18}},\ \bibinfo {pages} {024053} (\bibinfo {year} {2022})}\BibitemShut {NoStop}%
\bibitem [{\citenamefont {Wang}\ \emph {et~al.}(2024{\natexlab{a}})\citenamefont {Wang}, \citenamefont {John}, \citenamefont {Tidjani}, \citenamefont {Yu}, \citenamefont {Ivlev}, \citenamefont {D{\'e}prez}, \citenamefont {van Riggelen-Doelman}, \citenamefont {Woods}, \citenamefont {Hendrickx}, \citenamefont {Lawrie} \emph {et~al.}}]{wang2024operating}%
  \BibitemOpen
  \bibfield  {author} {\bibinfo {author} {\bibfnamefont {C.-A.}\ \bibnamefont {Wang}}, \bibinfo {author} {\bibfnamefont {V.}~\bibnamefont {John}}, \bibinfo {author} {\bibfnamefont {H.}~\bibnamefont {Tidjani}}, \bibinfo {author} {\bibfnamefont {C.~X.}\ \bibnamefont {Yu}}, \bibinfo {author} {\bibfnamefont {A.~S.}\ \bibnamefont {Ivlev}}, \bibinfo {author} {\bibfnamefont {C.}~\bibnamefont {D{\'e}prez}}, \bibinfo {author} {\bibfnamefont {F.}~\bibnamefont {van Riggelen-Doelman}}, \bibinfo {author} {\bibfnamefont {B.~D.}\ \bibnamefont {Woods}}, \bibinfo {author} {\bibfnamefont {N.~W.}\ \bibnamefont {Hendrickx}}, \bibinfo {author} {\bibfnamefont {W.~I.}\ \bibnamefont {Lawrie}}, \emph {et~al.},\ }\bibfield  {title} {\bibinfo {title} {Operating semiconductor quantum processors with hopping spins},\ }\href {https://doi.org/10.1126/science.ado5915} {\bibfield  {journal} {\bibinfo  {journal} {Science}\ }\textbf {\bibinfo {volume} {385}},\ \bibinfo {pages} {447} (\bibinfo {year} {2024}{\natexlab{a}})}\BibitemShut {NoStop}%
\bibitem [{\citenamefont {Unseld}\ \emph {et~al.}(2025)\citenamefont {Unseld}, \citenamefont {Undseth}, \citenamefont {Raymenants}, \citenamefont {Matsumoto}, \citenamefont {de~Snoo}, \citenamefont {Karwal}, \citenamefont {Pietx-Casas}, \citenamefont {Ivlev}, \citenamefont {Meyer}, \citenamefont {Sammak} \emph {et~al.}}]{unseld2025baseband}%
  \BibitemOpen
  \bibfield  {author} {\bibinfo {author} {\bibfnamefont {F.~K.}\ \bibnamefont {Unseld}}, \bibinfo {author} {\bibfnamefont {B.}~\bibnamefont {Undseth}}, \bibinfo {author} {\bibfnamefont {E.}~\bibnamefont {Raymenants}}, \bibinfo {author} {\bibfnamefont {Y.}~\bibnamefont {Matsumoto}}, \bibinfo {author} {\bibfnamefont {S.~L.}\ \bibnamefont {de~Snoo}}, \bibinfo {author} {\bibfnamefont {S.}~\bibnamefont {Karwal}}, \bibinfo {author} {\bibfnamefont {O.}~\bibnamefont {Pietx-Casas}}, \bibinfo {author} {\bibfnamefont {A.~S.}\ \bibnamefont {Ivlev}}, \bibinfo {author} {\bibfnamefont {M.}~\bibnamefont {Meyer}}, \bibinfo {author} {\bibfnamefont {A.}~\bibnamefont {Sammak}}, \emph {et~al.},\ }\bibfield  {title} {\bibinfo {title} {Baseband control of single-electron silicon spin qubits in two dimensions},\ }\href {https://doi.org/10.1038/s41467-025-60351-x} {\bibfield  {journal} {\bibinfo  {journal} {Nature Communications}\ }\textbf {\bibinfo {volume} {16}},\ \bibinfo {pages} {5605} (\bibinfo {year} {2025})}\BibitemShut
  {NoStop}%
\bibitem [{\citenamefont {Rimbach-Russ}\ \emph {et~al.}(2025)\citenamefont {Rimbach-Russ}, \citenamefont {John}, \citenamefont {van Straaten},\ and\ \citenamefont {Bosco}}]{rimbach2024spinless}%
  \BibitemOpen
  \bibfield  {author} {\bibinfo {author} {\bibfnamefont {M.}~\bibnamefont {Rimbach-Russ}}, \bibinfo {author} {\bibfnamefont {V.}~\bibnamefont {John}}, \bibinfo {author} {\bibfnamefont {B.}~\bibnamefont {van Straaten}},\ and\ \bibinfo {author} {\bibfnamefont {S.}~\bibnamefont {Bosco}},\ }\bibfield  {title} {\bibinfo {title} {Gapless single-spin qubit},\ }\href {https://doi.org/10.1103/mvtj-zhrl} {\bibfield  {journal} {\bibinfo  {journal} {Phys. Rev. Lett.}\ }\textbf {\bibinfo {volume} {135}},\ \bibinfo {pages} {197001} (\bibinfo {year} {2025})}\BibitemShut {NoStop}%
\bibitem [{\citenamefont {S{\'a}nchez}\ \emph {et~al.}(2014)\citenamefont {S{\'a}nchez}, \citenamefont {Granger}, \citenamefont {Gaudreau}, \citenamefont {Kam}, \citenamefont {Pioro-Ladri{\`e}re}, \citenamefont {Studenikin}, \citenamefont {Zawadzki}, \citenamefont {Sachrajda},\ and\ \citenamefont {Platero}}]{sanchez2014long}%
  \BibitemOpen
  \bibfield  {author} {\bibinfo {author} {\bibfnamefont {R.}~\bibnamefont {S{\'a}nchez}}, \bibinfo {author} {\bibfnamefont {G.}~\bibnamefont {Granger}}, \bibinfo {author} {\bibfnamefont {L.}~\bibnamefont {Gaudreau}}, \bibinfo {author} {\bibfnamefont {A.}~\bibnamefont {Kam}}, \bibinfo {author} {\bibfnamefont {M.}~\bibnamefont {Pioro-Ladri{\`e}re}}, \bibinfo {author} {\bibfnamefont {S.}~\bibnamefont {Studenikin}}, \bibinfo {author} {\bibfnamefont {P.}~\bibnamefont {Zawadzki}}, \bibinfo {author} {\bibfnamefont {A.}~\bibnamefont {Sachrajda}},\ and\ \bibinfo {author} {\bibfnamefont {G.}~\bibnamefont {Platero}},\ }\bibfield  {title} {\bibinfo {title} {Long-range spin transfer in triple quantum dots},\ }\href {https://doi.org/10.1103/PhysRevLett.112.176803} {\bibfield  {journal} {\bibinfo  {journal} {Physical Review Letters}\ }\textbf {\bibinfo {volume} {112}},\ \bibinfo {pages} {176803} (\bibinfo {year} {2014})}\BibitemShut {NoStop}%
\bibitem [{\citenamefont {Mills}\ \emph {et~al.}(2019)\citenamefont {Mills}, \citenamefont {Zajac}, \citenamefont {Gullans}, \citenamefont {Schupp}, \citenamefont {Hazard},\ and\ \citenamefont {Petta}}]{mills2019shuttling}%
  \BibitemOpen
  \bibfield  {author} {\bibinfo {author} {\bibfnamefont {A.}~\bibnamefont {Mills}}, \bibinfo {author} {\bibfnamefont {D.}~\bibnamefont {Zajac}}, \bibinfo {author} {\bibfnamefont {M.}~\bibnamefont {Gullans}}, \bibinfo {author} {\bibfnamefont {F.}~\bibnamefont {Schupp}}, \bibinfo {author} {\bibfnamefont {T.}~\bibnamefont {Hazard}},\ and\ \bibinfo {author} {\bibfnamefont {J.~R.}\ \bibnamefont {Petta}},\ }\bibfield  {title} {\bibinfo {title} {Shuttling a single charge across a one-dimensional array of silicon quantum dots},\ }\href {https://doi.org/10.1038/s41467-019-08970-z} {\bibfield  {journal} {\bibinfo  {journal} {Nature communications}\ }\textbf {\bibinfo {volume} {10}},\ \bibinfo {pages} {1063} (\bibinfo {year} {2019})}\BibitemShut {NoStop}%
\bibitem [{\citenamefont {Zwerver}\ \emph {et~al.}(2023)\citenamefont {Zwerver}, \citenamefont {Amitonov}, \citenamefont {De~Snoo}, \citenamefont {M{\k{a}}dzik}, \citenamefont {Rimbach-Russ}, \citenamefont {Sammak}, \citenamefont {Scappucci},\ and\ \citenamefont {Vandersypen}}]{zwerver2023shuttling}%
  \BibitemOpen
  \bibfield  {author} {\bibinfo {author} {\bibfnamefont {A.}~\bibnamefont {Zwerver}}, \bibinfo {author} {\bibfnamefont {S.}~\bibnamefont {Amitonov}}, \bibinfo {author} {\bibfnamefont {S.}~\bibnamefont {De~Snoo}}, \bibinfo {author} {\bibfnamefont {M.}~\bibnamefont {M{\k{a}}dzik}}, \bibinfo {author} {\bibfnamefont {M.}~\bibnamefont {Rimbach-Russ}}, \bibinfo {author} {\bibfnamefont {A.}~\bibnamefont {Sammak}}, \bibinfo {author} {\bibfnamefont {G.}~\bibnamefont {Scappucci}},\ and\ \bibinfo {author} {\bibfnamefont {L.}~\bibnamefont {Vandersypen}},\ }\bibfield  {title} {\bibinfo {title} {Shuttling an electron spin through a silicon quantum dot array},\ }\href {https://doi.org/10.1103/PRXQuantum.4.030303} {\bibfield  {journal} {\bibinfo  {journal} {PRX Quantum}\ }\textbf {\bibinfo {volume} {4}},\ \bibinfo {pages} {030303} (\bibinfo {year} {2023})}\BibitemShut {NoStop}%
\bibitem [{\citenamefont {van Riggelen-Doelman}\ \emph {et~al.}(2024)\citenamefont {van Riggelen-Doelman}, \citenamefont {Wang}, \citenamefont {de~Snoo}, \citenamefont {Lawrie}, \citenamefont {Hendrickx}, \citenamefont {Rimbach-Russ}, \citenamefont {Sammak}, \citenamefont {Scappucci}, \citenamefont {D{\'e}prez},\ and\ \citenamefont {Veldhorst}}]{van2024coherent}%
  \BibitemOpen
  \bibfield  {author} {\bibinfo {author} {\bibfnamefont {F.}~\bibnamefont {van Riggelen-Doelman}}, \bibinfo {author} {\bibfnamefont {C.-A.}\ \bibnamefont {Wang}}, \bibinfo {author} {\bibfnamefont {S.~L.}\ \bibnamefont {de~Snoo}}, \bibinfo {author} {\bibfnamefont {W.~I.}\ \bibnamefont {Lawrie}}, \bibinfo {author} {\bibfnamefont {N.~W.}\ \bibnamefont {Hendrickx}}, \bibinfo {author} {\bibfnamefont {M.}~\bibnamefont {Rimbach-Russ}}, \bibinfo {author} {\bibfnamefont {A.}~\bibnamefont {Sammak}}, \bibinfo {author} {\bibfnamefont {G.}~\bibnamefont {Scappucci}}, \bibinfo {author} {\bibfnamefont {C.}~\bibnamefont {D{\'e}prez}},\ and\ \bibinfo {author} {\bibfnamefont {M.}~\bibnamefont {Veldhorst}},\ }\bibfield  {title} {\bibinfo {title} {Coherent spin qubit shuttling through germanium quantum dots},\ }\href {https://doi.org/10.1038/s41467-024-49358-y} {\bibfield  {journal} {\bibinfo  {journal} {Nature Communications}\ }\textbf {\bibinfo {volume} {15}},\ \bibinfo {pages} {5716} (\bibinfo {year} {2024})}\BibitemShut
  {NoStop}%
\bibitem [{\citenamefont {K{\"u}nne}\ \emph {et~al.}(2024)\citenamefont {K{\"u}nne}, \citenamefont {Willmes}, \citenamefont {Oberl{\"a}nder}, \citenamefont {Gorjaew}, \citenamefont {Teske}, \citenamefont {Bhardwaj}, \citenamefont {Beer}, \citenamefont {Kammerloher}, \citenamefont {Otten}, \citenamefont {Seidler} \emph {et~al.}}]{kunne2024spinbus}%
  \BibitemOpen
  \bibfield  {author} {\bibinfo {author} {\bibfnamefont {M.}~\bibnamefont {K{\"u}nne}}, \bibinfo {author} {\bibfnamefont {A.}~\bibnamefont {Willmes}}, \bibinfo {author} {\bibfnamefont {M.}~\bibnamefont {Oberl{\"a}nder}}, \bibinfo {author} {\bibfnamefont {C.}~\bibnamefont {Gorjaew}}, \bibinfo {author} {\bibfnamefont {J.~D.}\ \bibnamefont {Teske}}, \bibinfo {author} {\bibfnamefont {H.}~\bibnamefont {Bhardwaj}}, \bibinfo {author} {\bibfnamefont {M.}~\bibnamefont {Beer}}, \bibinfo {author} {\bibfnamefont {E.}~\bibnamefont {Kammerloher}}, \bibinfo {author} {\bibfnamefont {R.}~\bibnamefont {Otten}}, \bibinfo {author} {\bibfnamefont {I.}~\bibnamefont {Seidler}}, \emph {et~al.},\ }\bibfield  {title} {\bibinfo {title} {The spinbus architecture for scaling spin qubits with electron shuttling},\ }\href {https://doi.org/10.1038/s41467-024-49182-4} {\bibfield  {journal} {\bibinfo  {journal} {Nature Communications}\ }\textbf {\bibinfo {volume} {15}},\ \bibinfo {pages} {4977} (\bibinfo {year} {2024})}\BibitemShut {NoStop}%
\bibitem [{\citenamefont {Langrock}\ \emph {et~al.}(2023)\citenamefont {Langrock}, \citenamefont {Krzywda}, \citenamefont {Focke}, \citenamefont {Seidler}, \citenamefont {Schreiber},\ and\ \citenamefont {Cywi{\'n}ski}}]{langrock2023blueprint}%
  \BibitemOpen
  \bibfield  {author} {\bibinfo {author} {\bibfnamefont {V.}~\bibnamefont {Langrock}}, \bibinfo {author} {\bibfnamefont {J.~A.}\ \bibnamefont {Krzywda}}, \bibinfo {author} {\bibfnamefont {N.}~\bibnamefont {Focke}}, \bibinfo {author} {\bibfnamefont {I.}~\bibnamefont {Seidler}}, \bibinfo {author} {\bibfnamefont {L.~R.}\ \bibnamefont {Schreiber}},\ and\ \bibinfo {author} {\bibfnamefont {{\L}.}~\bibnamefont {Cywi{\'n}ski}},\ }\bibfield  {title} {\bibinfo {title} {Blueprint of a scalable spin qubit shuttle device for coherent mid-range qubit transfer in disordered si/sige/sio 2},\ }\href {https://doi.org/10.1103/PRXQuantum.4.020305} {\bibfield  {journal} {\bibinfo  {journal} {PRX Quantum}\ }\textbf {\bibinfo {volume} {4}},\ \bibinfo {pages} {020305} (\bibinfo {year} {2023})}\BibitemShut {NoStop}%
\bibitem [{\citenamefont {Fern{\'a}ndez-Fern{\'a}ndez}\ \emph {et~al.}(2024)\citenamefont {Fern{\'a}ndez-Fern{\'a}ndez}, \citenamefont {Ban},\ and\ \citenamefont {Platero}}]{fernandez2024flying}%
  \BibitemOpen
  \bibfield  {author} {\bibinfo {author} {\bibfnamefont {D.}~\bibnamefont {Fern{\'a}ndez-Fern{\'a}ndez}}, \bibinfo {author} {\bibfnamefont {Y.}~\bibnamefont {Ban}},\ and\ \bibinfo {author} {\bibfnamefont {G.}~\bibnamefont {Platero}},\ }\bibfield  {title} {\bibinfo {title} {Flying spin qubits in quantum dot arrays driven by spin-orbit interaction},\ }\href {https://doi.org/10.22331/q-2024-11-21-1533} {\bibfield  {journal} {\bibinfo  {journal} {Quantum}\ }\textbf {\bibinfo {volume} {8}},\ \bibinfo {pages} {1533} (\bibinfo {year} {2024})}\BibitemShut {NoStop}%
\bibitem [{\citenamefont {Ginzel}\ \emph {et~al.}(2024)\citenamefont {Ginzel}, \citenamefont {Fellner}, \citenamefont {Ertler}, \citenamefont {Schreiber}, \citenamefont {Bluhm},\ and\ \citenamefont {Lechner}}]{ginzel2024scalable}%
  \BibitemOpen
  \bibfield  {author} {\bibinfo {author} {\bibfnamefont {F.}~\bibnamefont {Ginzel}}, \bibinfo {author} {\bibfnamefont {M.}~\bibnamefont {Fellner}}, \bibinfo {author} {\bibfnamefont {C.}~\bibnamefont {Ertler}}, \bibinfo {author} {\bibfnamefont {L.~R.}\ \bibnamefont {Schreiber}}, \bibinfo {author} {\bibfnamefont {H.}~\bibnamefont {Bluhm}},\ and\ \bibinfo {author} {\bibfnamefont {W.}~\bibnamefont {Lechner}},\ }\bibfield  {title} {\bibinfo {title} {Scalable parity architecture with a shuttling-based spin qubit processor},\ }\href {https://doi.org/10.1103/PhysRevB.110.075302} {\bibfield  {journal} {\bibinfo  {journal} {Physical Review B}\ }\textbf {\bibinfo {volume} {110}},\ \bibinfo {pages} {075302} (\bibinfo {year} {2024})}\BibitemShut {NoStop}%
\bibitem [{\citenamefont {Benito}\ \emph {et~al.}(2019)\citenamefont {Benito}, \citenamefont {Croot}, \citenamefont {Adelsberger}, \citenamefont {Putz}, \citenamefont {Mi}, \citenamefont {Petta},\ and\ \citenamefont {Burkard}}]{benito2019electric}%
  \BibitemOpen
  \bibfield  {author} {\bibinfo {author} {\bibfnamefont {M.}~\bibnamefont {Benito}}, \bibinfo {author} {\bibfnamefont {X.}~\bibnamefont {Croot}}, \bibinfo {author} {\bibfnamefont {C.}~\bibnamefont {Adelsberger}}, \bibinfo {author} {\bibfnamefont {S.}~\bibnamefont {Putz}}, \bibinfo {author} {\bibfnamefont {X.}~\bibnamefont {Mi}}, \bibinfo {author} {\bibfnamefont {J.~R.}\ \bibnamefont {Petta}},\ and\ \bibinfo {author} {\bibfnamefont {G.}~\bibnamefont {Burkard}},\ }\bibfield  {title} {\bibinfo {title} {Electric-field control and noise protection of the flopping-mode spin qubit},\ }\href {https://doi.org/10.1103/PhysRevB.100.125430} {\bibfield  {journal} {\bibinfo  {journal} {Physical Review B}\ }\textbf {\bibinfo {volume} {100}},\ \bibinfo {pages} {125430} (\bibinfo {year} {2019})}\BibitemShut {NoStop}%
\bibitem [{\citenamefont {Mutter}\ and\ \citenamefont {Burkard}(2021)}]{mutter2021natural}%
  \BibitemOpen
  \bibfield  {author} {\bibinfo {author} {\bibfnamefont {P.~M.}\ \bibnamefont {Mutter}}\ and\ \bibinfo {author} {\bibfnamefont {G.}~\bibnamefont {Burkard}},\ }\bibfield  {title} {\bibinfo {title} {Natural heavy-hole flopping mode qubit in germanium},\ }\href {https://doi.org/10.1103/PhysRevResearch.3.013194} {\bibfield  {journal} {\bibinfo  {journal} {Physical Review Research}\ }\textbf {\bibinfo {volume} {3}},\ \bibinfo {pages} {013194} (\bibinfo {year} {2021})}\BibitemShut {NoStop}%
\bibitem [{\citenamefont {Froning}\ \emph {et~al.}(2021{\natexlab{b}})\citenamefont {Froning}, \citenamefont {Ran{\v{c}}i{\'c}}, \citenamefont {Het{\'e}nyi}, \citenamefont {Bosco}, \citenamefont {Rehmann}, \citenamefont {Li}, \citenamefont {Bakkers}, \citenamefont {Zwanenburg}, \citenamefont {Loss}, \citenamefont {Zumb{\"u}hl} \emph {et~al.}}]{froning2021strong}%
  \BibitemOpen
  \bibfield  {author} {\bibinfo {author} {\bibfnamefont {F.}~\bibnamefont {Froning}}, \bibinfo {author} {\bibfnamefont {M.}~\bibnamefont {Ran{\v{c}}i{\'c}}}, \bibinfo {author} {\bibfnamefont {B.}~\bibnamefont {Het{\'e}nyi}}, \bibinfo {author} {\bibfnamefont {S.}~\bibnamefont {Bosco}}, \bibinfo {author} {\bibfnamefont {M.}~\bibnamefont {Rehmann}}, \bibinfo {author} {\bibfnamefont {A.}~\bibnamefont {Li}}, \bibinfo {author} {\bibfnamefont {E.~P.}\ \bibnamefont {Bakkers}}, \bibinfo {author} {\bibfnamefont {F.~A.}\ \bibnamefont {Zwanenburg}}, \bibinfo {author} {\bibfnamefont {D.}~\bibnamefont {Loss}}, \bibinfo {author} {\bibfnamefont {D.}~\bibnamefont {Zumb{\"u}hl}}, \emph {et~al.},\ }\bibfield  {title} {\bibinfo {title} {Strong spin-orbit interaction and g-factor renormalization of hole spins in ge/si nanowire quantum dots},\ }\href {https://doi.org/10.1103/PhysRevResearch.3.013081} {\bibfield  {journal} {\bibinfo  {journal} {Physical Review Research}\ }\textbf {\bibinfo {volume} {3}},\ \bibinfo {pages} {013081}
  (\bibinfo {year} {2021}{\natexlab{b}})}\BibitemShut {NoStop}%
\bibitem [{\citenamefont {Hu}\ \emph {et~al.}(2023)\citenamefont {Hu}, \citenamefont {Ma}, \citenamefont {Ni}, \citenamefont {Zhou}, \citenamefont {Chu}, \citenamefont {Liao}, \citenamefont {Kong}, \citenamefont {Cao}, \citenamefont {Wang}, \citenamefont {Li} \emph {et~al.}}]{hu2023flopping}%
  \BibitemOpen
  \bibfield  {author} {\bibinfo {author} {\bibfnamefont {R.-Z.}\ \bibnamefont {Hu}}, \bibinfo {author} {\bibfnamefont {R.-L.}\ \bibnamefont {Ma}}, \bibinfo {author} {\bibfnamefont {M.}~\bibnamefont {Ni}}, \bibinfo {author} {\bibfnamefont {Y.}~\bibnamefont {Zhou}}, \bibinfo {author} {\bibfnamefont {N.}~\bibnamefont {Chu}}, \bibinfo {author} {\bibfnamefont {W.-Z.}\ \bibnamefont {Liao}}, \bibinfo {author} {\bibfnamefont {Z.-Z.}\ \bibnamefont {Kong}}, \bibinfo {author} {\bibfnamefont {G.}~\bibnamefont {Cao}}, \bibinfo {author} {\bibfnamefont {G.-L.}\ \bibnamefont {Wang}}, \bibinfo {author} {\bibfnamefont {H.-O.}\ \bibnamefont {Li}}, \emph {et~al.},\ }\bibfield  {title} {\bibinfo {title} {Flopping-mode spin qubit in a si-mos quantum dot},\ }\bibfield  {journal} {\bibinfo  {journal} {Applied Physics Letters}\ }\textbf {\bibinfo {volume} {122}},\ \href {https://doi.org/10.1063/5.0137259} {10.1063/5.0137259} (\bibinfo {year} {2023})\BibitemShut {NoStop}%
\bibitem [{\citenamefont {Sen}\ \emph {et~al.}(2023)\citenamefont {Sen}, \citenamefont {Frank}, \citenamefont {Kolok}, \citenamefont {Danon},\ and\ \citenamefont {P{\'a}lyi}}]{sen2023classification}%
  \BibitemOpen
  \bibfield  {author} {\bibinfo {author} {\bibfnamefont {A.}~\bibnamefont {Sen}}, \bibinfo {author} {\bibfnamefont {G.}~\bibnamefont {Frank}}, \bibinfo {author} {\bibfnamefont {B.}~\bibnamefont {Kolok}}, \bibinfo {author} {\bibfnamefont {J.}~\bibnamefont {Danon}},\ and\ \bibinfo {author} {\bibfnamefont {A.}~\bibnamefont {P{\'a}lyi}},\ }\bibfield  {title} {\bibinfo {title} {Classification and magic magnetic field directions for spin-orbit-coupled double quantum dots},\ }\href {https://doi.org/10.1103/PhysRevB.108.245406} {\bibfield  {journal} {\bibinfo  {journal} {Physical Review B}\ }\textbf {\bibinfo {volume} {108}},\ \bibinfo {pages} {245406} (\bibinfo {year} {2023})}\BibitemShut {NoStop}%
\bibitem [{\citenamefont {Stastny}\ and\ \citenamefont {Burkard}(2025)}]{stastny2025singlet}%
  \BibitemOpen
  \bibfield  {author} {\bibinfo {author} {\bibfnamefont {S.}~\bibnamefont {Stastny}}\ and\ \bibinfo {author} {\bibfnamefont {G.}~\bibnamefont {Burkard}},\ }\bibfield  {title} {\bibinfo {title} {Singlet-triplet and exchange-only flopping-mode spin qubits},\ }\href {https://doi.org/10.1103/v1b6-r6jm} {\bibfield  {journal} {\bibinfo  {journal} {PRX Quantum}\ }\textbf {\bibinfo {volume} {6}},\ \bibinfo {pages} {030360} (\bibinfo {year} {2025})}\BibitemShut {NoStop}%
\bibitem [{\citenamefont {Teske}\ \emph {et~al.}(2023)\citenamefont {Teske}, \citenamefont {Butt}, \citenamefont {Cerfontaine}, \citenamefont {Burkard},\ and\ \citenamefont {Bluhm}}]{teske2023flopping}%
  \BibitemOpen
  \bibfield  {author} {\bibinfo {author} {\bibfnamefont {J.~D.}\ \bibnamefont {Teske}}, \bibinfo {author} {\bibfnamefont {F.}~\bibnamefont {Butt}}, \bibinfo {author} {\bibfnamefont {P.}~\bibnamefont {Cerfontaine}}, \bibinfo {author} {\bibfnamefont {G.}~\bibnamefont {Burkard}},\ and\ \bibinfo {author} {\bibfnamefont {H.}~\bibnamefont {Bluhm}},\ }\bibfield  {title} {\bibinfo {title} {Flopping-mode electron dipole spin resonance in the strong-driving regime},\ }\href {https://doi.org/10.1103/PhysRevB.107.035302} {\bibfield  {journal} {\bibinfo  {journal} {Physical Review B}\ }\textbf {\bibinfo {volume} {107}},\ \bibinfo {pages} {035302} (\bibinfo {year} {2023})}\BibitemShut {NoStop}%
\bibitem [{\citenamefont {Liles}\ \emph {et~al.}(2021)\citenamefont {Liles}, \citenamefont {Martins}, \citenamefont {Miserev}, \citenamefont {Kiselev}, \citenamefont {Thorvaldson}, \citenamefont {Rendell}, \citenamefont {Jin}, \citenamefont {Hudson}, \citenamefont {Veldhorst}, \citenamefont {Itoh} \emph {et~al.}}]{liles2021electrical}%
  \BibitemOpen
  \bibfield  {author} {\bibinfo {author} {\bibfnamefont {S.}~\bibnamefont {Liles}}, \bibinfo {author} {\bibfnamefont {F.}~\bibnamefont {Martins}}, \bibinfo {author} {\bibfnamefont {D.}~\bibnamefont {Miserev}}, \bibinfo {author} {\bibfnamefont {A.}~\bibnamefont {Kiselev}}, \bibinfo {author} {\bibfnamefont {I.}~\bibnamefont {Thorvaldson}}, \bibinfo {author} {\bibfnamefont {M.}~\bibnamefont {Rendell}}, \bibinfo {author} {\bibfnamefont {I.}~\bibnamefont {Jin}}, \bibinfo {author} {\bibfnamefont {F.}~\bibnamefont {Hudson}}, \bibinfo {author} {\bibfnamefont {M.}~\bibnamefont {Veldhorst}}, \bibinfo {author} {\bibfnamefont {K.}~\bibnamefont {Itoh}}, \emph {et~al.},\ }\bibfield  {title} {\bibinfo {title} {Electrical control of the g tensor of the first hole in a silicon mos quantum dot},\ }\href {https://doi.org/10.1103/PhysRevB.104.235303} {\bibfield  {journal} {\bibinfo  {journal} {Physical Review B}\ }\textbf {\bibinfo {volume} {104}},\ \bibinfo {pages} {235303} (\bibinfo {year} {2021})}\BibitemShut {NoStop}%
\bibitem [{\citenamefont {Abadillo-Uriel}\ \emph {et~al.}(2023)\citenamefont {Abadillo-Uriel}, \citenamefont {Rodr{\'\i}guez-Mena}, \citenamefont {Martinez},\ and\ \citenamefont {Niquet}}]{abadillo2023hole}%
  \BibitemOpen
  \bibfield  {author} {\bibinfo {author} {\bibfnamefont {J.~C.}\ \bibnamefont {Abadillo-Uriel}}, \bibinfo {author} {\bibfnamefont {E.~A.}\ \bibnamefont {Rodr{\'\i}guez-Mena}}, \bibinfo {author} {\bibfnamefont {B.}~\bibnamefont {Martinez}},\ and\ \bibinfo {author} {\bibfnamefont {Y.-M.}\ \bibnamefont {Niquet}},\ }\bibfield  {title} {\bibinfo {title} {Hole-spin driving by strain-induced spin-orbit interactions},\ }\href {https://doi.org/10.1103/PhysRevLett.131.097002} {\bibfield  {journal} {\bibinfo  {journal} {Physical Review Letters}\ }\textbf {\bibinfo {volume} {131}},\ \bibinfo {pages} {097002} (\bibinfo {year} {2023})}\BibitemShut {NoStop}%
\bibitem [{\citenamefont {Rodr{\'\i}guez-Mena}\ \emph {et~al.}(2023)\citenamefont {Rodr{\'\i}guez-Mena}, \citenamefont {Abadillo-Uriel}, \citenamefont {Veste}, \citenamefont {Martinez}, \citenamefont {Li}, \citenamefont {Skl{\'e}nard},\ and\ \citenamefont {Niquet}}]{rodriguez2023linear}%
  \BibitemOpen
  \bibfield  {author} {\bibinfo {author} {\bibfnamefont {E.~A.}\ \bibnamefont {Rodr{\'\i}guez-Mena}}, \bibinfo {author} {\bibfnamefont {J.~C.}\ \bibnamefont {Abadillo-Uriel}}, \bibinfo {author} {\bibfnamefont {G.}~\bibnamefont {Veste}}, \bibinfo {author} {\bibfnamefont {B.}~\bibnamefont {Martinez}}, \bibinfo {author} {\bibfnamefont {J.}~\bibnamefont {Li}}, \bibinfo {author} {\bibfnamefont {B.}~\bibnamefont {Skl{\'e}nard}},\ and\ \bibinfo {author} {\bibfnamefont {Y.-M.}\ \bibnamefont {Niquet}},\ }\bibfield  {title} {\bibinfo {title} {Linear-in-momentum spin orbit interactions in planar ge/gesi heterostructures and spin qubits},\ }\href {https://doi.org/10.1103/PhysRevB.108.205416} {\bibfield  {journal} {\bibinfo  {journal} {Physical Review B}\ }\textbf {\bibinfo {volume} {108}},\ \bibinfo {pages} {205416} (\bibinfo {year} {2023})}\BibitemShut {NoStop}%
\bibitem [{\citenamefont {Martinez}\ \emph {et~al.}(2022)\citenamefont {Martinez}, \citenamefont {Abadillo-Uriel}, \citenamefont {Rodr{\'\i}guez-Mena},\ and\ \citenamefont {Niquet}}]{martinez2022hole}%
  \BibitemOpen
  \bibfield  {author} {\bibinfo {author} {\bibfnamefont {B.}~\bibnamefont {Martinez}}, \bibinfo {author} {\bibfnamefont {J.~C.}\ \bibnamefont {Abadillo-Uriel}}, \bibinfo {author} {\bibfnamefont {E.~A.}\ \bibnamefont {Rodr{\'\i}guez-Mena}},\ and\ \bibinfo {author} {\bibfnamefont {Y.-M.}\ \bibnamefont {Niquet}},\ }\bibfield  {title} {\bibinfo {title} {Hole spin manipulation in inhomogeneous and nonseparable electric fields},\ }\href {https://doi.org/10.1103/PhysRevB.106.235426} {\bibfield  {journal} {\bibinfo  {journal} {Physical Review B}\ }\textbf {\bibinfo {volume} {106}},\ \bibinfo {pages} {235426} (\bibinfo {year} {2022})}\BibitemShut {NoStop}%
\bibitem [{\citenamefont {Martinez}\ and\ \citenamefont {Niquet}(2026)}]{martinez2025variability}%
  \BibitemOpen
  \bibfield  {author} {\bibinfo {author} {\bibfnamefont {B.}~\bibnamefont {Martinez}}\ and\ \bibinfo {author} {\bibfnamefont {Y.-M.}\ \bibnamefont {Niquet}},\ }\bibfield  {title} {\bibinfo {title} {Variability of hole-spin qubits in planar germanium},\ }\href {https://doi.org/10.1103/mtky-93p1} {\bibfield  {journal} {\bibinfo  {journal} {Phys. Rev. Appl.}\ }\textbf {\bibinfo {volume} {25}},\ \bibinfo {pages} {014018} (\bibinfo {year} {2026})}\BibitemShut {NoStop}%
\bibitem [{\citenamefont {Mi}\ \emph {et~al.}(2018)\citenamefont {Mi}, \citenamefont {Benito}, \citenamefont {Putz}, \citenamefont {Zajac}, \citenamefont {Taylor}, \citenamefont {Burkard},\ and\ \citenamefont {Petta}}]{mi2018coherent}%
  \BibitemOpen
  \bibfield  {author} {\bibinfo {author} {\bibfnamefont {X.}~\bibnamefont {Mi}}, \bibinfo {author} {\bibfnamefont {M.}~\bibnamefont {Benito}}, \bibinfo {author} {\bibfnamefont {S.}~\bibnamefont {Putz}}, \bibinfo {author} {\bibfnamefont {D.~M.}\ \bibnamefont {Zajac}}, \bibinfo {author} {\bibfnamefont {J.~M.}\ \bibnamefont {Taylor}}, \bibinfo {author} {\bibfnamefont {G.}~\bibnamefont {Burkard}},\ and\ \bibinfo {author} {\bibfnamefont {J.~R.}\ \bibnamefont {Petta}},\ }\bibfield  {title} {\bibinfo {title} {A coherent spin--photon interface in silicon},\ }\href {https://doi.org/10.1038/nature25769} {\bibfield  {journal} {\bibinfo  {journal} {Nature}\ }\textbf {\bibinfo {volume} {555}},\ \bibinfo {pages} {599} (\bibinfo {year} {2018})}\BibitemShut {NoStop}%
\bibitem [{\citenamefont {Samkharadze}\ \emph {et~al.}(2018)\citenamefont {Samkharadze}, \citenamefont {Zheng}, \citenamefont {Kalhor}, \citenamefont {Brousse}, \citenamefont {Sammak}, \citenamefont {Mendes}, \citenamefont {Blais}, \citenamefont {Scappucci},\ and\ \citenamefont {Vandersypen}}]{samkharadze2018strong}%
  \BibitemOpen
  \bibfield  {author} {\bibinfo {author} {\bibfnamefont {N.}~\bibnamefont {Samkharadze}}, \bibinfo {author} {\bibfnamefont {G.}~\bibnamefont {Zheng}}, \bibinfo {author} {\bibfnamefont {N.}~\bibnamefont {Kalhor}}, \bibinfo {author} {\bibfnamefont {D.}~\bibnamefont {Brousse}}, \bibinfo {author} {\bibfnamefont {A.}~\bibnamefont {Sammak}}, \bibinfo {author} {\bibfnamefont {U.}~\bibnamefont {Mendes}}, \bibinfo {author} {\bibfnamefont {A.}~\bibnamefont {Blais}}, \bibinfo {author} {\bibfnamefont {G.}~\bibnamefont {Scappucci}},\ and\ \bibinfo {author} {\bibfnamefont {L.}~\bibnamefont {Vandersypen}},\ }\bibfield  {title} {\bibinfo {title} {Strong spin-photon coupling in silicon},\ }\href {https://doi.org/10.1126/science.aar4054} {\bibfield  {journal} {\bibinfo  {journal} {Science}\ }\textbf {\bibinfo {volume} {359}},\ \bibinfo {pages} {1123} (\bibinfo {year} {2018})}\BibitemShut {NoStop}%
\bibitem [{\citenamefont {Yu}\ \emph {et~al.}(2023)\citenamefont {Yu}, \citenamefont {Zihlmann}, \citenamefont {Abadillo-Uriel}, \citenamefont {Michal}, \citenamefont {Rambal}, \citenamefont {Niebojewski}, \citenamefont {Bedecarrats}, \citenamefont {Vinet}, \citenamefont {Dumur}, \citenamefont {Filippone} \emph {et~al.}}]{yu2023strong}%
  \BibitemOpen
  \bibfield  {author} {\bibinfo {author} {\bibfnamefont {C.~X.}\ \bibnamefont {Yu}}, \bibinfo {author} {\bibfnamefont {S.}~\bibnamefont {Zihlmann}}, \bibinfo {author} {\bibfnamefont {J.~C.}\ \bibnamefont {Abadillo-Uriel}}, \bibinfo {author} {\bibfnamefont {V.~P.}\ \bibnamefont {Michal}}, \bibinfo {author} {\bibfnamefont {N.}~\bibnamefont {Rambal}}, \bibinfo {author} {\bibfnamefont {H.}~\bibnamefont {Niebojewski}}, \bibinfo {author} {\bibfnamefont {T.}~\bibnamefont {Bedecarrats}}, \bibinfo {author} {\bibfnamefont {M.}~\bibnamefont {Vinet}}, \bibinfo {author} {\bibfnamefont {{\'E}.}~\bibnamefont {Dumur}}, \bibinfo {author} {\bibfnamefont {M.}~\bibnamefont {Filippone}}, \emph {et~al.},\ }\bibfield  {title} {\bibinfo {title} {Strong coupling between a photon and a hole spin in silicon},\ }\href {https://doi.org/10.1038/s41565-023-01332-3} {\bibfield  {journal} {\bibinfo  {journal} {Nature Nanotechnology}\ }\textbf {\bibinfo {volume} {18}},\ \bibinfo {pages} {741} (\bibinfo {year} {2023})}\BibitemShut {NoStop}%
\bibitem [{\citenamefont {Noirot}\ \emph {et~al.}(2025)\citenamefont {Noirot}, \citenamefont {Yu}, \citenamefont {Abadillo-Uriel}, \citenamefont {Dumur}, \citenamefont {Niebojewski}, \citenamefont {Bertrand}, \citenamefont {Maurand},\ and\ \citenamefont {Zihlmann}}]{noirot2025coherence}%
  \BibitemOpen
  \bibfield  {author} {\bibinfo {author} {\bibfnamefont {L.}~\bibnamefont {Noirot}}, \bibinfo {author} {\bibfnamefont {C.~X.}\ \bibnamefont {Yu}}, \bibinfo {author} {\bibfnamefont {J.~C.}\ \bibnamefont {Abadillo-Uriel}}, \bibinfo {author} {\bibfnamefont {{\'E}.}~\bibnamefont {Dumur}}, \bibinfo {author} {\bibfnamefont {H.}~\bibnamefont {Niebojewski}}, \bibinfo {author} {\bibfnamefont {B.}~\bibnamefont {Bertrand}}, \bibinfo {author} {\bibfnamefont {R.}~\bibnamefont {Maurand}},\ and\ \bibinfo {author} {\bibfnamefont {S.}~\bibnamefont {Zihlmann}},\ }\bibfield  {title} {\bibinfo {title} {Coherence of a hole spin flopping-mode qubit in a circuit quantum electrodynamics environment},\ }\href {https://arxiv.org/abs/2503.10788} {\bibfield  {journal} {\bibinfo  {journal} {arXiv preprint arXiv:2503.10788}\ } (\bibinfo {year} {2025})}\BibitemShut {NoStop}%
\bibitem [{\citenamefont {Dijkema}\ \emph {et~al.}(2025)\citenamefont {Dijkema}, \citenamefont {Xue}, \citenamefont {Harvey-Collard}, \citenamefont {Rimbach-Russ}, \citenamefont {de~Snoo}, \citenamefont {Zheng}, \citenamefont {Sammak}, \citenamefont {Scappucci},\ and\ \citenamefont {Vandersypen}}]{dijkema2025cavity}%
  \BibitemOpen
  \bibfield  {author} {\bibinfo {author} {\bibfnamefont {J.}~\bibnamefont {Dijkema}}, \bibinfo {author} {\bibfnamefont {X.}~\bibnamefont {Xue}}, \bibinfo {author} {\bibfnamefont {P.}~\bibnamefont {Harvey-Collard}}, \bibinfo {author} {\bibfnamefont {M.}~\bibnamefont {Rimbach-Russ}}, \bibinfo {author} {\bibfnamefont {S.~L.}\ \bibnamefont {de~Snoo}}, \bibinfo {author} {\bibfnamefont {G.}~\bibnamefont {Zheng}}, \bibinfo {author} {\bibfnamefont {A.}~\bibnamefont {Sammak}}, \bibinfo {author} {\bibfnamefont {G.}~\bibnamefont {Scappucci}},\ and\ \bibinfo {author} {\bibfnamefont {L.~M.}\ \bibnamefont {Vandersypen}},\ }\bibfield  {title} {\bibinfo {title} {Cavity-mediated iswap oscillations between distant spins},\ }\href {https://doi.org/10.1038/s41567-024-02694-8} {\bibfield  {journal} {\bibinfo  {journal} {Nature Physics}\ }\textbf {\bibinfo {volume} {21}},\ \bibinfo {pages} {168} (\bibinfo {year} {2025})}\BibitemShut {NoStop}%
\bibitem [{\citenamefont {Reed}\ \emph {et~al.}(2016)\citenamefont {Reed}, \citenamefont {Maune}, \citenamefont {Andrews}, \citenamefont {Borselli}, \citenamefont {Eng}, \citenamefont {Jura}, \citenamefont {Kiselev}, \citenamefont {Ladd}, \citenamefont {Merkel}, \citenamefont {Milosavljevic} \emph {et~al.}}]{reed2016reduced}%
  \BibitemOpen
  \bibfield  {author} {\bibinfo {author} {\bibfnamefont {M.}~\bibnamefont {Reed}}, \bibinfo {author} {\bibfnamefont {B.}~\bibnamefont {Maune}}, \bibinfo {author} {\bibfnamefont {R.}~\bibnamefont {Andrews}}, \bibinfo {author} {\bibfnamefont {M.}~\bibnamefont {Borselli}}, \bibinfo {author} {\bibfnamefont {K.}~\bibnamefont {Eng}}, \bibinfo {author} {\bibfnamefont {M.}~\bibnamefont {Jura}}, \bibinfo {author} {\bibfnamefont {A.}~\bibnamefont {Kiselev}}, \bibinfo {author} {\bibfnamefont {T.}~\bibnamefont {Ladd}}, \bibinfo {author} {\bibfnamefont {S.}~\bibnamefont {Merkel}}, \bibinfo {author} {\bibfnamefont {I.}~\bibnamefont {Milosavljevic}}, \emph {et~al.},\ }\bibfield  {title} {\bibinfo {title} {Reduced sensitivity to charge noise in semiconductor spin qubits via symmetric operation},\ }\href {https://doi.org/10.1103/PhysRevLett.116.110402} {\bibfield  {journal} {\bibinfo  {journal} {Physical review letters}\ }\textbf {\bibinfo {volume} {116}},\ \bibinfo {pages} {110402} (\bibinfo {year} {2016})}\BibitemShut
  {NoStop}%
\bibitem [{\citenamefont {Michal}\ \emph {et~al.}(2023)\citenamefont {Michal}, \citenamefont {Abadillo-Uriel}, \citenamefont {Zihlmann}, \citenamefont {Maurand}, \citenamefont {Niquet},\ and\ \citenamefont {Filippone}}]{michal2023tunable}%
  \BibitemOpen
  \bibfield  {author} {\bibinfo {author} {\bibfnamefont {V.}~\bibnamefont {Michal}}, \bibinfo {author} {\bibfnamefont {J.}~\bibnamefont {Abadillo-Uriel}}, \bibinfo {author} {\bibfnamefont {S.}~\bibnamefont {Zihlmann}}, \bibinfo {author} {\bibfnamefont {R.}~\bibnamefont {Maurand}}, \bibinfo {author} {\bibfnamefont {Y.-M.}\ \bibnamefont {Niquet}},\ and\ \bibinfo {author} {\bibfnamefont {M.}~\bibnamefont {Filippone}},\ }\bibfield  {title} {\bibinfo {title} {Tunable hole spin-photon interaction based on g-matrix modulation},\ }\href {https://doi.org/10.1103/PhysRevB.107.L041303} {\bibfield  {journal} {\bibinfo  {journal} {Physical Review B}\ }\textbf {\bibinfo {volume} {107}},\ \bibinfo {pages} {L041303} (\bibinfo {year} {2023})}\BibitemShut {NoStop}%
\bibitem [{\citenamefont {Michal}\ \emph {et~al.}(2021)\citenamefont {Michal}, \citenamefont {Venitucci},\ and\ \citenamefont {Niquet}}]{michal2021longitudinal}%
  \BibitemOpen
  \bibfield  {author} {\bibinfo {author} {\bibfnamefont {V.~P.}\ \bibnamefont {Michal}}, \bibinfo {author} {\bibfnamefont {B.}~\bibnamefont {Venitucci}},\ and\ \bibinfo {author} {\bibfnamefont {Y.-M.}\ \bibnamefont {Niquet}},\ }\bibfield  {title} {\bibinfo {title} {Longitudinal and transverse electric field manipulation of hole spin-orbit qubits in one-dimensional channels},\ }\href {https://doi.org/10.1103/PhysRevB.103.045305} {\bibfield  {journal} {\bibinfo  {journal} {Physical Review B}\ }\textbf {\bibinfo {volume} {103}},\ \bibinfo {pages} {045305} (\bibinfo {year} {2021})}\BibitemShut {NoStop}%
\bibitem [{\citenamefont {Bosco}\ \emph {et~al.}(2021)\citenamefont {Bosco}, \citenamefont {Benito}, \citenamefont {Adelsberger},\ and\ \citenamefont {Loss}}]{bosco2021squeezed}%
  \BibitemOpen
  \bibfield  {author} {\bibinfo {author} {\bibfnamefont {S.}~\bibnamefont {Bosco}}, \bibinfo {author} {\bibfnamefont {M.}~\bibnamefont {Benito}}, \bibinfo {author} {\bibfnamefont {C.}~\bibnamefont {Adelsberger}},\ and\ \bibinfo {author} {\bibfnamefont {D.}~\bibnamefont {Loss}},\ }\bibfield  {title} {\bibinfo {title} {Squeezed hole spin qubits in ge quantum dots with ultrafast gates at low power},\ }\href {https://doi.org/10.1103/PhysRevB.104.115425} {\bibfield  {journal} {\bibinfo  {journal} {Physical Review B}\ }\textbf {\bibinfo {volume} {104}},\ \bibinfo {pages} {115425} (\bibinfo {year} {2021})}\BibitemShut {NoStop}%
\bibitem [{\citenamefont {Wang}\ \emph {et~al.}(2024{\natexlab{b}})\citenamefont {Wang}, \citenamefont {Sarkar}, \citenamefont {Liles}, \citenamefont {Saraiva}, \citenamefont {Dzurak}, \citenamefont {Hamilton},\ and\ \citenamefont {Culcer}}]{wang2024electrical}%
  \BibitemOpen
  \bibfield  {author} {\bibinfo {author} {\bibfnamefont {Z.}~\bibnamefont {Wang}}, \bibinfo {author} {\bibfnamefont {A.}~\bibnamefont {Sarkar}}, \bibinfo {author} {\bibfnamefont {S.}~\bibnamefont {Liles}}, \bibinfo {author} {\bibfnamefont {A.}~\bibnamefont {Saraiva}}, \bibinfo {author} {\bibfnamefont {A.}~\bibnamefont {Dzurak}}, \bibinfo {author} {\bibfnamefont {A.}~\bibnamefont {Hamilton}},\ and\ \bibinfo {author} {\bibfnamefont {D.}~\bibnamefont {Culcer}},\ }\bibfield  {title} {\bibinfo {title} {Electrical operation of hole spin qubits in planar mos silicon quantum dots},\ }\href {https://doi.org/10.1103/PhysRevB.109.075427} {\bibfield  {journal} {\bibinfo  {journal} {Physical Review B}\ }\textbf {\bibinfo {volume} {109}},\ \bibinfo {pages} {075427} (\bibinfo {year} {2024}{\natexlab{b}})}\BibitemShut {NoStop}%
\bibitem [{pse()}]{pseudofootnote}%
  \BibitemOpen
  \href@noop {} {}\bibinfo {note} {The localized $\hat{g}$-matrices of Fig.~\ref{fig:intro} are given by $\hat{g}_\mathrm{L}=\begin{pmatrix} 0.87 & 0.49 & 0 \\ -1.07 & 1.9 & 0 \\ 0 & 0 & g_{\mathrm{L},zz} \end{pmatrix}$ and $\hat{g}_\mathrm{R}=\begin{pmatrix} 0.86 & 0.33 & 0 \\ -0.81 & 2.1 & 0 \\ 0 & 0 & g_{\mathrm{R},zz} \end{pmatrix}$ where the vertical $g$-factors could not be determined experimentally~\cite{yu2023strong}.}\BibitemShut {Stop}%
\bibitem [{Note1()}]{Note1}%
  \BibitemOpen
  \bibinfo {note} {We assume here $\protect \mathcal {T}\mathinner {|{\uparrow }\rangle }=-\mathinner {|{\downarrow }\rangle }$ and $\protect \mathcal {T}\mathinner {|{\downarrow }\rangle }=\mathinner {|{\uparrow }\rangle }$, where $\protect \mathcal {T}$ is the time-reversal symmetry operator.}\BibitemShut {Stop}%
\bibitem [{Note2()}]{Note2}%
  \BibitemOpen
  \bibinfo {note} {We assume that the Hamiltonian transforms under the time-reversal symmetry as $\protect \mathcal {T}^{-1}H_\protect \text {eff}(\protect \mathbf {B})\protect \mathcal {T}=H_\protect \text {eff}(\protect \mathbf {-B})$, which implies that the vector potential $\protect \mathbf {A}$ must be odd with respect to the magnetic field $\protect \mathbf {B}$ [$\protect \mathbf {A}(\protect \mathbf {r}, -\protect \mathbf {B})=-\protect \mathbf {A}(\protect \mathbf {r}, \protect \mathbf {B})$]. This is the case for all usual gauge choices (symmetric, Landau, ...). We emphasize that this constraint on the gauge simplifies the shape of Eq.~\protect \eqref {eq:eff0} with no observable effect.}\BibitemShut {Stop}%
\bibitem [{Note3()}]{Note3}%
  \BibitemOpen
  \bibinfo {note} {Microscopically, $\protect \bm {\mu }_\protect \mathrm {T}$ may arise as a Peierls phase $t_c\rightarrow t_c\exp (i\protect \frac {e}{\hbar }\DOTSI \intop \ilimits@ ^{\protect \mathbf {x}_1}_{\protect \mathbf {x}_0} \protect \mathbf {A}\cdot d\protect \mathbf {r})$ expanded up to first order in $\protect \mathbf {B}$ in our linear-response theory, where $\protect \mathbf {A}$ is the vector potential, see Appendix~\ref {appendix:microscopic}}\BibitemShut {NoStop}%
\bibitem [{\citenamefont {Winkler}\ \emph {et~al.}(2001)\citenamefont {Winkler}, \citenamefont {Papadakis}, \citenamefont {De~Poortere},\ and\ \citenamefont {Shayegan}}]{winkler2001spin}%
  \BibitemOpen
  \bibfield  {author} {\bibinfo {author} {\bibfnamefont {R.}~\bibnamefont {Winkler}}, \bibinfo {author} {\bibfnamefont {S.}~\bibnamefont {Papadakis}}, \bibinfo {author} {\bibfnamefont {E.}~\bibnamefont {De~Poortere}},\ and\ \bibinfo {author} {\bibfnamefont {M.}~\bibnamefont {Shayegan}},\ }\bibfield  {title} {\bibinfo {title} {Spin-orbit coupling in two-dimensional electron and hole systems},\ }in\ \href@noop {} {\emph {\bibinfo {booktitle} {Advances in Solid State Physics}}}\ (\bibinfo  {publisher} {Springer},\ \bibinfo {year} {2001})\ pp.\ \bibinfo {pages} {211--223}\BibitemShut {NoStop}%
\bibitem [{\citenamefont {Venitucci}\ \emph {et~al.}(2018)\citenamefont {Venitucci}, \citenamefont {Bourdet}, \citenamefont {Pouzada},\ and\ \citenamefont {Niquet}}]{venitucci2018electrical}%
  \BibitemOpen
  \bibfield  {author} {\bibinfo {author} {\bibfnamefont {B.}~\bibnamefont {Venitucci}}, \bibinfo {author} {\bibfnamefont {L.}~\bibnamefont {Bourdet}}, \bibinfo {author} {\bibfnamefont {D.}~\bibnamefont {Pouzada}},\ and\ \bibinfo {author} {\bibfnamefont {Y.-M.}\ \bibnamefont {Niquet}},\ }\bibfield  {title} {\bibinfo {title} {Electrical manipulation of semiconductor spin qubits within the g-matrix formalism},\ }\href {https://doi.org/10.1103/PhysRevB.98.155319} {\bibfield  {journal} {\bibinfo  {journal} {Physical Review B}\ }\textbf {\bibinfo {volume} {98}},\ \bibinfo {pages} {155319} (\bibinfo {year} {2018})}\BibitemShut {NoStop}%
\bibitem [{Note4()}]{Note4}%
  \BibitemOpen
  \bibinfo {note} {Note that when we mention rotations about a spin axis, we are implicitly fixing a specific spin basis set. The observables, such as the effective $g$-factors of the eigenstates, remain the same independently of this choice.}\BibitemShut {Stop}%
\bibitem [{\citenamefont {Mauro}\ \emph {et~al.}(2025)\citenamefont {Mauro}, \citenamefont {Rodr{\'\i}guez-Mena}, \citenamefont {Martinez},\ and\ \citenamefont {Niquet}}]{mauro2025strain}%
  \BibitemOpen
  \bibfield  {author} {\bibinfo {author} {\bibfnamefont {L.}~\bibnamefont {Mauro}}, \bibinfo {author} {\bibfnamefont {E.~A.}\ \bibnamefont {Rodr{\'\i}guez-Mena}}, \bibinfo {author} {\bibfnamefont {B.}~\bibnamefont {Martinez}},\ and\ \bibinfo {author} {\bibfnamefont {Y.-M.}\ \bibnamefont {Niquet}},\ }\bibfield  {title} {\bibinfo {title} {Strain engineering in ge/ge-si spin-qubit heterostructures},\ }\href {https://doi.org/10.1103/PhysRevApplied.23.024057} {\bibfield  {journal} {\bibinfo  {journal} {Physical Review Applied}\ }\textbf {\bibinfo {volume} {23}},\ \bibinfo {pages} {024057} (\bibinfo {year} {2025})}\BibitemShut {NoStop}%
\bibitem [{\citenamefont {Kolok}\ and\ \citenamefont {P{\'a}lyi}(2024)}]{kolok2024protocols}%
  \BibitemOpen
  \bibfield  {author} {\bibinfo {author} {\bibfnamefont {B.}~\bibnamefont {Kolok}}\ and\ \bibinfo {author} {\bibfnamefont {A.}~\bibnamefont {P{\'a}lyi}},\ }\bibfield  {title} {\bibinfo {title} {Protocols to measure the non-abelian berry phase by pumping a spin qubit through a quantum-dot loop},\ }\href {https://doi.org/10.1103/PhysRevB.109.045438} {\bibfield  {journal} {\bibinfo  {journal} {Physical Review B}\ }\textbf {\bibinfo {volume} {109}},\ \bibinfo {pages} {045438} (\bibinfo {year} {2024})}\BibitemShut {NoStop}%
\bibitem [{\citenamefont {Lin}\ \emph {et~al.}(2026)\citenamefont {Lin}, \citenamefont {Steinacker}, \citenamefont {Feng}, \citenamefont {Dash}, \citenamefont {Serrano}, \citenamefont {Lim}, \citenamefont {Itoh}, \citenamefont {Hudson}, \citenamefont {Tanttu}, \citenamefont {Saraiva}, \citenamefont {Laucht}, \citenamefont {Dzurak}, \citenamefont {Goan},\ and\ \citenamefont {Yang}}]{lin2025interplayzeemansplittingtunnel}%
  \BibitemOpen
  \bibfield  {author} {\bibinfo {author} {\bibfnamefont {S.-C.}\ \bibnamefont {Lin}}, \bibinfo {author} {\bibfnamefont {P.}~\bibnamefont {Steinacker}}, \bibinfo {author} {\bibfnamefont {M.}~\bibnamefont {Feng}}, \bibinfo {author} {\bibfnamefont {A.}~\bibnamefont {Dash}}, \bibinfo {author} {\bibfnamefont {S.}~\bibnamefont {Serrano}}, \bibinfo {author} {\bibfnamefont {W.~H.}\ \bibnamefont {Lim}}, \bibinfo {author} {\bibfnamefont {K.~M.}\ \bibnamefont {Itoh}}, \bibinfo {author} {\bibfnamefont {F.~E.}\ \bibnamefont {Hudson}}, \bibinfo {author} {\bibfnamefont {T.}~\bibnamefont {Tanttu}}, \bibinfo {author} {\bibfnamefont {A.}~\bibnamefont {Saraiva}}, \bibinfo {author} {\bibfnamefont {A.}~\bibnamefont {Laucht}}, \bibinfo {author} {\bibfnamefont {A.~S.}\ \bibnamefont {Dzurak}}, \bibinfo {author} {\bibfnamefont {H.-S.}\ \bibnamefont {Goan}},\ and\ \bibinfo {author} {\bibfnamefont {C.~H.}\ \bibnamefont {Yang}},\ }\bibfield  {title} {\bibinfo {title} {Interplay of zeeman splitting and tunnel coupling in coherent
  spin-qubit shuttling},\ }\href {https://doi.org/10.1103/3d1t-pr7m} {\bibfield  {journal} {\bibinfo  {journal} {Phys. Rev. Appl.}\ ,\ } (\bibinfo {year} {2026})}\BibitemShut {NoStop}%
\bibitem [{\citenamefont {Gu{\'e}ry-Odelin}\ \emph {et~al.}(2019)\citenamefont {Gu{\'e}ry-Odelin}, \citenamefont {Ruschhaupt}, \citenamefont {Kiely}, \citenamefont {Torrontegui}, \citenamefont {Mart{\'\i}nez-Garaot},\ and\ \citenamefont {Muga}}]{guery2019shortcuts}%
  \BibitemOpen
  \bibfield  {author} {\bibinfo {author} {\bibfnamefont {D.}~\bibnamefont {Gu{\'e}ry-Odelin}}, \bibinfo {author} {\bibfnamefont {A.}~\bibnamefont {Ruschhaupt}}, \bibinfo {author} {\bibfnamefont {A.}~\bibnamefont {Kiely}}, \bibinfo {author} {\bibfnamefont {E.}~\bibnamefont {Torrontegui}}, \bibinfo {author} {\bibfnamefont {S.}~\bibnamefont {Mart{\'\i}nez-Garaot}},\ and\ \bibinfo {author} {\bibfnamefont {J.~G.}\ \bibnamefont {Muga}},\ }\bibfield  {title} {\bibinfo {title} {Shortcuts to adiabaticity: Concepts, methods, and applications},\ }\href@noop {} {\bibfield  {journal} {\bibinfo  {journal} {Reviews of Modern Physics}\ }\textbf {\bibinfo {volume} {91}},\ \bibinfo {pages} {045001} (\bibinfo {year} {2019})}\BibitemShut {NoStop}%
\bibitem [{\citenamefont {Bosco}\ \emph {et~al.}(2024)\citenamefont {Bosco}, \citenamefont {Zou},\ and\ \citenamefont {Loss}}]{bosco2024high}%
  \BibitemOpen
  \bibfield  {author} {\bibinfo {author} {\bibfnamefont {S.}~\bibnamefont {Bosco}}, \bibinfo {author} {\bibfnamefont {J.}~\bibnamefont {Zou}},\ and\ \bibinfo {author} {\bibfnamefont {D.}~\bibnamefont {Loss}},\ }\bibfield  {title} {\bibinfo {title} {High-fidelity spin qubit shuttling via large spin-orbit interactions},\ }\href {https://doi.org/10.1103/PRXQuantum.5.020353} {\bibfield  {journal} {\bibinfo  {journal} {PRX Quantum}\ }\textbf {\bibinfo {volume} {5}},\ \bibinfo {pages} {020353} (\bibinfo {year} {2024})}\BibitemShut {NoStop}%
\bibitem [{Note5()}]{Note5}%
  \BibitemOpen
  \bibinfo {note} {While the harmonic length depends on whether we have a heavy-hole or light-hole state through their in-plane effective masses $m^\parallel _{h,l}$, we find the difference to have no qualitative impact in our results except for complicating the expressions. Therefore, we approximate $L_x=L_x^{(h)}\approx L_x^{(l)}$ in the manuscript}\BibitemShut {NoStop}%
\bibitem [{Note6()}]{Note6}%
  \BibitemOpen
  \bibinfo {note} {We go beyond this approximation to include detuning-dependent corrections.}\BibitemShut {Stop}%
\bibitem [{\citenamefont {Graf}\ and\ \citenamefont {Vogl}(1995)}]{graf1995electromagnetic}%
  \BibitemOpen
  \bibfield  {author} {\bibinfo {author} {\bibfnamefont {M.}~\bibnamefont {Graf}}\ and\ \bibinfo {author} {\bibfnamefont {P.}~\bibnamefont {Vogl}},\ }\bibfield  {title} {\bibinfo {title} {Electromagnetic fields and dielectric response in empirical tight-binding theory},\ }\href {https://doi.org/10.1103/PhysRevB.51.4940} {\bibfield  {journal} {\bibinfo  {journal} {Physical Review B}\ }\textbf {\bibinfo {volume} {51}},\ \bibinfo {pages} {4940} (\bibinfo {year} {1995})}\BibitemShut {NoStop}%
\bibitem [{\citenamefont {Ismail-Beigi}\ \emph {et~al.}(2001)\citenamefont {Ismail-Beigi}, \citenamefont {Chang},\ and\ \citenamefont {Louie}}]{ismail2001coupling}%
  \BibitemOpen
  \bibfield  {author} {\bibinfo {author} {\bibfnamefont {S.}~\bibnamefont {Ismail-Beigi}}, \bibinfo {author} {\bibfnamefont {E.~K.}\ \bibnamefont {Chang}},\ and\ \bibinfo {author} {\bibfnamefont {S.~G.}\ \bibnamefont {Louie}},\ }\bibfield  {title} {\bibinfo {title} {Coupling of nonlocal potentials to electromagnetic fields},\ }\href {https://doi.org/10.1103/PhysRevLett.87.087402} {\bibfield  {journal} {\bibinfo  {journal} {Physical Review Letters}\ }\textbf {\bibinfo {volume} {87}},\ \bibinfo {pages} {087402} (\bibinfo {year} {2001})}\BibitemShut {NoStop}%
\bibitem [{\citenamefont {Voisin}\ \emph {et~al.}(2014)\citenamefont {Voisin}, \citenamefont {Nguyen}, \citenamefont {Renard}, \citenamefont {Jehl}, \citenamefont {Barraud}, \citenamefont {Triozon}, \citenamefont {Vinet}, \citenamefont {Duchemin}, \citenamefont {Niquet}, \citenamefont {de~Franceschi},\ and\ \citenamefont {Sanquer}}]{Voisin_2014}%
  \BibitemOpen
  \bibfield  {author} {\bibinfo {author} {\bibfnamefont {B.}~\bibnamefont {Voisin}}, \bibinfo {author} {\bibfnamefont {V.-H.}\ \bibnamefont {Nguyen}}, \bibinfo {author} {\bibfnamefont {J.}~\bibnamefont {Renard}}, \bibinfo {author} {\bibfnamefont {X.}~\bibnamefont {Jehl}}, \bibinfo {author} {\bibfnamefont {S.}~\bibnamefont {Barraud}}, \bibinfo {author} {\bibfnamefont {F.}~\bibnamefont {Triozon}}, \bibinfo {author} {\bibfnamefont {M.}~\bibnamefont {Vinet}}, \bibinfo {author} {\bibfnamefont {I.}~\bibnamefont {Duchemin}}, \bibinfo {author} {\bibfnamefont {Y.-M.}\ \bibnamefont {Niquet}}, \bibinfo {author} {\bibfnamefont {S.}~\bibnamefont {de~Franceschi}},\ and\ \bibinfo {author} {\bibfnamefont {M.}~\bibnamefont {Sanquer}},\ }\bibfield  {title} {\bibinfo {title} {Few-electron edge-state quantum dots in a silicon nanowire field-effect transistor},\ }\href {https://doi.org/10.1021/nl500299h} {\bibfield  {journal} {\bibinfo  {journal} {Nano Letters}\ }\textbf {\bibinfo {volume} {14}},\ \bibinfo {pages} {2094}
  (\bibinfo {year} {2014})}\BibitemShut {NoStop}%
\bibitem [{Note7()}]{Note7}%
  \BibitemOpen
  \bibinfo {note} {Specifically, $\alpha =\protect \mathrm {Tr}(D_R)/\protect \sqrt {\protect \mathrm {Tr}(D_L)^2+\protect \mathrm {Tr}(D_R)^2}$ and $\beta =\protect \mathrm {Tr}(D_L)/\protect \sqrt {\protect \mathrm {Tr}(D_L)^2+\protect \mathrm {Tr}(D_R)^2}$, where $\protect \mathrm {Tr}$ is the trace and $D_L$, $D_R$ are the matrices of $v_L$, $v_R$ in the $\{\mathinner {|{0}\rangle },\mathinner {|{1}\rangle },\mathinner {|{2}\rangle },\mathinner {|{3}\rangle }\}$ basis set.}\BibitemShut {Stop}%
\bibitem [{\citenamefont {Hu}\ and\ \citenamefont {O'Connell}(1996)}]{hu1996analytical}%
  \BibitemOpen
  \bibfield  {author} {\bibinfo {author} {\bibfnamefont {G.}~\bibnamefont {Hu}}\ and\ \bibinfo {author} {\bibfnamefont {R.~F.}\ \bibnamefont {O'Connell}},\ }\bibfield  {title} {\bibinfo {title} {Analytical inversion of symmetric tridiagonal matrices},\ }\href@noop {} {\bibfield  {journal} {\bibinfo  {journal} {Journal of Physics A: Mathematical and General}\ }\textbf {\bibinfo {volume} {29}},\ \bibinfo {pages} {1511} (\bibinfo {year} {1996})}\BibitemShut {NoStop}%
\end{thebibliography}%

\end{document}